\documentclass[12pt,a4paper]{article}
\usepackage{amssymb}
\usepackage{amsmath}
\usepackage{graphicx}

\begin{document}

\title{V-Langevin Equations, Continuous Time Random Walks and Fractional
Diffusion.}
\author{R.\ Balescu$^{\dagger}$ \footnote{%
$^{\dagger}$Radu BALESCU (July 18, 1932 - June 1$^{st.}$2006).\ This manuscript has been
dated by the author on \textbf{February 2, 2006} and later edited by his
collaborators J.H.\ Misguich, M.\ Negrea, F.\ Spineanu, M.\ Vlad and B.\
Weyssow (Universit\'{e} Libre de Bruxelles, Belgium, %
{\it bweyssow@ulb.ac.be}. 
} \\
%EndAName
Association Euratom-Etat Belge,\\
Universit\'{e} Libre de Bruxelles, CP\ 231, \\
Campus Plaine ULB, Bd du Triomphe, \\
1050 Bruxelles, Belgium}
\date{February 2, 2006.}
\maketitle

\begin{abstract}
The following question is addressed: under what conditions can a strange diffusive process, defined by a semi-dynamical V-Langevin equation or its associated Hybrid kinetic equation (HKE), be described by an equivalent purely stochastic process, defined by a Continuous Time Random Walk (CTRW) or by a Fractional Differential Equation (FDE)? More specifically, does there exist a class of V-Langevin equations with long-range (algebraic) velocity temporal correlation, that leads to a time-fractional superdiffusive process? The answer is always affirmative in one dimension. It is always negative in two dimensions: any algebraically decaying temporal velocity correlation (with a Gaussian spatial correlation) produces a normal diffusive process. General conditions relating the diffusive nature of the process to the temporal exponent of the Lagrangian velocity correlation (in Corrsin approximation) are derived.
\end{abstract}

\section{Introduction\label{Intro}}

\textbf{Strange Transport} has been the object of intense studies in recent
years (A very recent qualitative review is \cite{Klaf-Sokol}). (We use the
terminology "Strange transport" \cite{RB} rather than "Anomalous transport"
which is customary in Dynamical Systems theory, but has a different meaning
in Plasma transport theory). The concept first appeared in the theory of
stochastic processes, especially the theory of Continuous Time Random Walks (%
\textbf{CTRW)} \cite{Mont-Weiss}, \cite{Mont-Shles}.

Consider a \textit{disordered }system (\textit{e.g.} a turbulent fluid or
plasma). The position $\mathbf{x(}t)$ at time $t$ of one of its particles is
determined by its interactions with the other particles and/or with external
sources. In a strongly disordered system these interactions are modelled by
a random field.\ The statistical description of the system thus involves the
definition of an ensemble of realizations. One of the important quantities
describing the system is the \textit{mean square deviation} (\textit{MSD})
of the random variable $\mathbf{x}(t)$.\ In many cases, this is represented
asymptotically ($t\rightarrow \infty $) by a simple increasing function of
time:

\begin{equation}
\left\langle \delta x^{2}(t)\right\rangle =M~t^{\mu },  \label{eeq.1}
\end{equation}

where $\delta \mathbf{x}(t)=\mathbf{x}(t)-\left\langle \mathbf{x}%
(t)\right\rangle .$ The value of the "\textit{diffusion exponent"} $\mu $
characterizes the diffusion regime of the system\footnote{%
There exist superdiffusive regimes with $\mu >2$, but they will not be
considered here.}:

\begin{equation}
\begin{tabular}{|c|c|}
\hline
Diffusion exponent & Regime \\ \hline
$0<\mu <1$ & \ \ SUBDIFFUSIVE \ \  \\ \hline
$\mu =1$ & \ \ (NORMAL) DIFFUSIVE \ \  \\ \hline
$1<\mu \leq 2$ & \ \ SUPERDIFFUSIVE \ \  \\ \hline
\end{tabular}
\label{eeq.2}
\end{equation}

The subdiffusive and the superdiffusive regimes are the ones called
"STRANGE". It will be seen in the forthcoming sections that the concept of
"strange transport" involves much more than a simple statement about the
behaviour of the MSD.

All present theories of strange transport are of stochastic nature.\ From
the very voluminous literature we only cite here a few of the more
comprehensive, physically oriented review papers and books: \cite{Mont-Weiss}%
\textsc{\ - }\cite{Gor-Main 2005}\textsc{,} where additional references will
be found.

As stated in \cite{RB}, transport theories (e.g., for fluids or plasmas) can
be constructed on three levels.\bigskip

$\alpha $) A purely statistical mechanical theory would be based on the
kinetic equation for a set of interacting charged particles, combined with
Maxwell's equations. This would be the most fundamental description, but
becomes impossibly complicated in practice.\bigskip\ 

$\beta $) The next level of description would be a compromise, based on a
"semi-dynamical" model of particles moving according to the laws of
mechanics (Newton's equation), but under the action of a \textit{fluctuating
field}, representing the action of the turbulent environment.\ This leads to
stochastic ordinary differential equations of the Langevin\textbf{\ }type.
More specifically, we consider \textbf{V-Langevin equations }\cite{RB},%
\textbf{\ }\textit{i.e.} equations for the position of a "particle":

\begin{equation}
\frac{d\mathbf{x}(t)}{dt}=\mathbf{v}[\mathbf{x}(t),t],\ \ \ \ \ \ \ \mathbf{x%
}(0)=0  \label{eeq.3}
\end{equation}

where the right hand side represents a fluctuating velocity field, defined
by its statistical properties. We always consider a \textit{divergenceless}
velocity field:

\begin{equation}
\mathbf{\nabla \cdot v(x},t)=0.  \label{eeq.4}
\end{equation}

Associated with (\ref{eeq.3}) is a kinetic-type first-order partial
differential equation for the fluctuating particle distribution function $f(%
\mathbf{x},t)$ (whose average is the \textit{density profile}), called a 
\textbf{hybrid kinetic equation (HKE) }\cite{RB}. By definition, the
characteristics of this equation are precisely the V-Langevin equations (\ref%
{eeq.3}):

\begin{equation}
\partial _{t}f(\mathbf{x},t)+\mathbf{v(x},t)\cdot \mathbf{\nabla }f(\mathbf{x%
},t)=0.  \label{eeq.5}
\end{equation}

Due to the property (\ref{eeq.4}) this equation is equivalent to the
following:

\begin{equation}
\partial _{t}f(\mathbf{x},t)+\mathbf{\nabla \cdot }\left[ \mathbf{v(x},t)~f(%
\mathbf{x},t)\right] =0.  \label{eeq.6}
\end{equation}

This equation ensures the conservation of the number of particles, and is
rightly interpreted as a kinetic equation.\bigskip

$\gamma $) The last level of description is the level of \textbf{continuous
time random walk} \textbf{(CTRW) }theories and \textbf{fractional diffusion
equations (FDE)}, in which the deterministic dynamical laws are completely
given up, and replaced by a purely random process.\bigskip

All these concepts will be defined and used in the forthcoming sections.
These three levels are, of course, interrelated: each one of them should be
justifiable as an approximation of the more fundamental one.\bigskip

The strange transport theories associated with CTRW and with FDE have been
very successful in recent years in modelling many peculiar behaviors
observed, among other fields, in plasma and fusion physics. There is,
however, still a serious gap in justifying these purely stochastic models on
the basis of a molecular theory, \textit{i.e.}, theories of the type $\alpha 
$), or at least $\beta $).

In \cite{RB} we treated several models exhibiting strange transport, showing
that a stochastic CTRW can be associated with a semi-dynamical V-Langevin
equation under certain limiting conditions.\ All these models (diffusion in
a fluctuating electrostatic field or in a fluctuating magnetic field) were
either diffusive, or \textit{subdiffusive}. One of the purposes of the
present work is to try to determine whether there exist V-Langevin equations
leading to \textit{temporal superdiffusive behavior} and associating them
with a CTRW\ or with a fractional diffusion equation.

A first work following the same philosophy was done by West et al. \cite%
{West-Grigo}. In that paper a specially simplified model of a
one-dimensional, two-state CTRW was used.\ This means that a "particle"
performs a CTRW in one dimension, moving with a velocity that is constant in
absolute value: $\pm V$. The particle moves during a given (random) time,
after which it suddenly jumps and reverses its velocity.\ The velocity
autocorrelation function is given as an algebraically decaying function of
time. In a first step of their work, the jumps are supposed to be
instantaneous events.\ This model leads already to possible superdiffusion
of time-fractional type. The authors are, however, looking for a
superdiffusive regime of L\'{e}vy type (\textit{space-fractional diffusion)}%
.\ They then modify their model. The jump PDF and the waiting time PDF of
the CTRW are now \textit{linked} by assuming that the particles move during
their "jump" with final velocity $\pm V$: thus longer jumps are "penalized"
by longer durations ("\textit{L\'{e}vy walk}" rather than "L\'{e}vy flight", 
\cite{Metz-Klaf 2000}).

It appears to us that the limitation to two states makes the model unadapted
to describing the motion of a physical particle; it is, however, necessary
for the definition of the L\'{e}vy walk in their model. In the next section
we show that the assumption about the two states can be dropped, and the
velocity can be considered as a general random function of time: $v(t)$
[Secs. \ref{V(t)} - \ref{Algeb Correl}].

The model can be further extended by considering a \textit{velocity field}: $%
\mathbf{v}(\mathbf{x},t)$ (Secs. \ref{V-field} and following).\ But this
case leads to additional difficulties, because it requires the determination
of the \textit{Lagrangian velocity correlation}, a problem that is well
known as a basic difficulty of turbulence theory. Here we shall consider
only the crudest approximation for the latter quantity, but the application
of more precise methods that were recently developed in this field can be
envisaged.

\section{CTRW and Fractional Calculus\label{CTRW FDE}}

In the forthcoming text, the abbreviation "\textit{PDF}" denotes a
"probability distribution function"; the symbol may equivalently be
translated as "probability density function".

In the \textit{Continuous Time Random Walk} model one considers a particle
which at time zero makes an instantaneous jump from position $\mathbf{x}_{0}$
to $\mathbf{x}_{0}+\mathbf{x}$, then waits at the new position during time $%
t $, then jumps again to a new position, where it waits, etc. Both $\mathbf{x%
}$ and $t$ are random quantities. In all forthcoming calculations, we assume
that the random processes are \textit{symmetrical (centred)}, \textit{i.e.}, 
$\left\langle \mathbf{x}(t)\right\rangle =0$, $\forall t$; hence [see Eq. (%
\ref{eeq.1})], $\delta \mathbf{x(}t)=\mathbf{x}(t).$ In its standard form, a
CTRW is defined by two functions:

\begin{center}
$f(\mathbf{x})$: the PDF of a jump defined by the vector $\mathbf{x}$; \
Fourier transform: \ $\widehat{f}(\mathbf{k}),$

$\psi (t)$: the waiting time PDF; \ Laplace transform: \ $\widetilde{\psi }%
(s).$
\end{center}

We only consider in the present work "classical" CTRW's, for which $f(%
\mathbf{x})$ and $\psi (t)$ are independent quantities.\ This means that $f(%
\mathbf{x})$ is independent of $t$, and $\psi (t)$ is independent of $%
\mathbf{x}$.\ Processes such as L\'{e}vy walks \cite{Metz-Klaf 2000} or the
vMSC\ model \cite{van Mil} are not discussed here.

A derived function that plays an important role is the \textit{memory kernel}
$\widetilde{\phi }(s)$ (in Laplace representation):

\begin{center}
\begin{equation}
\widetilde{\phi }(s)=\frac{s~\widetilde{\psi }(s)}{1-\widetilde{\psi }(s)}
\label{eeq.7}
\end{equation}
\end{center}

This standard CTRW has been solved exactly by \cite{Mont-Weiss} in 1965.\
Let $n(\mathbf{x},t)$ be the PDF of finding the walker at position $\mathbf{x%
}$ at time $t$, given that it started certainly at the origin $\mathbf{x}=0$
at time $t=0$ [\textit{i.e.}, $n(\mathbf{x},0)=\delta (\mathbf{x})$ and $n(%
\mathbf{x},t)\rightarrow 0$ as $|\mathbf{x}|\rightarrow \infty $]. Clearly,
from this definition, $n(\mathbf{x},t)$ is the \textit{propagator} (or 
\textit{Green's function}) of the CTRW process. It is given, in
Fourier-Laplace representation, by

\begin{equation}
\widehat{\widetilde{n}}(\mathbf{k},s)=\frac{1-\widetilde{\psi }(s)}{s}~\frac{%
1}{1-\widetilde{\psi }(s)~\widehat{f}(\mathbf{k})}  \label{eeq.8}
\end{equation}

This is the celebrated \textit{Montroll-Weiss equation}. An equivalent form
is:

\begin{equation}
\widehat{\widetilde{n}}(\mathbf{k},s)=\frac{1}{s+\widetilde{\phi }(s)~\left[
1-\widehat{f}(\mathbf{k})\right] }  \label{eeq.9}
\end{equation}

The latter equation is easily transformed into:

\begin{equation}
s~\widehat{\widetilde{n}}(\mathbf{k},s)-1=-\widetilde{\phi }(s)~\left[ 1-%
\widehat{f}(\mathbf{k})\right] ~\widehat{\widetilde{n}}(\mathbf{k},s)
\label{eeq.10}
\end{equation}

This is the Fourier-Laplace transform of the integro-differential equation
obeyed by the propagator:

\begin{equation}
\partial _{t}n(\mathbf{x},t)=\int_{0}^{t}dt^{\prime }~\phi (t-t^{\prime })~%
\left[ -n(\mathbf{x},t^{\prime })+\int d\mathbf{x}^{\prime }~f(\mathbf{x}-%
\mathbf{x}^{\prime })~n(\mathbf{x}^{\prime },t^{\prime })\right]
\label{eeq.11}
\end{equation}

This \textit{Montroll-Shlesinger (MS) Master equation} is \textbf{non-local}%
, both in space and in time.\ It shows that the rate of change at point $%
\mathbf{x}$ and time $t$ is determined both by the past history, whose
influence is measured by $\phi (t)$, and by the spatial environment
characterized by $f(\mathbf{x})$.

Although Eqs. (\ref{eeq.8}) and (\ref{eeq.9}) are explicit solutions for the
propagator (or the \textit{resolvent}, as the Fourier-Laplace transform of
the propagator is usually called), the expression of the \textit{inverse }%
Fourier-Laplace transform in terms of known functions of $\mathbf{x}$ and $t$
is, in general, difficult or impossible.\bigskip

It has been shown in a series of important recent works that this problem
can be solved under rather general conditions in the so-called \textbf{fluid
limit, }which is actually an \textbf{asymptotic} limit of large distances
and times [\cite{Metz-Klaf 2000}, \cite{Metz-Klaf 2004}, \cite{Gor-Main 2005}%
, \cite{delCast 2004}, \cite{RB}]. (It should be noted that, in many
interesting cases, the notion of "long" is ambiguous when there is no
intrinsic time scale!) \ The fluid limit is defined as follows (using
dimensionless variables $\mathbf{k},~s,~x,~t$) [$d$ is the dimensionality of
the system]:

\begin{eqnarray}
\widehat{f}(\mathbf{k}) &=&1-|\mathbf{k}|^{\alpha }...,\ \ \ \ 0<\alpha \leq
2,\ \ \ \ k\rightarrow 0  \label{eeq.12} \\
\widetilde{\psi }(s) &=&1-s^{\beta }...,\ \ \ \ \ 0<\beta \leq 1,\ \ \ \
s\rightarrow 0  \label{eeq.13}
\end{eqnarray}

In the same limit, the memory function is:

\begin{equation}
\widetilde{\phi }(s)=s^{1-\beta }  \label{eeq.14}
\end{equation}

This is the same as Eq. (17.12) of \cite{RB}, describing subdiffusion ($%
\beta <1$) or normal diffusion ($\beta =1$) [see Sec. \ref{Classification}].
Later in the present work we will need an extension to the range $1<\beta <2$%
. This extension is not trivial and requires a closer discussion, which is
developed in Sec. \ref{V(t)}. Thus, in the fluid limit, \textit{the CTRW is
entirely defined by the two exponents }$\alpha $\textit{\ and }$\beta $.

The (F-L) equation of evolution takes the form:

\begin{equation}
s~\widehat{\widetilde{n}}(\mathbf{k},s)-1=-s^{1-\beta }~\left\vert \mathbf{k}%
\right\vert ^{\alpha }~\widehat{\widetilde{n}}(\mathbf{k},s).  \label{eeq.15}
\end{equation}

Its solution, which is the resolvent of the process, is:

\begin{equation}
\widehat{\widetilde{n}}(\mathbf{k},s)=\frac{s^{\beta -1}}{s^{\beta }+|%
\mathbf{k}|^{\alpha }}  \label{eeq.16}
\end{equation}

Eq. (\ref{eeq.15}) is, in F-L form, a \textbf{fractional differential
equation (FDE)}, corresponding to:

\begin{eqnarray}
-\left\vert \mathbf{k}\right\vert ^{\alpha } &\rightarrow &\frac{\partial
^{\alpha }}{\partial |\mathbf{x}|^{\alpha }}\ \equiv D_{|\mathbf{x}%
|}^{\alpha }:\text{Riesz fractional derivative }  \notag \\
s^{\beta }\widehat{g}(s)-s^{\beta -1}g(t &=&0)\rightarrow
~_{0}^{C}D_{t}^{\beta }~g(t):\text{Caputo fractional derivative }\ \ 
\label{eeq.17}
\end{eqnarray}

A primer on fractional calculus, including the definition of these objects
is given in the Appendix A. Thus, Eq. (\ref{eeq.15}) can be rewritten as:

\begin{equation}
s^{\beta }~\widehat{\widetilde{n}}(\mathbf{k},s)-s^{\beta -1}=-|\mathbf{k}%
|^{\alpha }~\widehat{\widetilde{n}}(k,s),  \label{eeq.18}
\end{equation}

which is the F-L transform of the FDE:

\begin{equation}
~_{0}^{C}D_{t}^{\beta }~n(\mathbf{x},t)=D_{|\mathbf{x}|}^{\alpha }~n(\mathbf{%
x},t),  \label{eeq.19}
\end{equation}

with the initial condition:

\begin{equation}
n(\mathbf{x},t=0)=\delta (\mathbf{x})  \label{eeq.20}
\end{equation}

Thus, $n(\mathbf{x},t)$ is the \textbf{propagator} (or \textit{Green's
function}, or fundamental solution) \textbf{of the FDE\ }(\ref{eeq.19}%
).\bigskip

Although the Fourier-Laplace inversion of Eq. (\ref{eeq.16}) is difficult,
an important property, \textit{i.e.}, the \textit{scaling of the solution in 
}$\mathbf{x}-t$\textit{\ space} can be derived without performing explicitly
the latter operation.\ The following argument is due to Mainardi \cite{Main
2001}, and is much more compact than the one presented in \cite{Ball}, \cite%
{Bouch-G} and \cite{RB}. It is based on the well-known properties of the
Fourier- and Laplace transformations:

\begin{equation}
f(a\mathbf{x})\overset{\mathcal{F}}{\longleftrightarrow }\frac{1}{a}~%
\widehat{f}\left( \frac{\mathbf{k}}{a}\right) ,\ \ \ \ \ f(bt)\overset{%
\mathcal{L}}{\longleftrightarrow }\frac{1}{b}~\widetilde{f}\left( \frac{s}{b}%
\right) ,  \label{eeq.21}
\end{equation}

where the double arrow connects a pair of Fourier (Laplace) transforms.
Applying this property to Eq. (\ref{eeq.16}), we find, after a short
calculation:

\begin{equation}
n_{\alpha ,\beta }(a\mathbf{x},bt)\longleftrightarrow \frac{1}{ab}~\widehat{%
\widetilde{n}}_{\alpha ,\beta }\left( \frac{\mathbf{k}}{a},\frac{s}{b}%
\right) =\frac{1}{a}~~\widehat{\widetilde{n}}_{\alpha ,\beta }\left( \frac{%
b^{\beta /\alpha }}{a}\mathbf{k},s\right)  \label{eeq.22}
\end{equation}

Applying the relations (\ref{eeq.21}) to the last form we find:

\begin{equation*}
\frac{1}{a}~~\widehat{\widetilde{n}}_{\alpha ,\beta }\left( \frac{b^{\beta
/\alpha }}{a}\mathbf{k},s\right) \longleftrightarrow \frac{1}{b^{\beta
/\alpha }}~n_{\alpha ,\beta }\left( \frac{a}{b^{\beta /\alpha }}\mathbf{x,~}%
t\right)
\end{equation*}

Thus, finally:

\begin{equation}
n_{\alpha ,\beta }(a\mathbf{x},bt)=\frac{1}{b^{\beta /\alpha }}~n_{\alpha
,\beta }\left( \frac{a}{b^{\beta /\alpha }}\mathbf{x,~}t\right) .
\label{eeq.23}
\end{equation}

Let us assume now that the density profile has the scaling form:

\begin{equation}
n_{\alpha ,\beta }(\mathbf{x},t)=t^{-\gamma }~K_{\alpha ,\beta }\left( \frac{%
\mathbf{x}}{t^{\gamma }}\right)  \label{eeq.24}
\end{equation}

Under what condition is this form compatible with Eq. (\ref{eeq.23})? The
left hand side of the latter yields:

\begin{equation*}
n_{\alpha ,\beta }(a\mathbf{x},bt)=\frac{1}{b^{\gamma }t^{\gamma }}%
~K_{\alpha ,\beta }\left( \frac{a\mathbf{x}}{b^{\gamma }t^{\gamma }}\right) ;
\end{equation*}

and the right hand side yields:

\begin{equation*}
\frac{1}{b^{\beta /\alpha }}t^{-\gamma }~K_{\alpha ,\beta }\left( \frac{%
\frac{a}{b^{\beta /\alpha }}\mathbf{x}}{t^{\gamma }}\right) =\frac{1}{%
b^{\beta /\alpha }t^{\gamma }}~K_{\alpha ,\beta }\left( \frac{a\mathbf{x}}{%
b^{\beta /\alpha }t^{\gamma }}\right)
\end{equation*}

These two expressions are equal, provided that $\gamma =\beta /\alpha $. We
thus conclude:

\begin{equation}
n_{\alpha ,\beta }(\mathbf{x},t)=t^{-\beta /\alpha }~K_{\alpha ,\beta
}\left( \frac{\mathbf{x}}{t^{\beta /\alpha }}\right) ,  \label{eeq.25}
\end{equation}

This is the important \textit{\textbf{scaling relation}} obeyed by the
density profile.\ It tells us, in particular, that it depends on space only
through the combination $\mathbf{p=(x}/t^{\beta /\alpha })$, called the 
\textit{\textbf{similarity variable}.} From this result we immediately
deduce the scaling of the MSD:

\begin{equation*}
\left\langle x^{2}(t)\right\rangle _{\alpha ,\beta }=\int d\mathbf{x~}%
x^{2}~t^{-\beta /\alpha }~K_{\alpha ,\beta }\left( \frac{\mathbf{x}}{%
t^{\beta /\alpha }}\right)
\end{equation*}

With the substitution \ $\mathbf{p=x}/t^{\beta /\alpha }$ we find:

\begin{equation}
\left\langle x^{2}(t)\right\rangle _{\alpha ,\beta }=M_{\alpha ,\beta
}~t^{2\beta /\alpha },  \label{eeq.26}
\end{equation}

identical to Eq. (\ref{eeq.1}), with the following definition of the
constant:

\begin{equation}
M_{\alpha ,\beta }=\int d\mathbf{p~}p^{2}~K_{\alpha ,\beta }(\mathbf{p).}
\label{eeq.27}
\end{equation}%
\bigskip

Eq. (\ref{eeq.26}) provides us with a criterion of "strangeness" expressed
in terms of the two parameters $\alpha ,~\beta $:

\begin{equation}
\begin{tabular}{|c|c|}
\hline
Diffusion exponent & Regime \\ \hline
$2\beta <\alpha $ & \ \ SUBDIFFUSIVE \ \  \\ \hline
$2\beta =\alpha $ & \ \ (NORMAL) DIFFUSIVE \ \  \\ \hline
$2\beta >\alpha $ & \ \ SUPERDIFFUSIVE \ \  \\ \hline
\end{tabular}
\label{eeq.28}
\end{equation}

The dimensional argument leading to Eq. (\ref{eeq.26}) does not tell us
anything about the constant $M_{\alpha ,\beta }$: we do not even know wheter
it is finite or infinite! The determination of $M_{\alpha ,\beta }$ requires
the explicit form of the solution through the function $K_{\alpha ,\beta }(%
\mathbf{p})$.

We proceed here in a more direct way \cite{Gor-Abd}.\ The inverse Laplace
transform of Eq.\ (\ref{eeq.16}) is known \cite{Podlubny}, \cite{Main 2001}:

\begin{equation}
\widehat{n}_{\alpha ,\beta }(\mathbf{k},t)=E_{\beta }(-|k|^{\alpha }t^{\beta
}),  \label{eeq.29}
\end{equation}

where $E_{\beta }(z)$ is the \textit{Mittag-Leffler function} \cite{Erdelyi}%
, a natural (fractional!) generalization of the exponential:

\begin{equation}
E_{\beta }(z)=\sum_{n=0}^{\infty }~\frac{z^{n}}{\Gamma (\beta n+1)}.
\label{eeq.30}
\end{equation}

The first terms of the expansion of the Fourier-density profile are thus:

\begin{equation}
\widehat{n}_{\alpha ,\beta }(\mathbf{k},t)=1+\frac{-|k|^{\alpha }t^{\beta }}{%
\Gamma (\alpha +1)}+\frac{|k|^{2\alpha }t^{2\beta }}{\Gamma (2\alpha +1)}+...
\label{eeq.31}
\end{equation}

The MSD can be determined directly from the Fourier-density profile by the
well-known formula:

\begin{equation}
\left\langle x^{2}(t)\right\rangle _{\alpha ,\beta }=-\left\{ \frac{\partial 
}{\partial \mathbf{k}}\cdot \frac{\partial }{\partial \mathbf{k}}~\widehat{n}%
_{\alpha ,\beta }(\mathbf{k},t)\right\} _{\mathbf{k=0}}.  \label{eeq.32}
\end{equation}

When applied to the Mittag-Leffler function, this yields (in $d$ dimensions):

\begin{eqnarray}
\left\langle x^{2}(t)\right\rangle _{\alpha ,\beta } &=&\left\{ \alpha
(\alpha -2+d)\frac{|k|^{\alpha -2}t^{\beta }}{\Gamma (\beta +1)}-2\alpha
(2\alpha -2+d)\frac{|k|^{2\alpha -2}t^{2\beta }}{\Gamma (2\beta +1)}%
+...\right\} _{\mathbf{k=0}}  \notag \\
&&  \label{eeq.33}
\end{eqnarray}

Two cases are to be distinguished:\bigskip

a) \ $\mathbf{\alpha =2}$.

In this case, the first term has a finite positive value in $\mathbf{k=0}$;
all the other terms in the series vanish in this point. Hence:

\begin{equation}
\left\langle x^{2}(t)\right\rangle _{\alpha ,\beta }=\frac{2d}{\Gamma (\beta
+1)}~t^{\beta },\ \ \ \ \alpha =2,\ \ \ \forall \beta .  \label{eeq.34}
\end{equation}

This completes Eq. (\ref{eeq.26}) [Note: $(2\beta /\alpha )=\beta $]\bigskip
.

b) $\mathbf{0<\alpha <2}$.

The first term in the right hand side of Eq. (\ref{eeq.34}) contains the
factor \ $|k|^{\alpha -2}$ which diverges in $\mathbf{k=0}$. The remaining
terms are either \ $0$ \ or \ $\infty $ \ in this range of $\alpha $.\ Thus,
the constant in Eq. (\ref{eeq.26}) is:

\begin{equation}
M_{\alpha ,\beta }=\infty ,\ \ \ \ 0<\alpha <2,\ \ \ \ \forall \beta .
\label{eeq.35}
\end{equation}

This means that the density profiles of the CTRW in the present range of $%
\alpha $ have long tails, whatever the value of $\beta .$Restricting the
concept of superdiffusion to the sole discussion of Eq. (\ref{eeq.26}) would
put all the processes with $0<\alpha <2$ on the same rank. The distinction
among them requires a deeper discussion of the density profile. These
questions will be further discussed in the next Section.

\section{Classification of Strange Transport Regimes\label{Classification}}

The very clear papers of Mainardi and coworkers \cite{Gor-Main 1997}, \cite%
{Main 2001}, \cite{Main 2003} contain an exhaustive study of the Green's
function of the FDE (\ref{eeq.19}) in one dimension, $d=1$. In the
forthcoming work, we take full advantage of the fact that the general
solution $n_{\alpha ,\beta }(\mathbf{x},t)$ (for all $\alpha ,\beta $) of
Eq. (\ref{eeq.19}) has been obtained by these authors.. This solution was
also obtained by \cite{Metz-Klaf 2000} in a different form (in terms of Fox
functions). We shall first discuss some particular cases (for $d=1$%
).\bigskip\ \ \ \ \ \ \ \ \ \ \ \ \ \ \ \ \ \ \ \ \ \ \ \ \ \ \ \ \ \ \ \ \ 
\begin{equation*}
\begin{tabular}{c}
$%
\begin{tabular}{c}
$\mathbf{\alpha =2,\ \ \ \ \beta =1}$%
\end{tabular}%
$%
\end{tabular}%
\end{equation*}

This is the classical case of \textbf{normal diffusion}. Eq. (\ref{eeq.15})
becomes in this case:

\begin{equation}
s~\widehat{\widetilde{n}}(k,s)-1=-~\left\vert k\right\vert ^{2}~\widehat{%
\widetilde{n}}(k,s),  \label{eeq.37}
\end{equation}

which is simply the F-L transform of the ordinary diffusion equation (with a
diffusion coefficient $D=1$):

\begin{equation}
\partial _{t}n(x,t)=\nabla ^{2}n(x,t)  \label{eeq.38}
\end{equation}

It is well known that in this case the MSD is:

\begin{equation}
\left\langle x^{2}(t)\right\rangle =2t,  \label{eeq.39}
\end{equation}

in agreement with Eq. (\ref{eeq.26}). The solution of Eq. (\ref{eeq.37}) is:

\begin{equation}
\widehat{\widetilde{n}}(k,s)=\frac{1}{s+k^{2}}  \label{eeq.40}
\end{equation}

The \textit{propagator} associated with Eq.\ (\ref{eeq.38}) is the \textit{%
Gaussian wave packet}:

\begin{equation}
n(x,t)=\frac{1}{2(\pi t)^{1/2}}~\exp \left( -\frac{x^{2}}{4t}\right)
\label{eeq.41}
\end{equation}

Note that\ Eq. (\ref{eeq.41}) is a special case of the \textit{scaling
relation} (\ref{eeq.25}):

\begin{equation}
n_{2,1}(x,t)=t^{-(1/2)}~F(xt^{-1/2})  \label{eeq.42}
\end{equation}

The Gaussian scaling function $F(p)$ in the present case is a consequence of
the Central Limit theorem of probability theory. "Most" physical systems
behave in this normal diffusive way. But it is the \textit{strange behavior}%
, which deviates from this Gaussian form, that will be the main object of
our study.\bigskip

\begin{center}
\begin{tabular}{c}
$%
\begin{array}{l}
\mathbf{\alpha =2,\ \ \ 0<\beta <1}%
\end{array}%
$%
\end{tabular}%
\bigskip
\end{center}

This case was studied in detail in \cite{RB}, and more completely in \cite%
{Gor-Main 1997} and \cite{Main 2001}.\ It was called the \textit{Standard
Long Tail CTRW} (\textbf{SLT-CTRW}) in \cite{RB}, and \textit{%
time-fractional subdiffusion} in \cite{Main 2001}. The main result in this
connection may be stated as follows.\ Eq. (\ref{eeq.26}) shows that the
SLT-CTRW is a \textbf{subdiffusive }process (because $2\beta <\alpha $), for
all values of $\beta $ in the range ($0,\ 1$). The propagator is
non-Gaussian, and has a long tail of \textit{stretched exponential }form:

\begin{equation}
n_{2,\beta }(x,t)=A_{2,\beta }~t^{(\beta -1)/(2-\beta )}~\exp \left(
-b_{\beta }~p^{2/(2-\beta )}\right) ,\ \ \ \ p\gg 1  \label{eeq.43}
\end{equation}

The values of the constants $A_{\alpha ,\beta }$ and $b_{\beta }$ \cite{Main
2001} are irrelevant here.

Eq. (\ref{eeq.43}) has again a scaling form, with a (rather complicated)
finite constant $A_{2,\beta }.$ This function has a single maximum in \ $x=0$%
, and has a finite MSD.\bigskip

\begin{equation*}
\begin{array}{l}
\begin{tabular}{c}
$\mathbf{0<\alpha <2,\ \ \ \ \beta =1}$%
\end{tabular}%
\end{array}%
\end{equation*}%
$\bigskip $

This is the case of \textit{space-fractional diffusion} of Mainardi \cite%
{Main 2001} [see also Sec. 12.3 of \cite{RB}]. The corresponding resolvent
is:

\begin{equation}
\widehat{\widetilde{n}}(k,s)=\frac{1}{s+|k|^{\alpha }}.  \label{eeq.44}
\end{equation}

Its inverse Laplace transform is:

\begin{equation}
\widehat{n}(k,t)=\exp (-\left\vert k\right\vert ^{\alpha }t).  \label{eeq.45}
\end{equation}

This is the Fourier transform of a \textit{L\'{e}vy stable distribution} (%
\cite{RB}, Appendix D). It is well known that the inverse Fourier transform
of these distributions is in general not expressible in terms of simple
functions, except for a few values of $\alpha $.\ The general propagator
(for arbitrary $\alpha $) can, however, be expressed as a \textit{Mainardi
function}, defined by a convergent power series [see Eq. (\ref{eeq.77})
below]. The most striking property of the L\'{e}vy distributions is their
"fat" long tail, \textit{i.e., an algebraic power law decay for }$p\gg 1$:

\begin{equation}
F(p)\sim \frac{1}{\left\vert p\right\vert ^{1+\alpha }},\ \ \ \ p\gg 1.
\label{eeq.46}
\end{equation}

This law of decay is much slower than the stretched exponential decay of Eq.
(\ref{eeq.43}). As a result, the MSD of all the L\'{e}vy distributions is
infinite, as shown in Sec.\ \ref{CTRW FDE}. The three cases discussed above
are illustrated in the following figure by the corresponding Mainardi
functions. In these figures $t=1$, hence $p=x$.

Red (solid) line: \ subdiffusive SLT: \ $\alpha =2,\ \ \beta =1/2$

Blue (dotted) line: \ \ superdiffusive L\'{e}vy flight: \ $\alpha =3/2,\ \ \
\beta =1.$

Black (dashed) line: \ normal diffusive Gaussian process: \ $\alpha =2,\ \ \
\beta =1$

\begin{figure}[tbph]
\centerline{\includegraphics[height=7.5cm]{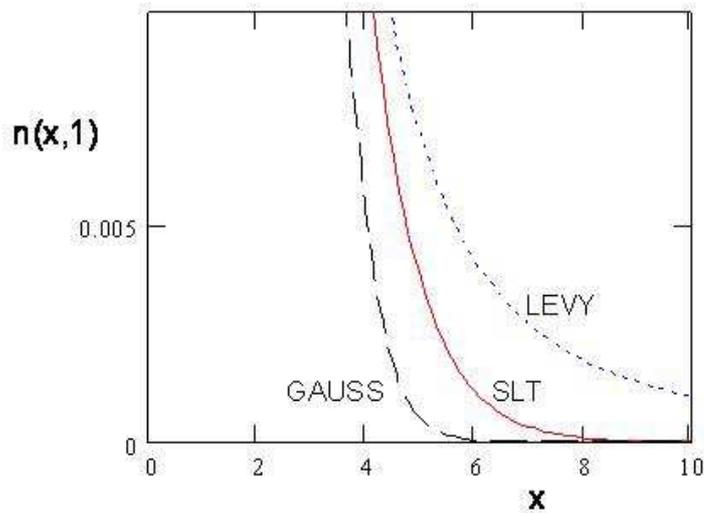}}
\caption{Propagators (Green functions) $%
G(x)=n(x,t=1)$ for three cases of fractional diffusion.\ Dotted blue: L\'{e}%
vy superdiffusion ($\protect\alpha =1.5,\ \protect\beta =1$); Solid red: SLT
subdiffusion ($\protect\alpha =2,~\protect\beta =0.5$), Dashed black:
Gaussian diffusion ($\protect\alpha =2,\ \protect\beta =1$).}
\label{LFDE 1}
\end{figure}

We represent in Fig. \ref{LFDE 1} the propagators for these three cases.

%%%%%%%%%%%%%%%%%%%%%%%%%%%%%%%%%%%%%%%%%%%%%%%%%%%%%%%%%%%%%%%%%%%%% 
%%%%%%%%%%%%%%%%%%%%%%%%%%%%%%%%%%%%%%%%%%%%%%%%%%%%%%%%%%%%%%%%%%%%%

\begin{figure}[tbph]
\centerline{\includegraphics[height=7.5cm]{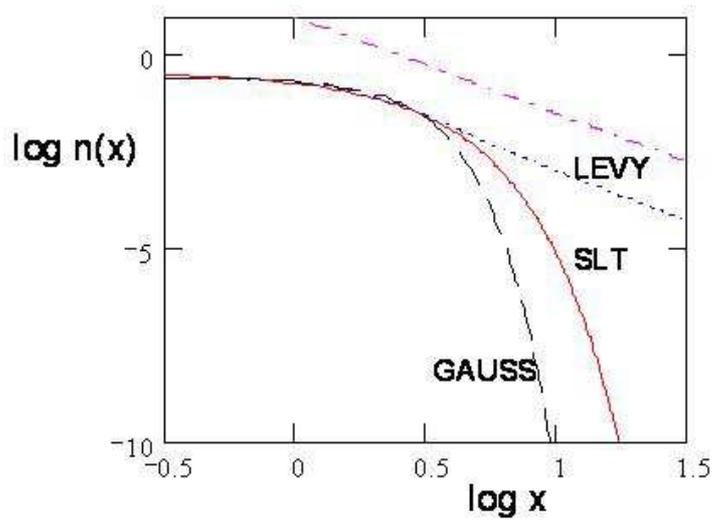}}
\caption{Same data as in Fig. \protect\ref%
{LFDE 1}, in log-log representation.}
\label{LFDE 2}
\end{figure}

The asymptotic behavior is most strikingly seen in log-log representation
(Fig. \ref{LFDE 2}). The upper reference straight line has a slope [$%
-(1+\alpha )=-2.5$].

%%%%%%%%%%%%%%%%%%%%%%%%%%%%%%%%%%%%%%%%%%%%%%%%%%%%%%%%%%%%%%%%%%%%% 
%%%%%%%%%%%%%%%%%%%%%%%%%%%%%%%%%%%%%%%%%%%%%%%%%%%%%%%%%%%%%%%%%%%%%

We thus see that there are two types of long tails:

\begin{itemize}
\item \textit{The stretched exponential tail}, characteristic, typically, of
the subdiffusive \ SLT

\item \textit{The algebraic tail}, characteristic of the superdiffusive L%
\'{e}vy process.
\end{itemize}

Note that both are "fat" tails, \textit{i.e.} the decay at large $x$ is
slower than the corresponding diffusive Gaussian. The second type is the
most slowly decaying one. Warning: it should not be concluded that all
subdiffusive processes have exponential tails, whereas all superdiffusive
ones have algebraic tails!! (see below).\bigskip

\begin{center}
\begin{tabular}{c}
$%
\begin{array}{l}
\mathbf{\alpha =2,\ \ \ \ 1<\beta <2}%
\end{array}%
$%
\end{tabular}%
\bigskip
\end{center}

This case corresponds to \textit{\textbf{time-fractional superdiffusion}}.
This very interesting case will be treated in detail in Sec. \ref{Algeb
Correl}.\bigskip

\begin{center}
\begin{tabular}{c}
$%
\begin{array}{l}
\mathbf{0<\alpha <2,\ \ \ \ 0<\beta <2}%
\end{array}%
$%
\end{tabular}%
\bigskip
\end{center}

This is the generic, mixed spatial and temporal fractional diffusion.\ In
this case there is a competition between super- and subdiffusion, expressed
by the ratio $2\beta /\alpha $ [see Eq. (\ref{eeq.26})]. In the work of
del-Castillo Negrete et al \cite{delCast 2004}, a model with $\alpha =3/4$, $%
\beta =0.5$ has been successfully compared with numerical simulations of a
set of plasmadynamical equations. In a recent paper\ \cite{delCast 2005},
additional values of $\alpha <2$ were explored.

In the present work we do not consider the case of space-fractional (L\'{e}%
vy) superdiffusion.\ We thus fix in all subsequent sections:

\begin{equation}
\alpha =2  \label{eeq.47}
\end{equation}

\section{Random Time-dependent Velocity\label{V(t)}}

We consider a particle moving in one dimension; its position $X(T)$ at time $%
T$ is determined "semi-dynamically" by a V-Langevin equation (\cite{RB},
Ch.\ 11): \footnote{%
Capital notations $X,T,V,\kappa ,S$ denote \textit{dimensional quantities.}
An exception is made for the wave vector $\kappa $, because the symbol $K$
will be used later for the Kubo number, Eq. (91).}

\begin{equation}
\frac{dX(T)}{dT}=V(T),  \label{eeq.48}
\end{equation}

where the random velocity $V(T)$ (\textit{depending solely on time)} is
defined statistically by specifying the first two moments of the ensemble of
its realizations, supposed to be Gaussian, homogeneous and stationary:

\begin{equation}
\left\langle V(T)\right\rangle =0,\ \ \ \ \ \left\langle
V(0)~V(T)\right\rangle =V_{0}^{2}~\mathcal{T}(T).  \label{eeq.49}
\end{equation}

The function $\mathcal{T}(T)$ need not yet be specified explicitly, except
by requiring $\mathcal{T}(0)=1$. $V_{0}$, the initial velocity rms is
supposed to be a given parameter.

Let $f(X,T)$ be the (fluctuating) distribution function of the particle.\ It
is represented as usual as:

\begin{equation}
f(X,T)=n(X,T)+\delta f(X,T)  \label{eeq.50}
\end{equation}

where the ensemble-average $n(X,T)\equiv \left\langle f(X,T)\right\rangle $
is called the \textit{density profile, }and $\delta f(X,T)$ is the
fluctuation. The distribution function obeys the \textit{Hybrid Kinetic
Equation (HKE)}:

\begin{equation}
\partial _{T}f(X,T)+V(T)~\partial _{X}f(X,T)=0.  \label{eeq.51}
\end{equation}

The characteristic equation of this first-order partial differential
equation coincides with the V-Langevin equation (\ref{eeq.48}). This
equation is split, as usual \cite{RB}, into two components by the averaging
operation:

\begin{equation}
\partial _{T}n(X,T)+\partial _{X}~\left\langle V(T)~\delta
f(X,T)\right\rangle =0  \label{eeq.52}
\end{equation}

\begin{equation}
\partial _{T}\delta f(X,T)+V(T)~\partial _{X}~\delta f(X,T)+V(T)~\partial
_{X}~n(X,T)-\partial _{X}\left\langle V(T)~\delta f(X,T\right\rangle =0
\label{eeq.53}
\end{equation}

The latter equation is solved by using the propagator:

\begin{equation}
g(X,T|X^{\prime },T^{\prime })=\delta \left[ X-X^{\prime }-\int_{T^{\prime
}}^{T}dT_{1}~V(T_{1})\right]  \label{eeq.54}
\end{equation}

Substituting its solution into (\ref{eeq.52}), we find:

\begin{eqnarray}
\partial _{T}n(X,T) &=&\partial _{X}^{2}~\int_{0}^{T}dT_{1}~\left\langle
V(T)V(T_{1})~n\left[ X-\int_{T_{1}}^{T}dT_{2}~V(T_{2}),\ T_{1}\right]
\right\rangle  \notag \\
&&+\partial _{X}~\left\langle V(T)~\delta f\left[ X-%
\int_{0}^{T}dT_{1}V(T_{1}),\ 0\right] \right\rangle  \label{eeq.55}
\end{eqnarray}

This is the exact solution of the problem.\ It is expressed by an
integro-differential equation that is non-local both in time (non-Markovian)
and in space (non-local). We do not repeat here the arguments justifying the%
\textbf{\ local approximation}, by which the fluctuating displacements $%
\int_{T_{1}}^{T}dT_{2}~V(T_{2}),$ $\int_{0}^{T}dT_{1}~V(T_{1})$ in the
arguments of $n$ and of $\delta f$ are neglected, and by which the second
term of the right hand side, containing the initial fluctuation, is also
neglected \cite{RB}.\ The result is an (approximate) \textit{local, but
non-markovian} diffusion equation:

\begin{equation}
\partial _{T}n(X,T)=\partial _{X}^{2}~\int_{0}^{T}dT_{1}~\left\langle
V(T)V(T_{1})\right\rangle ~n(X,T_{1}).  \label{eeq.56}
\end{equation}

Using Eq. (\ref{eeq.49}), this becomes:

\begin{equation}
\partial _{T}n(X,T)=\partial _{X}^{2}~\int_{0}^{T}dT_{1}~V_{0}^{2}~\mathcal{T%
}(T-T_{1})~n(X,T_{1}).  \label{eeq.57}
\end{equation}%
\bigskip

Consider, on the other hand, a particle performing a one-dimensional CTRW
characterized, in the \textit{fluid limit}, by a Gaussian jump PDF:

\begin{equation}
\widehat{f}(\kappa )=1-\sigma ^{2}~\kappa ^{2}+...  \label{eeq.58}
\end{equation}

where $\sigma $ is a characteristic length and $\kappa $ is the dimensional
wave vector. The form of the waiting time PDF is not specified at this stage.

The density profile\footnote{%
The density profile considered in all forthcoming developments is the
solution of the evolution equation with the initial value: $n(X,0)=\delta
(X).$ In other words, $n(X,T)$ is the \textit{propagator} of the equation.
We shall continue, however, to call this function the "\textit{density
profile}", which has a more physical connotation.} (in F-L representation)
obeys Eq. (\ref{eeq.10}) which, written with a dimensional Laplace variable $%
S$, combined with (\ref{eeq.58}) reduces to:

\begin{equation}
S~\widehat{\widetilde{n}}(\kappa ,S)-1=-\widetilde{\phi }(S)~\sigma
^{2}\kappa ^{2}~\widehat{\widetilde{n}}(\kappa ,S)  \label{eeq.59}
\end{equation}

Its inverse F-L transform is:

\begin{equation}
\partial _{T}n(X,T)=\sigma ^{2}~\partial _{X}^{2}~\int_{0}^{T}dT_{1}~\phi
(T-T_{1})~n(X,T_{1})~  \label{eeq.60}
\end{equation}

This equation constitutes the main result of the first part of our work.\
Comparing it with Eq. (\ref{eeq.57}), we see that the density profile
produced (in the local approximation) by the semi-dynamical V-Langevin
equation with a time-dependent $V(t)$, and the density profile produced by a
CTRW\ with $\alpha =2$, - or, more specifically, \textit{by the fractional
diffusion equation (\ref{eeq.59})} - obey formally identical equations of
evolution. This statement will be made more precise in the discussion of
Sec.\ \ref{Algeb Correl}. The quantitative correspondence is obtained by 
\textit{identifying the velocity correlation function of the V-Langevin
process with the memory function:}

\begin{equation}
V_{0}^{2}~\mathcal{T}(T)\rightarrow \sigma ^{2}~\phi (T),  \label{eeq.61}
\end{equation}

i.e., in F-L representation

\begin{equation}
\widetilde{\mathcal{T}}\mathcal{(S)=}\frac{\sigma ^{2}}{V_{0}^{2}}~%
\widetilde{\phi }(S).  \label{eeq.62}
\end{equation}

We note that Eqs. (\ref{eeq.57}) and (\ref{eeq.59}) were also obtained in
Ref. \cite{West-Grigo} in a quite different manner, which is only valid for
the 2-state CTRW.\ The relation (\ref{eeq.62}) is much more general and was
obtained in a simpler direct way.

\section{Algebraic Velocity Correlation\label{Algeb Correl}}

We now make the problem explicit, by adopting the same form of the velocity
correlation as in Ref.\ \cite{West-Grigo}:

\begin{equation}
V_{0}^{2}\mathcal{T}(T)=\frac{A}{T^{\gamma }},\ \ \ \ 0<\gamma <1,
\label{eeq.63}
\end{equation}

where $A$ is a quantity having dimension [length]$^{2}$[time]$^{-2+\gamma }$%
. We then associate with the V-Langevin equation (for $d=1$) (\ref{eeq.48})
[or the Hybrid kinetic equation (\ref{eeq.51})] a CTRW\ with memory function
given, in the fluid limit, by Eq. (\ref{eeq.61}), leading to the fractional
diffusion equation \textbf{(FDE) }(\ref{eeq.59}):

\begin{equation}
S~\widehat{\widetilde{n}}(\kappa ,S)-1=-A~\kappa ^{2}~S^{\gamma -1}~\widehat{%
\widetilde{n}}(\kappa ,S),\ \ \ \ \ ~.  \label{eeq.64}
\end{equation}

This relation agrees with Eq. (33) of \cite{West-Grigo}.\ It is trivially
equivalent to:

\begin{equation}
S^{2-\gamma }~\widehat{\widetilde{n}}(\kappa ,S)-S^{1-\gamma }=-A~\kappa
^{2}~\widehat{\widetilde{n}}(\kappa ,S)  \label{eeq.64a}
\end{equation}

The F-L density profile is obtained from Eq. (\ref{eeq.64}) in the form:

\begin{equation}
\widehat{\widetilde{n}}(\kappa ,S)=\frac{S^{-1-\gamma }}{S^{2-\gamma
}+A~\kappa ^{2}},\ \ \ \ 0<\gamma <1.  \label{eeq.65}
\end{equation}

Comparing this result with Eq. (\ref{eeq.16}), we immediately see that the
two expressions become identical by taking the following value for the
waiting time exponent $\beta $:

\begin{equation}
\beta =2-\gamma .  \label{eeq.66}
\end{equation}

Thus, the process corresponds to \textit{\textbf{time-fractional
superdiffusion }}as defined in Sec. \ref{Classification}:

\begin{equation}
1<\beta <2.  \label{eeq.67}
\end{equation}

At this point there appears a subtle difficulty (the following argument
results from a remark of J.\ Klafter \cite{Klaf priv}). The FDE\ (\ref%
{eeq.64}) is \textit{formally} derived as the fluid limit of a CTRW with $%
\widetilde{\psi }(S)=1-S^{\beta }$, and $\beta $ in the range (\ref{eeq.66}%
): this is an \textit{extension} of the range defined in Eq. (\ref{eeq.13}).

The knowledge of the Laplace transform $\widetilde{\psi }(S)$ allows us to
calculate the \textit{average waiting time} between two jumps in the CTRW,
by using a well-known formula:

\begin{equation}
\left\langle T\right\rangle =-\left. \frac{\partial \widetilde{\psi }(S)}{%
\partial S}\right\vert _{S=0}=\left. \beta ~S^{\beta -1}\right\vert _{S=0}
\label{eeq.68}
\end{equation}

For $0<\beta <1$ the average waiting time of the CTRW is infinite. This
denotes a waiting time PDF with a long tail, leading to strange transport, 
\textit{i.e.}, \textit{subdiffusion} (Sec. \ref{Classification} and \cite{RB}%
). For $\beta =1$ the average waiting time is finite, corresponding to
normal diffusion.

For $1<\beta <2$, t\textit{he average waiting time is zero}, thus leading to
a paradoxical situation. Recalling that $T$ is a semi-definite positive
variable, it follows that $\psi (T)\sim \delta (T)$.\ Thus, any random
walker would remain trapped forever at its starting point, in clear
contradiction with \textit{superdiffusion}, as predicted by the diffusion
exponent $\mu =2\beta /\alpha =\beta >1$ (Sec. \ref{Classification}). The
only other possibility of obtaining a zero average waiting time would be to
have $\psi (T)$ negative in part of the range $(0,~\infty )$, but then $\psi
(T)$ does not qualify for a probability distribution function.

The conclusion of this argument is the following. \textit{The function }$%
\widehat{\widetilde{n}}(\kappa ,S)$ \textit{defined by Eq. (\ref{eeq.65})
is, indeed, the resolvent of the \textbf{fractional differential equation
(FDE)} (\ref{eeq.64}), but the latter cannot be derived as the fluid limit
of a \textbf{CTRW}. }

We are now faced with a new subtlety\footnote{%
We are indebted to F.\ Mainardi for calling our attention on this point and
on Ref. \cite{Main 2003}.}. The function $\widehat{\widetilde{n}}(\kappa ,S)$
is the solution of the linear algebraic equation (\ref{eeq.64a}).\ But, 
\textit{is the latter equation the F-L transform of a fractional diffusion
equation?} One is tempted to associate Eq. (\ref{eeq.64a}) with the FDE in $%
X-T$ representation:

\begin{equation}
_{0}^{C}D_{T}^{2-\gamma }n(X,T)=A~D_{|X|}^{2}~n(X,T),\ \ \ \ \ 0<\gamma <1,
\label{eeq.69}
\end{equation}

with the initial condition:

\begin{equation}
n(X,0)=\delta (X).  \label{eeq.70}
\end{equation}

This fractional equation is an "intermediate" between a diffusion equation ($%
\gamma =1$) and a wave equation ($\gamma =0$) \cite{Main 1996}, \cite%
{Gor-Main 1997}, \cite{Main 2003}. But, using the general formula for the
Laplace transform of the Caputo fractional derivative \cite{Podlubny}, \cite%
{Main 2003} and the Appendix of the present paper, the F-L transform of Eq. (%
\ref{eeq.69}) is:

\begin{equation}
S^{2-\gamma }~\widehat{\widetilde{n}}(\kappa ,S)-S^{1-\gamma }\widehat{n}%
(\kappa ,T=0)-S^{-\gamma }\widehat{n}_{T}(\kappa ,T=0)=-A~\kappa ^{2}~%
\widehat{\widetilde{n}}(\kappa ,S),\ \ \ 0<\gamma <1  \label{eeq.71}
\end{equation}

It thus appears that the general solution of the Cauchy problem for Eq. (\ref%
{eeq.69}) requires \textit{two initial conditions} (like the wave equation): 
$n(X,0)$ and $n_{T}(X,0)\equiv \lbrack \partial n(X,T)/\partial T]_{T=0}$.\
As a result, the general solution is expressed in terms of \textit{two}
propagators; indeed, the solution of Eq. (\ref{eeq.71}) is:

\begin{equation}
\widehat{\widetilde{n}}(\kappa ,S)=\frac{S^{-1-\gamma }}{S^{2-\gamma
}+A~\kappa ^{2}}\widehat{n}(\kappa ,0)+\frac{S^{-\gamma }}{S^{2-\gamma
}+A~\kappa ^{2}}\widehat{n}_{T}(\kappa ,0).  \label{eeq.72}
\end{equation}

. We refer to Ref. \cite{Main 2003} for a complete solution and discussion
of this problem. Here we shall only consider the simpler special case (also
considered in \cite{Main 2001}):

\begin{equation}
n(X,T=0)=\delta (X),\ \ \ \ n_{T}(X,T=0)=0  \label{eeq.73}
\end{equation}

\begin{equation}
\widehat{n}(\kappa ,T=0)=1,\ \ \ \ \widehat{n}_{T}(\kappa ,T=0)=0
\label{eeq.74}
\end{equation}

In this case, Eq. (\ref{eeq.71}) reduces to Eq. (\ref{eeq.64a}), and Eq. (%
\ref{eeq.65}) represents the F-L propagator of Eq. (\ref{eeq.69}) equipped
with the initial conditions (\ref{eeq.73}). The explicit solution for the
propagator of (\ref{eeq.69}) (i.e., the F-L inversion of Eq. (\ref{eeq.65})
was obtained analytically by Mainardi \cite{Main 1996}, \cite{Main 2001}
(and also by Metzler and Klafter \cite{Metz-Klaf 2000} in terms of Fox
H-functions):

\begin{equation}
n(X,T;\gamma )=\frac{1}{2\sqrt{A}~T^{1-(\gamma /2)}}~~M_{1-(\gamma
/2)}\left( \frac{X}{\sqrt{A}~T^{1-(\gamma /2)}}\right) ,\ \ \ \ \ \ 0<\gamma
<1  \label{eeq.75}
\end{equation}

This has precisely the \textit{scaling form} (\ref{eeq.25}), with the
similarity variable:

\begin{equation}
p=\frac{X}{\sqrt{A}T^{1-(\gamma /2)}}  \label{eeq.76}
\end{equation}

The special function $M_{1-(\gamma /2)}(p)$ is called the \textit{Mainardi
function}: it is defined for any order $\nu \in (0,1)$ as:

\begin{equation}
M_{\nu }(p)=\sum\limits_{k=0}^{\infty }\frac{(-p)^{k}}{k!~\Gamma \lbrack
-\nu k+(1-\nu )]},\ \ \ 0<\nu <1  \label{eeq.77}
\end{equation}

This is a special case of a \textit{Wright function}, defined as follows (%
\cite{Erdelyi}):

\begin{equation}
\phi (\alpha ,\beta ;p)=\sum\limits_{k=0}^{\infty }\frac{p^{k}}{k!~\Gamma
(\alpha k+\beta )},\ \ \ \ \alpha ,~\beta ~>0.  \label{eeq.78}
\end{equation}

The series (\ref{eeq.77}) is convergent, but the convergence is very slow.
In practice, for the numerical calculations displayed below, we take $100$
terms in the series. \textsc{Mainardi} et al 2001 \cite{Main 2001} also
showed that the \textit{asymptotic behavior} of the propagator is given by a 
\textit{stretched exponential}, which for the value $\beta =2-\gamma $ is:

\begin{equation}
n(X,T;\gamma )=\frac{1}{2\sqrt{A}T^{1-(\gamma /2)}}~B~p^{a}~\exp (-bp^{c}),\
\ \ \ p\gg 1,  \label{eeq.79}
\end{equation}

with:

\begin{eqnarray}
B &=&\left[ 2\pi \gamma ~2^{(2-\gamma )/\gamma }~(2-\gamma )^{2(\gamma
-1)/\gamma }\right] ^{-1/2},  \notag \\
a &=&\frac{1-\gamma }{\gamma },  \notag \\
b &=&\gamma \left[ \frac{(2-\gamma )^{2-\gamma }}{4}\right] ^{1/\gamma } 
\notag \\
c &=&\frac{2}{\gamma },\ \ \ \ \ \ \ \ \ \ \ \ \ \ \ \ \ \ \ \ \ \ \ \ \ \ \
\ \ \ 0<\gamma <1  \label{eeq.80}
\end{eqnarray}

The important point here is that the present density profile decays
asymptotically according to a stretched exponential law, with exponent $%
2/\gamma >2$. This implies that \textit{the propagator decays asymptotically 
\textbf{faster} than the Gaussian (contrary to the subdiffusive case }$%
0<\beta <1,$ discussed in Sec$.$\ref{Classification}\textit{.} Mainardi
calls this feature a "\textit{thin tail}". It thus appears clearly that 
\textit{superdiffusion is not necessarily associated with a long ("fat")
tail of the propagator.}

\begin{figure}[tbph]
\centerline{\includegraphics[height=3.1774in]{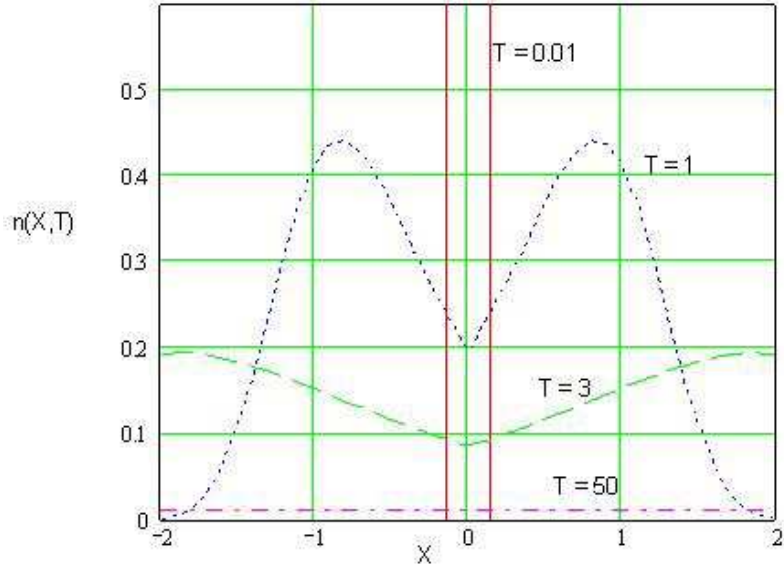}}
\caption{Propagator $n(X,T)$ at four
different times.\ $\protect\gamma =0.5,\ A=1.$}
\label{LFDE 3}
\end{figure}

Another important property of the propagator (\ref{eeq.75}) [\textit{i.e.},
of the Mainardi function $M_{\nu }(p)$ for $1/2<\nu <1,$ \textit{i.e.}, $%
1<\beta <2$] is its possessing for $T>0$ two symmetrical maxima that move
away from the origin, while becoming wider. As pointed out above, the
fractional equation (\ref{eeq.69}), written in a simpler notation ($\partial
^{\beta }/\partial T^{\beta }\equiv ~_{0}^{C}D_{T}^{\beta }$) as:

\begin{equation}
\frac{\partial ^{\beta }}{\partial T^{\beta }}n(X,T)=A\nabla ^{2}n(X,T),\ \
\ \ \ 1<\beta <2  \label{eeq.81}
\end{equation}

is a kind of interpolation between a diffusion equation ($\beta =1$) and a
wave equation ($\beta =2$). Thus, the solution represents a wave propagating
in both directions away from the origin, and damping out on the way 
\footnote{%
In two dimensions, this is a very familiar picture: it represents a wave
propagating radially on the surface of a lake, when a pebble is thrown into
the water.}. We illustrate these properties by the plot of Fig. \ref{LFDE 3}%
, corresponding to $\gamma =0.5$, hence $\beta =1.5$. For $T\rightarrow 0$,
the propagator tends toward $n(X,0)=\delta (X)$, as it should (\ref{eeq.70}).

%%%%%%%%%%%%%%%%%%%%%%%%%%%%%%%%%%%%%%%%%%%%%%%%%%%%%%%%%%%%%%%%%%%%% 
%%%%%%%%%%%%%%%%%%%%%%%%%%%%%%%%%%%%%%%%%%%%%%%%%%%%%%%%%%%%%%%%%%%%%

\begin{figure}[tbph]
\centerline{\includegraphics[height=2.9533in]{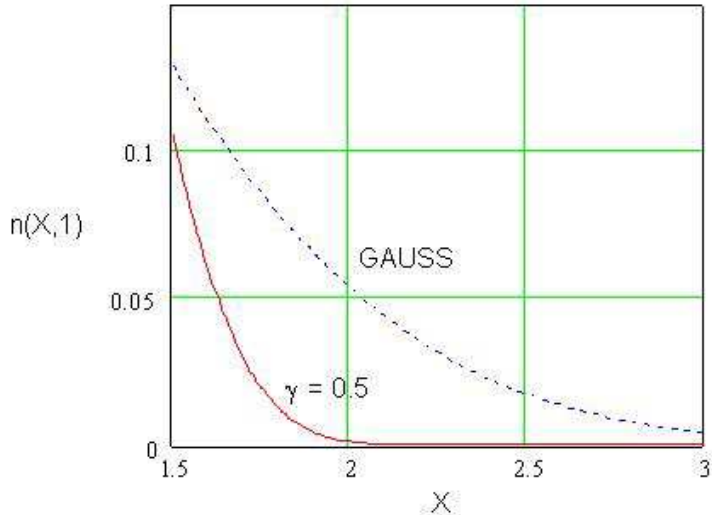}}
\caption{Tail of the propagator for $%
\protect\gamma =0.5$, compared to the corresponding Gaussian packet. $T=1.$}
\label{LFDE 4}
\end{figure}

We show in Fig. \ref{LFDE 4} the tail of the propagator (for $T=1$) compared
to the tail of the corresponding Gaussian.\ This situation is to be
contrasted with Figs.\ \ref{LFDE 1} and \ref{LFDE 2}: both the
time-fractional \textit{subdiffusive} propagator and the space-fractional 
\textit{superdiffusive} (L\'{e}vy) propagator have \textit{"fat" tails}.

%%%%%%%%%%%%%%%%%%%%%%%%%%%%%%%%%%%%%%%%%%%%%%%%%%%%%%%%%%%%%%%%%%%%% 
%%%%%%%%%%%%%%%%%%%%%%%%%%%%%%%%%%%%%%%%%%%%%%%%%%%%%%%%%%%%%%%%%%%%%
Figs. \ref{LFDE 3} and \ref{LFDE 4}
allow us to understand how superdiffusion is possible in spite of the thin
exponential tail.\ The dispersion of the matter is produced not only by the
broadening of the initial distribution (as in ordinary diffusion) but also
by the symmetric wave-like outward motion of the maximum.\bigskip

The results of the present section are not really new: \textsc{A}ll these
results were also obtained in Ref. \cite{West-Grigo}, using, however, the
2-state model. We showed here that the results are easily extended to an
arbitrary fluctuating velocity depending solely on time. Our main result can
be summarized as follows.

\textit{The motion of a particle in one dimension, with a random velocity
depending solely on time, obeying a semi-dynamical V-Langevin equation
(V-LE) (\ref{eeq.48}), can be equivalently described by a time-fractional
differential equation (FDE). There is, however, no CTRW whose fluid limit is
the FDE (\ref{eeq.74}).} The V-LE is characterized by a velocity correlation
depending (for all positive times!) algebraically on time, $\mathcal{T}%
(T)\sim T^{-\gamma }$, with an exponent $\gamma \in (0,1)$.\ The equivalent
FDE is characterized by a spatial exponent $\alpha =2$ and a waiting time
exponent $\beta =2-\gamma ;$ thus $1<\beta <2$. This FDE leads to \textit{%
time-fractional superdiffusion}, with diffusion exponent $\mu =2-\gamma >1$.
The density profile has two maxima that are moving symmetrically away from
each other, while broadening with increasing time.\ The asymptotic decay is
exponential, decreasing faster than a Gaussian in space ("thin tail").

The model V-LE considered here, though allowing an exact analytical
solution, has a physical drawback. The algebraic law of the velocity
correlation $\mathcal{T}(T)=T^{-\gamma }$ cannot be valid for short times.\
For an obvious physical reason, $\underset{T\rightarrow 0}{\lim }\mathcal{T}%
(T)=1$ [as required by Eq. (\ref{eeq.49})], instead of diverging as in the
present model. The model should certainly be improved by using a correlation
function that satisfies this constraint.

We now generalize our model by considering, instead of a velocity depending
on a single time variable, a \textit{velocity field in two dimensions, for
which the velocity depends on both time and space }$V(X,T)$.

\section{Random Velocity Field with Algebraic Time-correlation\label{V-field}%
}

We now return to Eq. (\ref{eeq.3}), and consider a true velocity field, 
\textit{i.e.}, a function both of space and time.\ We immediately note that
this generalization introduces nothing new in one dimension.\ We recall the
important condition of zero divergence (\ref{eeq.3}) which ensures the
correct physical status of the hybrid kinetic equation, \textit{i.e.}, the
equivalence of Eqs. (\ref{eeq.4}) and (\ref{eeq.5}). Clearly, the constraint
(\ref{eeq.3}) reduces, in one dimension, to $\partial _{X}V(X,T)=0$, hence
the velocity can only depend on time, and we are brought back to the
situation of Sec. \ref{V(t)}. In order to study a non-trivial problem, we
must consider $d\geq 2$. Specifically, we shall consider in the rest of the
present work \textit{two-dimensional systems}, $d=2$. As will be shown
below, such systems are actually of much greater physical interest than
one-dimensional ones.

We thus consider a particle whose position is specified by the vector $%
\mathbf{X}=(X,Y)$ and the velocity $\mathbf{V}=(V_{x},V_{y})$. The
corresponding (vectorial) V-Langevin equation is:

\begin{equation}
\frac{d\mathbf{X}(T)}{dT}=\mathbf{V[X}(T),T],\ \ \ \ \ \ \ \ \mathbf{X}(0)=%
\mathbf{0}.  \label{eeq.82}
\end{equation}

As a result of the previous discussion, we may associate with this
V-Langevin equation a hybrid kinetic equation, \textit{provided that the
velocity field has zero divergence:}

\begin{equation}
\mathbf{\nabla \cdot V(X},T)\equiv \partial _{X}V_{x}(\mathbf{X},T)+\partial
_{Y}V_{y}(\mathbf{X},T)=0.  \label{eeq.83}
\end{equation}

Contrary to the one-dimensional case, this condition can be realized with a
non-trivial velocity \textit{field}. The associated hybrid kinetic equation
possesses both required properties: it conserves the normalization of $f$
and its characteristic equations are Eqs. (\ref{eeq.82}).\ Thus, the two
following forms are equivalent:

\begin{equation}
\partial _{T}f(\mathbf{X},T)+\mathbf{\nabla \cdot }\left[ \mathbf{V(X},T)~f(%
\mathbf{X},T)\right] =0,  \label{eeq.84}
\end{equation}

\begin{equation}
\partial _{T}f(\mathbf{X},T)+\mathbf{V(X},T)\cdot \mathbf{\nabla }f(\mathbf{X%
},T)=0  \label{eeq.85}
\end{equation}

We note that the condition of zero divergence of a two-dimensional velocity
field is realized, in particular, by the intensely studied model of a
charged particle moving perpendicularly to a strong constant magnetic field $%
\mathbf{B}$, in presence of a fluctuating electrostatic field (see, \textit{%
e.g.} \cite{RB}, where many other references are given]. The coordinates of
its guiding centre are animated by the well-known electrostatic drift
velocity (in Gaussian units):

\begin{eqnarray}
\frac{dX(T)}{dT} &=&-\frac{c}{B}~\left. \frac{\partial \Phi \lbrack X,Y,T]}{%
\partial Y}\right\vert _{\mathbf{X=X(}T)},  \notag \\
\frac{dY(T)}{dT} &=&\frac{c}{B}~\left. \frac{\partial \Phi \lbrack X,Y,T]}{%
\partial X}\right\vert _{\mathbf{X=X(}T)},  \label{eeq.86}
\end{eqnarray}

where $\Phi (\mathbf{X},T)$ is the (fluctuating) scalar potential. The
statistical properties of the velocity field are thus derived from the
statistical definition of the potential. We assume that the latter is a
Gaussian centred, homogeneous, isotropic and stationary random process,
defined by a vanishing first moment and by the following factorized form of
the second moment:

\begin{eqnarray}
\left\langle \Phi (\mathbf{X},T)\right\rangle &=&0,  \notag \\
\left\langle \Phi (\mathbf{0},0)~\Phi (\mathbf{X},T)\right\rangle &=&%
\mathcal{E}(\mathbf{X})~\mathcal{T}(T).  \label{eeq.87}
\end{eqnarray}

The second relation defines the \textbf{Eulerian potential correlation}, 
\textit{i.e.}, the correlation between the values of the potential at the
origin and at a \textit{fixed} point $\mathbf{X}$ at time $T$. In the
present work we consider only the case where the spatial part of the
correlation has a Gaussian form:

\begin{equation}
\mathcal{E}(\mathbf{X})=\exp \left( -\frac{X^{2}+Y^{2}}{2\lambda ^{2}}%
\right) ,  \label{eeq.88}
\end{equation}

where $\lambda $ is the so-called \textit{correlation length}. The temporal
part $\mathcal{T}(T)$ will not yet be specified explicitly. It should,
however, satisfy the condition: $\mathcal{T}(0)=1$, and is supposed to
depend only on the absolute value of $T$.

It is convenient to introduce from here on dimensionless quantities (denoted
by lower case letters), defined as follows:

\begin{equation}
x=\frac{X}{\lambda },\ \ y=\frac{Y}{\lambda },\ \ t=\frac{T}{T_{c}},\ \ \phi
=\varepsilon ^{-1}\Phi .  \label{eeq.89}
\end{equation}

Here $\varepsilon $ is a measure of the intensity of the potential
fluctuations, \textit{e.g.}, the root mean square value of the potential
fluctuations. $T_{c}$ is an intrinsic characteristic time, to be defined
later [Sec. \ref{Algeb Gen}]. The V-Langevin equations of motion (\ref%
{eeq.86}) reduce to:

\begin{gather}
\frac{d\mathbf{x}(t)}{dt}=K~\mathbf{v[x}(t),t],  \notag \\
v_{x}(\mathbf{x},t)=-\frac{\partial \phi (\mathbf{x},t)}{\partial y},\ \ \
v_{y}(\mathbf{x},t)=\frac{\partial \phi (\mathbf{x},t)}{\partial x}.
\label{eeq.90}
\end{gather}

The dimensionless constant $K$ is called the \textit{Kubo number:}

\begin{equation}
K=\frac{c}{B}~\frac{\varepsilon T_{c}}{\lambda ^{2}}.  \label{eeq.91}
\end{equation}

Given the form of Eqs. (\ref{eeq.90}) it is even more convenient to use a
further reduction of the time variable:

\begin{equation}
\theta =Kt.  \label{eeq.92}
\end{equation}

The V-Langevin equations then reduce to:

\begin{equation}
\frac{d\mathbf{x}(\theta )}{d\theta }=\mathbf{v[x}(\theta ),\theta ],\ \ \ \ 
\mathbf{x}(0)=\mathbf{0}.  \label{eeq.93}
\end{equation}

The relevant dimensionless Eulerian correlations are now defined as follows:

\begin{eqnarray}
\left\langle \phi (\mathbf{0},0)~\phi (\mathbf{x,}\theta )\right\rangle &=&%
\mathcal{E(}\mathbf{x)~}\mathcal{T(}K^{-1}\theta ),  \notag \\
\left\langle v_{j}(\mathbf{0},0)~v_{n}(\mathbf{x,}\theta )\right\rangle &=&%
\mathcal{E}_{jn}\mathcal{(}\mathbf{x)~}\mathcal{T(}K^{-1}\theta ),
\label{eeq.94}
\end{eqnarray}

with (\ref{eeq.88}):

\begin{eqnarray}
\mathcal{E}_{xx}(x,y) &=&-\frac{\partial ^{2}}{\partial y^{2}}\mathcal{E}%
(x,y)=(1-y^{2})\mathcal{E}(x,y),  \notag \\
\mathcal{E}_{yy}(x,y) &=&-\frac{\partial ^{2}}{\partial x^{2}}\mathcal{E}%
(x,y)=(1-x^{2})\mathcal{E}(x,y),  \notag \\
\mathcal{E}_{xy} &=&\mathcal{E}_{yx}=\frac{\partial ^{2}}{\partial x\partial
y}\mathcal{E(}x,y)=xy\mathcal{E(}x,y).  \label{eeq.95}
\end{eqnarray}

A first, well-known general result is obtained from the V-Langevin equation (%
\ref{eeq.93}): the \textit{mean square deviation (MSD)} in the $x$-direction
is expressed as follows:

\begin{equation}
\left\langle x^{2}(\theta \right\rangle =2\int_{0}^{\theta }d\tau ~(\theta
-\tau )~\mathcal{L}_{xx}(\tau ).  \label{eeq.96}
\end{equation}

Here $\mathcal{L}_{xx}(\theta )$ is the \textbf{Lagrangian velocity
correlation}, calculated along the trajectory $\mathbf{x}(\theta )$ of the
particle, \textit{i.e.}, using the solution of Eq. (\ref{eeq.93}):

\begin{equation}
\mathcal{L}_{xx}(\theta )=\left\langle v_{x}(\mathbf{0},0)~v_{x}[\mathbf{x}%
(\theta ),\theta ]\right\rangle .  \label{eeq.97}
\end{equation}

The MSD\ is related to the dimensionless \textit{running diffusion
coefficient} $\mathcal{D}_{x}(\theta )$ in the $x$-direction by the
well-known Einstein relation:\footnote{%
In the general case, the factor $1/2$ in Eq. (\ref{eeq.98}) is to be
replaced by $1/2d$. In the present situation, although the system is
two-dimensional, we are looking for the diffusion in the single direction $x$%
, hence we must take $d=1$. If we calculated $\left\langle r^{2}(\theta
)\right\rangle ,$ we should take $d=2$.
\par
{}}

\begin{equation}
\mathcal{D}_{x}(\theta )=\frac{1}{2}~\frac{d\left\langle x^{2}(\theta
)\right\rangle }{d\theta }=\int_{0}^{\theta }d\tau ~\mathcal{L}_{xx}(\tau ).
\label{eeq.98}
\end{equation}

We now consider the evolution of the density profile $n(\mathbf{x},\theta
)=\left\langle f(\mathbf{x,}\theta \right\rangle $. The distribution
function is governed by the hybrid kinetic equation (\ref{eeq.84}), or
equivalently, (\ref{eeq.85}). The latter is treated by a straightforward
generalization of the reasoning leading from Eq. (\ref{eeq.51}) to (\ref%
{eeq.56}); it will not be repeated here. The final result differs from the
latter equation in that the simple one-dimensional velocity correlation $%
\left\langle v(\theta _{1})~v(\theta )\right\rangle $ is replaced by the 
\textit{Lagrangian velocity correlation tensor} $\left\langle v_{j}[\mathbf{x%
}(\theta _{1}),\theta _{1}]~v_{k}[\mathbf{x}(\theta ),\theta ]\right\rangle =%
\mathcal{L}_{jk}(\theta -\theta _{1})$:

\begin{equation}
\partial _{\theta }n(\mathbf{x},\theta )=\nabla _{j}\int_{0}^{\theta }d\tau ~%
\mathcal{L}_{jk}(\tau )~\nabla _{k}~n(\mathbf{x},\theta -\tau ).
\label{eeq.99}
\end{equation}

Clearly, the relation (\ref{eeq.96}) can be obtained directly from Eq. (\ref%
{eeq.99}).

\section{The Corrsin Approximation\label{Corrsin}}

The study of transport is now reduced to the (difficult) determination of
the Lagrangian velocity correlation. More precisely, we want to find a
relation between the (given) Eulerian velocity correlation and the
Lagrangian one. In the present work we will use only the simplest
approximation method for this problem: the \textit{Corrsin approximation }%
\cite{Corrsin}, \cite{RB}.

We rewrite the exact Lagrangian velocity correlation as follows:

\begin{gather}
\mathcal{L}_{jn}(\theta )=\left\langle v_{j}(\mathbf{0},0)~v_{n}[\mathbf{x}%
(\theta ),\theta ]\right\rangle  \notag \\
=\int d\mathbf{x}~\left\langle v_{j}(\mathbf{0},0)~v_{n}(\mathbf{x},\theta
)~\delta \lbrack \mathbf{x}-\mathbf{x}(\theta )]\right\rangle
\label{eeq.100}
\end{gather}

In the Corrsin approximation, the average in the integrand is assumed to be
factorized as follows:

\begin{equation}
\mathcal{L}_{jn}(\theta )\approx \int d\mathbf{x}~\left\langle v_{j}(\mathbf{%
0},0)~v_{n}(\mathbf{x},\theta )\right\rangle ~~\left\langle \delta \lbrack 
\mathbf{x}-\mathbf{x}(\theta )]\right\rangle  \label{eeq.101}
\end{equation}

The first factor in the integrand is recognized as the Eulerian correlation
tensor, assumed to be of the form (\ref{eeq.87}), thus:

\begin{equation}
\mathcal{L}_{jn}(\theta )\approx \int d\mathbf{x}~\mathcal{E}_{jn}(\mathbf{x}%
)~\mathcal{T}(\theta /K)~\left\langle \delta \lbrack \mathbf{x}-\mathbf{x}%
(\theta )]\right\rangle .  \label{eeq.102}
\end{equation}

The last factor is evaluated as follows, in the second cumulant
approximation:

\begin{eqnarray}
\left\langle \delta \lbrack \mathbf{x}-\mathbf{x}(\theta )]\right\rangle
&=&\int d\mathbf{k}~e^{i\mathbf{k\cdot x}}~\left\langle e^{-i\mathbf{k\cdot x%
}(\theta )}\right\rangle  \notag \\
&\approx &\int d\mathbf{k}~e^{i\mathbf{k\cdot x}}~\exp \left[ -~\tfrac{1}{2}%
~k^{2}\left\langle x^{2}(\theta )\right\rangle \right]  \notag \\
&=&\frac{1}{2\pi ~\left\langle x^{2}(\theta )\right\rangle }~\exp \left[ -~%
\frac{x^{2}+y^{2}}{2~\left\langle x^{2}(\theta )\right\rangle }\right] .
\label{eeq.103}
\end{eqnarray}

Thus:

\begin{equation}
\mathcal{L}_{jn}(\theta )=\frac{\mathcal{T}(\theta /K)}{2\pi ~\left\langle
x^{2}(\theta )\right\rangle }~\int d\mathbf{x}~\mathcal{E}_{jn}(\mathbf{x}%
)\exp \left[ -~\frac{x^{2}+y^{2}}{2~\left\langle x^{2}(\theta )\right\rangle 
}\right] .  \label{eeq.104}
\end{equation}

We now use the explicit forms (\ref{eeq.95}) for the Eulerian correlation.\
It is immediately seen by symmetry that the non-diagonal components of the
Lagrangian tensor vanish, and the two diagonal ones are equal to each other.

\begin{equation}
\mathcal{L}_{jn}(\theta )=\delta _{jn}~\mathcal{L}(\theta ),  \label{eeq.105}
\end{equation}

with:

\begin{equation*}
\mathcal{L}(\theta )=\frac{\mathcal{T}(\theta /K)}{2\pi \left\langle
x^{2}(\theta )\right\rangle }~\int dx~dy~(1-y^{2})~\exp \left[ -\left( 1+%
\frac{1}{\left\langle x^{2}(\theta )\right\rangle }\right) \frac{x^{2}+y^{2}%
}{2}\right] .
\end{equation*}

The integration over $x$ and $y$ is elementary.\ The result is combined with
Eq. (\ref{eeq.96}) to yield:

\begin{equation}
\mathcal{L}(\theta )=\frac{\mathcal{T}(\theta /K)}{\left[ 1+2\int_{0}^{%
\theta }d\tau ~(\theta -\tau )~\mathcal{L}(\tau )\right] ^{2}}
\label{eeq.106}
\end{equation}

We thus ended with an \textit{integral equation} for the Lagrangian velocity
correlation in the Corrsin approximation \cite{Wang}, \cite{RB}. This
equation will be solved in Secs. \ref{Determ Lagr 1} and \ref{Determ Lagran
2}.

The diagonal character of the Lagrangian velocity correlation introduces a
significant simplification in the equation for the density profile, which
reduces to the following non-Markovian diffusion equation:

\begin{equation}
\partial _{\theta }n(\mathbf{x},\theta )=\int_{0}^{\theta }d\tau ~\mathcal{L}%
(\tau )~\nabla ^{2}n(\mathbf{x},\theta -\tau ).  \label{eeq.107}
\end{equation}

The diffusion coefficients in the $x$- and $y$-directions are equal to each
other. We may therefore drop the subscripts in Eq. (\ref{eeq.98}) and write
simply:

\begin{equation}
\mathcal{D}(\theta )=\int_{0}^{\theta }d\tau ~\mathcal{L}(\tau ).
\label{eeq.108}
\end{equation}

Whenever this integral converges, its limit for $\theta \rightarrow \infty $
defines the ordinary diffusion constant: $\mathcal{D=D}(\infty )$. Whenever $%
\mathcal{D}$ is finite and positive, the regime is normal diffusive.

\section{Algebraic Time Correlation: General Properties\label{Algeb Gen}}

We now make the problem more explicit. We consider a class of V-Langevin
equations (and associated hybrid kinetic equations) for which the temporal
part of the Eulerian velocity correlation, $\mathcal{T}(\theta /K)$, is a
long-tailed function that decays algebraically for long times.\ It is of the
general form (\ref{eeq.63}), but we take now some additional precautions to
ensure a correct behavior at short times. We must, in particular, avoid the
divergence at $T\rightarrow 0$. Physically, any correct (dimensional)
correlation function should have the property: $\mathcal{T}(T=0)=1$. We thus
define a characteristic cut-off time $T_{c}$ as the value of $T$ at which
the right hand side of Eq.\ (\ref{eeq.63}) equals $1$.\ For all times $0\leq
T\leq T_{c}$ we put $\mathcal{T}(T)=1$. Note that a correlation of the form (%
\ref{eeq.63}) has no intrinsic time scale; but the adoption of a cut-off
introduces a characteristic time $T_{c}$. The latter allows us to complete
the definitions of the dimensionless time $t=T/T_{c}$ (\ref{eeq.89}), of the
Kubo number (\ref{eeq.91}) and of the dimensionless variable $\theta $ (\ref%
{eeq.92}). We thus adopt the final form of the dimensionless temporal
Eulerian velocity correlation as:

\begin{equation}
\mathcal{T(}K^{-1}\theta )=\left\{ 
\begin{array}{c}
1,\ \ \ \ \ 0\leq \theta \leq K \\ 
\dfrac{K^{\gamma }}{\theta ^{\gamma }},\ \ \ \ \ \theta \geq K~~\ \ 
\end{array}%
\right. ,\ \ \ 0<\gamma <1  \label{eeq.109}
\end{equation}

For convenience, $\gamma $ will be called the \textbf{Eulerian exponent}; by
definition it will be limited to the range ($0,1$). Anticipating the results
of the next Section, it will be shown that the Lagrangian correlation
function $\mathcal{L}(\theta )$ has the same general algebraic form, with
different values of the exponent and of the coefficients:

\begin{equation}
\mathcal{L(}\theta )=\left\{ 
\begin{array}{c}
1,\ \ \ \ \ \ \ \ \ \ \ \ \ \ \ \ \ 0\leq \theta \leq \theta _{L}, \\ 
\dfrac{h}{\theta ^{\Lambda }},\ \ \ h>0,\ \ \ \ \ \theta >\theta _{L}.~~\ \ 
\end{array}%
\right.  \label{eeq.110}
\end{equation}

In order to ensure a correct connection of the two parts, we must take:

\begin{equation}
\theta _{L}=h^{1/\Lambda }.  \label{eeq.111}
\end{equation}

$\Lambda $ will be called the \textbf{Lagrangian exponent}: its value
depends on the Eulerian exponent $\gamma $ and on $K$. It will be determined
in the next sections.

Before going into specific calculations, we wish to derive an important
property relating the nature of the macroscopic transport to the value of
the Lagrangian exponent. Combining the definition of the dimensionless
running diffusion coefficient (\ref{eeq.108}) with the form (\ref{eeq.110})
we obtain:

\begin{equation}
\mathcal{D}(\theta )=\left\{ 
\begin{array}{c}
\int_{0}^{\theta }d\tau ~1,\ \ \ \ \ \ \ \ \ \ \ \ \ \ \ \ \ \ 0\leq \theta
\leq \theta _{L}, \\ 
\int_{0}^{\theta _{L}}d\tau ~1+\int_{\theta _{L}}^{\theta }d\tau ~h~\theta
^{-\Lambda },\ \ \ \ \theta >\theta _{L}.%
\end{array}%
\right.  \label{eeq.112}
\end{equation}

The integrals are easily evaluated:

\begin{equation}
\mathcal{D(\theta )=}\left\{ 
\begin{array}{c}
\theta ,\ \ \ \ \ \ \ \ \ \ \ \ \ \ \ \ \ \ \ \ \ \ \ \ \ \ \ \ \ \ \ \ \ \
\ \ \ \ 0\leq \theta \leq \theta _{L,} \\ 
\theta _{L}+\dfrac{h}{1-\Lambda }\left( \theta ^{-\Lambda +1}-\theta
_{L}^{-\Lambda +1}\right) ,\ \ \ \ \theta >\theta _{L}.%
\end{array}%
\right. ,\ \ \ \Lambda \neq 1.  \label{eeq.113}
\end{equation}

The more interesting long-time part is rewritten as the sum of a constant
and of a time-dependent term; using also Eq. (\ref{eeq.111}) we find:

\begin{equation}
\mathcal{D}(\theta )=\frac{\Lambda }{\Lambda -1}~h^{1/\Lambda }+\frac{h}{%
1-\Lambda }~\theta ^{1-\Lambda },\ \ \ \ \ \ \theta >\theta _{L},\ \ \ \
\Lambda \neq 1.  \label{eeq.114}
\end{equation}

We now see the existence of two sharply separated ranges of the Lagrangian
correlation exponent $\Lambda .\bigskip $

A) $\mathbf{0<\Lambda <1}$. In this case the second term in the right hand
side of (\ref{eeq.114}) is a monotonically \textit{increasing function of
time. }Hence, for sufficiently long times this term strongly dominates the
first constant term, and the running diffusion coefficient behaves
asymptotically as:

\begin{equation}
\mathcal{D}(\theta )\underset{\theta \rightarrow \infty }{\sim }\theta
^{1-\Lambda },\ \ \ \ \ \Lambda <1.  \label{eeq.115}
\end{equation}

The regime is thus (time-fractional) \textbf{superdiffusive}.\bigskip

B) $\mathbf{\Lambda =1.}$ Eq. (\ref{eeq.113}) is no longer valid, but the
integral in (\ref{eeq.112}) can be evaluated as:

\begin{equation}
\mathcal{D}(\theta )=h^{1/\Lambda }+h~\ln \left( \frac{\theta }{h^{1/\Lambda
}}\right) ,\ \ \ \ \Lambda =1.  \label{eeq.116}
\end{equation}

The regime is "\textbf{weakly superdiffusive}".\bigskip

C) $\mathbf{\Lambda >1}$. The behavior is radically different.\ The second
term in the right hand side of Eq. (\ref{eeq.114}) is now a monotonically 
\textit{decreasing function of time.} For long times this term decreases to
zero.\ Hence the running diffusion coefficient tends toward a \textit{%
positive constant}:

\begin{equation}
\mathcal{D(\theta )}\underset{\theta \rightarrow \infty }{\sim }\mathcal{D}=%
\frac{\Lambda }{\Lambda -1}~h^{1/\Lambda }>0,\ \ \ \ \ \Lambda >1.
\label{eeq.117}
\end{equation}

The regime is now \textbf{normal diffusive} for all values of $\Lambda >1$.\
The asymptotic $\mathcal{D}$ is the ordinary diffusion constant. \bigskip

\begin{figure}[tbph]
\centerline{\includegraphics[height=2.1698in]{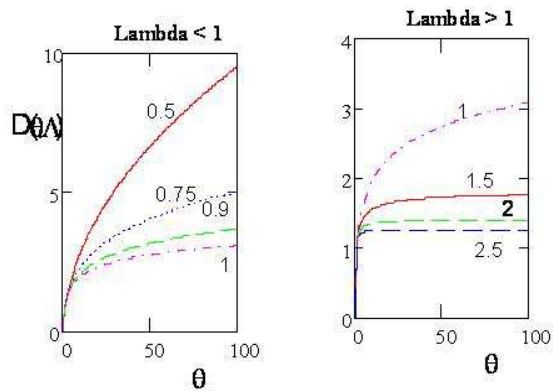}}
\caption{Running diffusion coefficient $%
\mathcal{D}(\protect\theta ,\Lambda )$ obtained from the kinetic equation,
for various values of $\Lambda <1$ and $\Lambda >1$.}
\label{LFDE 5}
\end{figure}

Clearly, the value $\Lambda =1$ is a very special one: as $\Lambda $ reaches 
$1$ from below, the regime changes abruptly from a superdiffusive to a
normal diffusive one, and remains so for all values of $\Lambda >1$. This
phenomenon is illustrated in Fig. \ref{LFDE 5}.

%%%%%%%%%%%%%%%%%%%%%%%%%%%%%%%%%%%%%%%%%%%%%%%%%%%%%%%%%%%%%%%%%%%%% 
%%%%%%%%%%%%%%%%%%%%%%%%%%%%%%%%%%%%%%%%%%%%%%%%%%%%%%%%%%%%%%%%%%%%%

\begin{figure}[tbph]
\centerline{\includegraphics[height=2.655in]{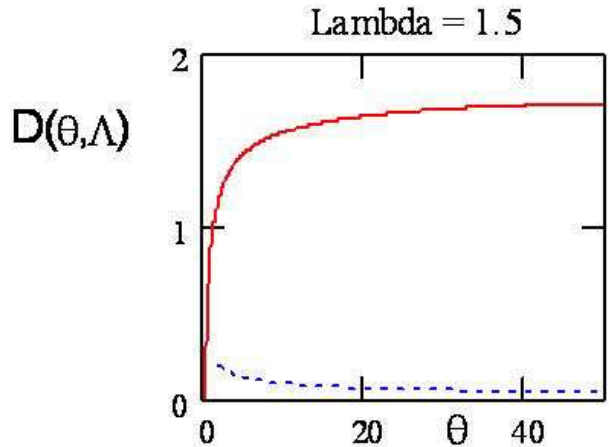}}
\caption{Running diffusion coefficient for 
$\Lambda =1.5$, as obtained from the kinetic equation (solid red) and as
predicted from the CTRW (dotted blue).}
\label{LFDE 6}
\end{figure}

The difference between the normal diffusive regime obtained from the kinetic
equation and the subdiffusive regime predicted by the CTRW for the same
value of $\Lambda >1$ appears clearly in Fig. \ref{LFDE 6}.

%%%%%%%%%%%%%%%%%%%%%%%%%%%%%%%%%%%%%%%%%%%%%%%%%%%%%%%%%%%%%%%%%%%%% 
%%%%%%%%%%%%%%%%%%%%%%%%%%%%%%%%%%%%%%%%%%%%%%%%%%%%%%%%%%%%%%%%%%%%%

In order to obtain a clearer picture of what happens at this point, we study
the density profile. More specifically, we investigate whether its evolution
can be described by an equivalent FDE, as was found in Sec. \ref{CTRW FDE}.

As shown in Sec. \ref{V(t)}, there exists a \textit{formal }equivalence
between, on one hand, the non-Markovian diffusion equation (\ref{eeq.107})
obtained from the hybrid kinetic equation and on the other hand, the
Montroll-Shlesinger equation for a CTRW in the fluid limit. It requires the
identification (\ref{eeq.61}) of the Lagrangian velocity correlation with
the memory function of the CTRW (in dimensionless variables):

\begin{equation}
\mathcal{L}(\theta )\rightarrow \phi (\theta ),  \label{eeq.118}
\end{equation}

or, in Laplace representation:

\begin{equation}
\widetilde{\mathcal{L}}(s)\rightarrow \widetilde{\phi }(s).  \label{eeq.119}
\end{equation}

The Laplace transform of the Lagrangian correlation (\ref{eeq.110}) is, for
long times:

\begin{equation}
\widetilde{\mathcal{L}}(s)\sim s^{\Lambda -1}.  \label{eeq.120}
\end{equation}

Comparing with Eq. (\ref{eeq.14}) we find that the temporal exponent of the
"formally equivalent" FDE is:

\begin{equation}
\beta =2-\Lambda .  \label{eeq.121}
\end{equation}

Substituting this value into Eq. (\ref{eeq.16}) we find a result similar to (%
\ref{eeq.65}), with $\Lambda $ replacing $\gamma .$\ The resolvent
given by the Montroll-Weiss equation is:

\begin{equation}
\widehat{\widetilde{n}}(k,s)=\frac{s^{1-\Lambda }}{s^{2-\Lambda }+\frac{1}{2}%
k^{2}}  \label{eeq.122}
\end{equation}

The difference with (\ref{eeq.65}) is that we should not limit consideration
to $\Lambda <1$. We consider, instead the various cases discussed
above.\bigskip

A) $\mathbf{0<\Lambda <1}$. The situation is the same as in Sec. \ref{Algeb
Correl}.\ There is indeed a complete identity between the kinetic result and
the FDE with $2-\Lambda $, \textit{i.e., }$1<\beta <2$. As shown, however,
in Sec. \ref{Algeb Correl}, there is no equivalent CTRW in this case. The
regime is one of \textit{time-fractional superdiffusion}.\ The density
profile is provided explicitly by the Mainardi function (\ref{eeq.75})%
\footnote{%
Because of the presence of the cut-off in the present problem, only the
asymptotic form (\ref{eeq.79}) is really relevant here.}. It has two moving
maxima, as represented in Fig. \ref{LFDE 3} and decays asymptotically as a
stretched exponential. The diffusion exponent $\mu =2-\Lambda $. All this is
in agreement with the kinetic result.\bigskip

B) $\mathbf{\Lambda =1}$. The resolvent (\ref{eeq.122}) of the FDE reduces
to:

\begin{equation}
\widehat{\widetilde{n}}(k,s)=\frac{1}{s+k^{2}},\ \ \ \ (\Lambda =1)
\label{eeq.123}
\end{equation}

This is the Fourier-Laplace transform of a \textit{Gaussian packet},
characteristic of a \textit{normal diffusive regime}. On the other hand, the
result (\ref{eeq.116}) shows that the running diffusion coefficient obtained
from the kinetic equation grows, slowly but indefinitely in time. The FDE is
no longer equivalent to kinetic theory: $\Lambda =1$ appears as a \textit{%
bifurcation point}. \bigskip

C) $\mathbf{1<\Lambda <2}$. The propagator (\ref{eeq.122}) corresponds now
to a FDE with $\alpha =2,\ 0<\beta <1$. It describes a \textit{%
time-fractional subdiffusion,} which can be derived as the fluid limit of a
CTRW. The corresponding propagator has a single maximum at the origin, and
decays according to a stretched exponential [see Eq. (\ref{eeq.43})]. This
regime has been studied in detail in \cite{RB}. On the other hand, the
kinetic theory predicts a \textit{normal diffusive} regime throughout this
range of $\Lambda $ [Eqs. (\ref{eeq.114}), (\ref{eeq.117})]. There appears
now a \textit{divorce between CTRW, FDE and kinetic theory. }$\Lambda =1$
has indeed the property of a bifurcation point: as the parameter $\Lambda $
increases from $0$; the propagator first follows the "FDE branch". At $%
\Lambda =1$ it leaves this branch and follows the completely different
"diffusive branch".

The kinetic equation for the propagator in this regime is obtained from Eqs.
(\ref{eeq.99}) and (\ref{eeq.110}):

\begin{equation}
\partial _{\theta }n(x,\theta )=\theta _{L}\nabla ^{2}n(x,\theta
)+~h\int_{\theta _{L}}^{\theta }d\tau ~\tau ^{-\Lambda }~\nabla
^{2}n(x,\theta -\tau ),\ \ \ \theta >\theta _{L}  \label{eeq.124}
\end{equation}

For long enough time, the equation can be Markovianized by neglecting the
retardation in the right hand side.\ This approximation is justified by our
knowledge that a constant diffusion coefficient exists. The final asymptotic
equation is an ordinary diffusion equation:

\begin{equation}
\partial _{\theta }n(x,\theta )=\mathcal{D}~\nabla ^{2}n(x,\theta ),
\label{eeq.125}
\end{equation}

with $\mathcal{D}$ given by Eq. (\ref{eeq.117}). Nevertheless, for any
finite time this Markovianization is not reliable, because of the
non-exponential Lagrangian correlation, hence the density profile is in
general non-Gaussian.\bigskip

D) $\mathbf{\Lambda >2.}$ In this range the divorce between kinetic theory
and FDE is complete.\ The latter looses its meaning.\ Indeed, it would
correspond to $\beta <0$, which would describe a waiting time distribution
that is an increasing function of time. Alternatively, its Laplace transform
diverges: $\widetilde{\psi }(s)\sim s^{-|\beta |}\rightarrow \infty $ as $%
s\rightarrow 0$. On the other hand, the Markovianization of the kinetic
equation (\ref{eeq.124}) is more and more justified as $\mathcal{L}(\theta )$
decreases faster for large $\theta $.

\section{Lagrangian Velocity Correlation:\protect\bigskip\ First
Approximation \label{Determ Lagr 1}}

\begin{figure}[tbph]
\centerline{\includegraphics[height=2.5872in]{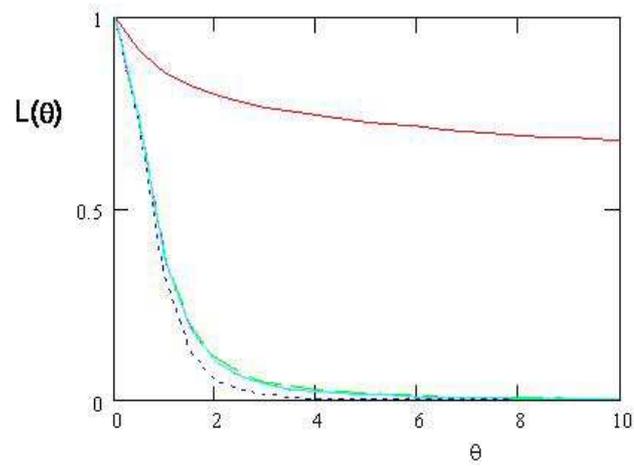}}
\caption{Solution of Eq. (\protect\ref%
{eeq.106}) for the Lagrangian velocity correlation.\ Red solid: starting
(Eulerian) correlation; Blue dotted: First iteration.\ The second, third and
fourth iterations are superposed on the green curve. The parameters are: $%
K=0.2$, $\protect\gamma =0.1.$}
\label{LFDE 7}
\end{figure}

We now determine the Lagrangian velocity correlation as a function of time
and of the parameters $\gamma $ and $K$ of the input Eulerian velocity
correlation, by solving the integral equation (\ref{eeq.106}).\ The
procedure used is one of successive numerical iterations $\mathcal{L}%
^{(n)}(\theta )$, starting with the natural Eulerian trial function $%
\mathcal{L}^{(0)}(\theta )=\mathcal{T}(\theta /K)$. We checked the
convergence of the procedure.\ As can be seen in Fig. \ref{LFDE 7}, the
second iteration provides already an excellent approximation.

%%%%%%%%%%%%%%%%%%%%%%%%%%%%%%%%%%%%%%%%%%%%%%%%%%%%%%%%%%%%%%%%%%%%% 
%%%%%%%%%%%%%%%%%%%%%%%%%%%%%%%%%%%%%%%%%%%%%%%%%%%%%%%%%%%%%%%%%%%%%

The first very important qualitative result is the following. An input
Eulerian correlation (\ref{eeq.109}) with an algebraic tail produces a
Lagrangian correlation having the same type of algebraic tail.\ This is
clearly seen in a log-log representation of the result, as shown in Fig. \ref%
{LFDE 8}. We thus confirm the anticipated form (\ref{eeq.111}). Our main
task will be the determination of the Lagrangian exponent $\Lambda $, of the
corresponding "intensity" $h$ and of the cut-off time $\theta _{L}$ as
functions of the input Eulerian exponent $\gamma $ and of the Kubo number $K$%
.

%%%%%%%%%%%%%%%%%%%%%%%%%%%%%%%%%%%%%%%%%%%%%%%%%%%%%%%%%%%%%%%%%%%%% 
\begin{figure}[tbph]
\centerline{\includegraphics[height=2.5872in]{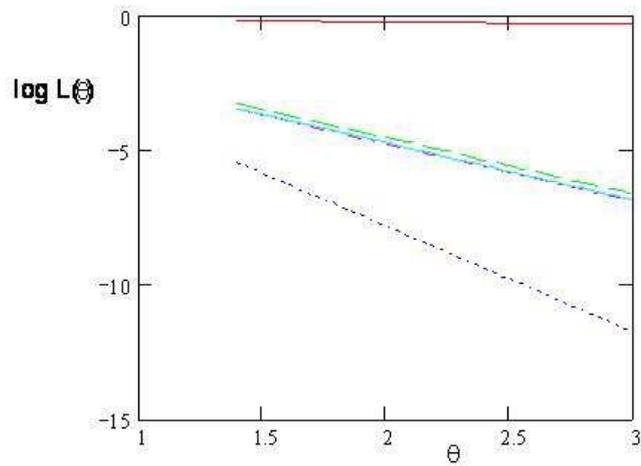}}
\caption{Same data as in Fig.  \protect
\ref{LFDE 7}, in log-log representation.}
\label{LFDE 8}
\end{figure}
%%%%%%%%%%%%%%%%%%%%%%%%%%%%%%%%%%%%%%%%%%%%%%%%%%%%%%%%%%%%%%%%%%%%%

A first glance at Fig. \ref{LFDE 7} shows that, even for relatively small
values of $K$, the Lagrangian correlation deviates very quickly from the
Eulerian one: this is clearly seen in Fig. \ref{LFDE 8}. Thus, the
nonlinearity, measured by $K,$ is far from producing just a small
perturbation.

%%%%%%%%%%%%%%%%%%%%%%%%%%%%%%%%%%%%%%%%%%%%%%%%%%%%%%%%%%%%%%%%%%%%% 
\begin{figure}[tbph]
\centerline{\includegraphics[height=2.7735in]{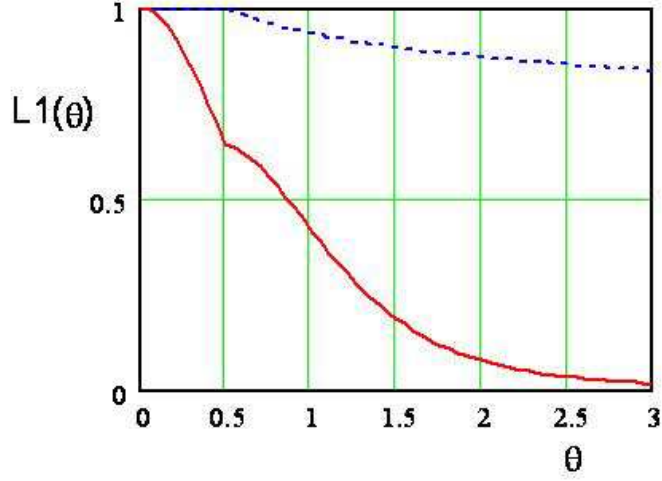}}
\caption{First iterate of the Lagrangian
velocity correlation $\mathcal{L}^{(1)}(\protect\theta )$, (solid red)
compared to the Eulerian correlation $\mathcal{T(\protect\theta )}$ (dotted
blue) at short times.\ $K=0.5,\ \protect\gamma =0.1.$}
\label{LFDE 9}
\end{figure}
%%%%%%%%%%%%%%%%%%%%%%%%%%%%%%%%%%%%%%%%%%%%%%%%%%%%%%%%%%%%%%%%%%%%%

The first iteration is not quantitatively accurate (Fig. \ref{LFDE 7}), but
it has the advantage that it can be evaluated analytically.\ It is therefore
interesting to study explicitly some of its properties which provide a
global qualitative picture. The explicit form is (\ref{eeq.106}):

\begin{equation}
\mathcal{L}^{(1)}(\theta )=\frac{\mathcal{T(\theta }/K)}{\left[
1+2\int_{0}^{\theta }d\tau ~(\theta -\tau )~\mathcal{T}(\tau /K)\right] ^{2}}%
.  \label{eeq.126}
\end{equation}

The integral in the denominator, combined with the definition (\ref{eeq.109}%
), can be written in terms of two other integrals as follows:

\begin{equation}
\int_{0}^{\theta }d\tau ~(\theta -\tau )~\mathcal{T}(\tau /K)=\left\{ 
\begin{array}{c}
I_{1}(\theta ),\ \ \ \ \ \ \ \ \ \ \ \ \theta <K \\ 
I_{1}(K)+I_{2}(\theta ),\ \ \ \ \ \theta >K\ \ \ \ \ \ \ \ 
\end{array}%
\right.  \label{eeq.127}
\end{equation}

where:

\begin{equation}
I_{1}(\theta )=\int_{0}^{\theta }d\tau ~(\theta -\tau )=\tfrac{1}{2}~\theta
^{2},  \label{eeq.128}
\end{equation}

\begin{gather}
I_{2}(\theta )=\int_{K}^{\theta }d\tau ~(\theta -\tau )~\frac{K^{\gamma }}{%
\tau ^{\gamma }}  \notag \\
=\frac{1}{(1-\gamma )(2-\gamma )}~K^{\gamma }\theta ^{2-\gamma }-\frac{1}{%
1-\gamma }~K\theta +\frac{1}{2-\gamma }~K^{2}.  \label{eeq.129}
\end{gather}

Combining these partial results, we finally obtain:

\begin{equation}
\mathcal{L}^{(1)}(\theta )=\left( 1+\theta ^{2}\right) ^{-2},\ \ \ \ \ \
\theta <K,  \label{eeq.130}
\end{equation}

\begin{eqnarray}
\mathcal{L}^{(1)}(\theta ) &=&\dfrac{K^{\gamma }}{\theta ^{\gamma }\left[ 1+%
\dfrac{2}{(1-\gamma )(2-\gamma )}K^{\gamma }\theta ^{2-\gamma }-\dfrac{2}{%
(1-\gamma )}K\theta +\dfrac{4-\gamma }{(2-\gamma )}K^{2}\right] ^{2}},\ \ \
\theta >K.  \notag \\
&&  \label{eeq.131}
\end{eqnarray}

The behavior of this function for \textit{short times} is very simple:
starting from $1$, it decreases parabolically:

\begin{equation}
\mathcal{L}^{(1)}(\theta )\sim 1-2\theta ^{2},\ \ \ \ \theta \ll K
\label{eeq.132}
\end{equation}

As $\theta $ reaches $K$, we find, of course, that the two branches (\ref%
{eeq.130}) and (\ref{eeq.131}) join each other at the same point.\ But the
derivative at $\theta =K$ is discontinuous.\ Indeed, consider the two
integrals (\ref{eeq.128}) and (\ref{eeq.129}); one easily finds:

\begin{equation*}
\left. \frac{d}{d\theta }I_{1}(\theta )\right\vert _{\theta =K}=K,\ \ \ \ \
\left. \frac{d}{d\theta }I_{2}(\theta )\right\vert _{\theta =K}=0.
\end{equation*}

Thus, $\mathcal{L}^{(1)}(\theta )$ exhibits a kink at $\theta =K$, as can be
seen in Fig. \ref{LFDE 9}.

The \textit{asymptotic behavior for large }$\theta $ is more interesting.\
Here we must distinguish two ranges of the Eulerian exponent $\gamma $.
\bigskip

a) $\mathbf{\gamma <1}$.\ In this case the term proportional to $\theta
^{2-\gamma }$ dominates all others in the denominator of Eq. (\ref{eeq.131})
for large $\theta $.\ We thus find:

\begin{equation*}
\mathcal{L}^{(1)}(\theta )\sim \frac{K^{\gamma }}{\theta ^{\gamma }~\frac{4}{%
(1-\gamma )^{2}(2-\gamma )^{2}}~K^{2\gamma }\theta ^{4-2\gamma }}.
\end{equation*}

We thus find, indeed, for very small $\gamma $, an algebraic tail of the
form (\ref{eeq.110}):

\begin{equation}
\mathcal{L}^{(1)}(\theta )\sim \frac{h^{(1)}(K,\gamma )}{\theta ^{4-\gamma }}%
,\ \ \ \theta \gg K,\ \ \ \ \gamma \ll 1,  \label{eeq.133}
\end{equation}

with:

\begin{equation}
h^{(1)}(K,\gamma )=\left[ \frac{(1-\gamma )(2-\gamma )}{2}\right]
^{2}K^{-\gamma }.  \label{eeq.134}
\end{equation}

Next, we note that the Lagrangian exponent in this approximation is:

\begin{equation}
\Lambda ^{(1)}(\gamma )=4-\gamma ,\ \ \ \gamma \ll 1  \label{eeq.135}
\end{equation}

Although this value is not accurate (as can be seen from Fig. \ref{LFDE 8}),
it has the important property (which will be confirmed) that \textit{it only
depends on }$\gamma $,\textit{\ not on }$K$ (for very small $\gamma )$%
.\bigskip

b) $\mathbf{\gamma >1}$. In this case the dominant term at large $\theta $
in the denominator of Eq. (\ref{eeq.131}) is the term proportional to $%
\theta $. The asymptotic behavior is thus different:

\begin{equation*}
\mathcal{L}^{(1)}(\theta )\sim \frac{K^{\gamma }}{\theta ^{\gamma }~\frac{4}{%
(1-\gamma )^{2}}~K^{2}\theta ^{2}},
\end{equation*}

hence:

\begin{equation}
\mathcal{L}^{(1)}(\theta )\sim \frac{k^{(1)}(K,\gamma )}{\theta ^{2+\gamma }}%
,\ \ \ \ \theta \gg K,\ \ \ \ \gamma \gg 1,  \label{eeq.136}
\end{equation}

where $k^{(1)}(K,\gamma )$ is found from this equation. The important point
is that the Lagrangian exponent takes a different dependence on $\gamma $
for larger values of this exponent:

\begin{equation}
\Lambda ^{(1)}(\gamma )\sim 2+\gamma ,\ \ \ \ \ \ \gamma \gg 1.
\label{eeq.137}
\end{equation}

Thus, $\Lambda ^{(1)}(\gamma )$, which was a decreasing function, starts
increasing with $\gamma $ beyond $\gamma =1$. We may calculate analytically
the exact dependence of $\Lambda ^{(1)}(\gamma )$ by evaluating the slope of
the log-log graph of $\mathcal{L}^{(1)}(\theta )$ (see Fig. \ref{LFDE 8}):

\begin{equation}
\Lambda ^{(1)}(\gamma )=\log [\mathcal{L}^{(1)}(100,K,\gamma )]-\log [%
\mathcal{L}^{(1)}(1000,K,\gamma )]  \label{eeq.138}
\end{equation}

\begin{figure}[tbph]
\centerline{\includegraphics[height=2.655in]{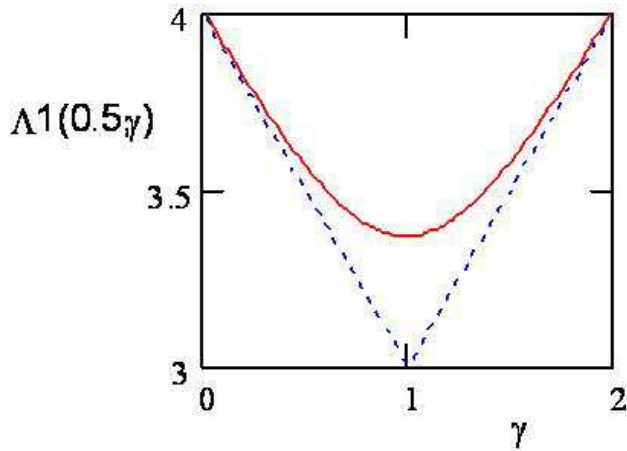}}
\caption{Lagrangian exponent in the first
iteration, as a function of the Eulerian exponent $\protect\gamma .$ $K=0.5$%
. The left (right) dotted straight lines correspond to Eqs. ( \protect\ref%
{eeq.135}) [resp. (\protect\ref{eeq.137})].}
\label{LFDE 10}
\end{figure}

The result is shown in Fig. \ref{LFDE 10}.\ Comparing this type of figures
for different values of $K$, one finds a very weak dependence of $\Lambda $
on $K$ for $\gamma $ close to $1$.

%%%%%%%%%%%%%%%%%%%%%%%%%%%%%%%%%%%%%%%%%%%%%%%%%%%%%%%%%%%%%%%%%%%%% 
%%%%%%%%%%%%%%%%%%%%%%%%%%%%%%%%%%%%%%%%%%%%%%%%%%%%%%%%%%%%%%%%%%%%%

The important consequence of this new "bifurcation" is the fact that, for
all values of $\gamma $ we find:

\begin{equation}
\Lambda ^{(1)}(\gamma )>3,\ \ \ \ \ \forall \gamma .  \label{eeq.139}
\end{equation}

The first iteration thus predicts, in accordance with the discussion at the
end of Sec. \ref{Algeb Gen}, that the first iterate of the Lagrangian
correlation leads to a \textbf{diffusive regime} \textit{for all values of }$%
\gamma $. More than that: the Lagrangian exponent is in the range D), where
there is no equivalence between kinetic theory and CTRW or FDE. This
statement, however, has to be confirmed by the final solution.

\section{Lagrangian Velocity Correlation. Final Numerical Calculation.\label%
{Determ Lagran 2}}

As shown in the previous section, the second iterate of Eq. (\ref{eeq.106})
yields already an excellent approximation, while not requiring long
numerical calculations (we therefore omit the superscript $^{(2)})$.The
short-time ($\theta <K$) part can even be calculated analytically. We thus
have:

\begin{equation}
\mathcal{L(\theta )=}\frac{1}{(1+\theta \arctan \theta )^{2}},\ \ \ \ \theta
<K,  \label{eeq.140}
\end{equation}

\begin{equation}
\mathcal{L}(\theta )=\frac{K^{\gamma }}{\theta ^{\gamma }\left[ 1+K\arctan
K+2\int_{K}^{\theta }d\tau ~(\theta -\tau )~\mathcal{L}^{(1)}(\tau )\right]
^{2}},\ \ \ \ \theta >K.  \label{eeq.141}
\end{equation}

Eq. (\ref{eeq.131}) is substituted in the integral, and the expression is
evaluated numerically.

For short times, the Lagrangian correlation behaves as:

\begin{equation}
\mathcal{L}(\theta )\sim (1+\theta ^{2})^{-2}\sim 1-2\theta ^{2},\ \ \ \ \
\theta \ll K,  \label{eeq.142}
\end{equation}

\begin{figure}[tbph]
\centerline{\includegraphics[height=2.7224in]{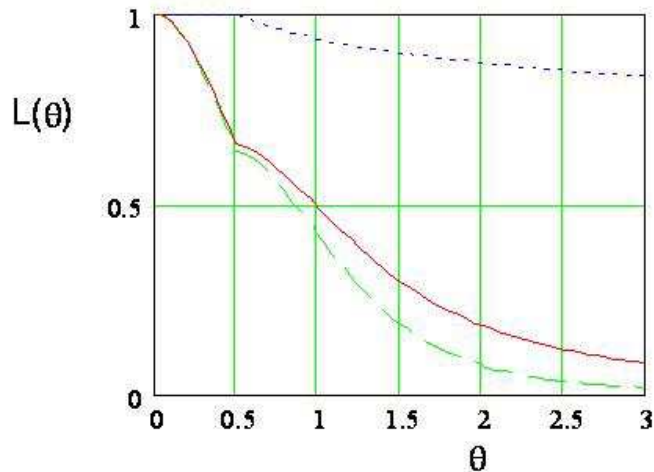}}
\caption{Second iterate of the
Lagrangian velocity correlation $\mathcal{L}^{(2)}(\protect\theta )=\mathcal{%
L}(\protect\theta )$, (solid red) compared to the first iterate $\mathcal{L}%
^{(1)}(\protect\theta )$ (dashed green) and to the Eulerian correlation $%
\mathcal{T(\protect\theta )}$ (dotted blue) at short times.\ $K=0.5,\ 
\protect\gamma =0.1.$}
\label{LFDE 11}
\end{figure}

\textit{i.e.}, exactly as in the first approximation (\ref{eeq.132}). As can
be seen in Fig. \ref{LFDE 11}, the kink which appeared at $\theta =K$ in the
first approximation is still present in the final expression.

%%%%%%%%%%%%%%%%%%%%%%%%%%%%%%%%%%%%%%%%%%%%%%%%%%%%%%%%%%%%%%%%%%%%% 
%%%%%%%%%%%%%%%%%%%%%%%%%%%%%%%%%%%%%%%%%%%%%%%%%%%%%%%%%%%%%%%%%%%%%

The long-time behavior is, of course, more interesting. The first important
point, that is seen in Fig. \ref{LFDE 8}, is that the Lagrangian velocity
correlation behaves asymptotically as a \textit{power law,} (\ref{eeq.110}),
thus confirming our previous anticipation. We now study its parameters.

\begin{figure}[tbph]
\centerline{\includegraphics[height=2.5364in]{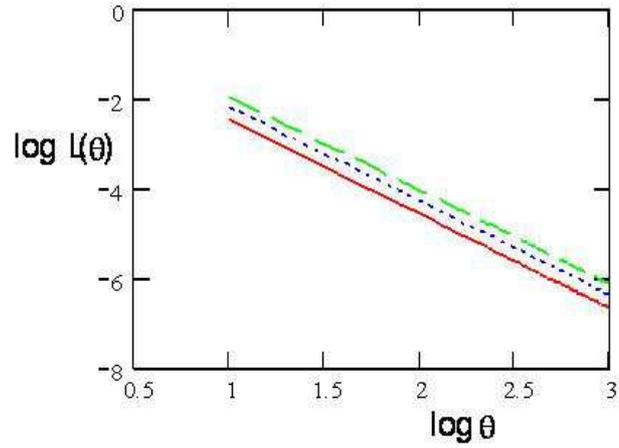}}
\caption{Lagrangian velocity correlation 
\textit{vs.} $\protect\theta $, in log-log representation. The curves
correspond to $K=0.1$ (solid red), $K=0.5$ (dotted blue) and $K=0.7$ (dashed
green). $\protect\gamma =0.1.$}
\label{LFDE 12}
\end{figure}

The \textit{Lagrangian exponent} $\Lambda $ is in all cases smaller than the
one given by the first iterate, but much larger than the Eulerian exponent $%
\gamma $, as seen in Fig. \ref{LFDE 8}. In Fig. \ref{LFDE 12} we plotted $%
\log \mathcal{L}(\theta )$ vs $\log \theta $ for given $\gamma =0.1$ and
three values of $K=0.1,~0.5,0.7$. For all these relatively small values of $%
K $ \footnote{%
We know that the Corrsin approximation is not valid for large Kubo numbers.}
the exponent $\Lambda $ is strictly the same: $\Lambda (\gamma =0.1)=2.10,$ $%
\forall K$.

%%%%%%%%%%%%%%%%%%%%%%%%%%%%%%%%%%%%%%%%%%%%%%%%%%%%%%%%%%%%%%%%%%%%% 
%%%%%%%%%%%%%%%%%%%%%%%%%%%%%%%%%%%%%%%%%%%%%%%%%%%%%%%%%%%%%%%%%%%%%

\begin{figure}[tbph]
\centerline{\includegraphics[height=2.4076in]{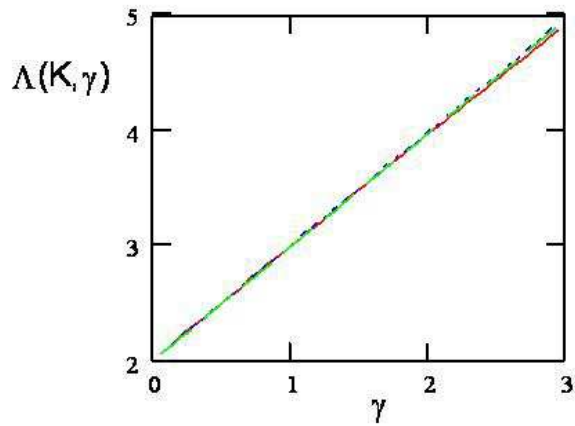}}
\caption{Lagrangian exponent $\Lambda $ 
\textit{vs}.\ Eulerian exponent $\protect\gamma $. The curves for $K=0.1,\
0.5,\ 0.9$ are all superposed.}
\label{LFDE 13}
\end{figure}

Next, we measured the dependence of $\Lambda $ on the Eulerian exponent.\
The result is a very simple linear dependence, shown in Fig.\ \ref{LFDE 13}.
This dependence is very accurately fitted by the formula:

\begin{equation}
\Lambda (\gamma )=2+\gamma ,\ \ \ \ \ \ 0<\gamma <1  \label{eeq.143}
\end{equation}

%%%%%%%%%%%%%%%%%%%%%%%%%%%%%%%%%%%%%%%%%%%%%%%%%%%%%%%%%%%%%%%%%%%%% 
%%%%%%%%%%%%%%%%%%%%%%%%%%%%%%%%%%%%%%%%%%%%%%%%%%%%%%%%%%%%%%%%%%%%%

We note already here the following very important point: \textit{the
Lagrangian exponent }$\Lambda (\gamma )>2,\ \forall \gamma $. Hence \textit{%
the process described by the V-Langevin equation (\ref{eeq.93}), with the
Eulerian velocity correlation (\ref{eeq.110}) in the Corrsin approximation,
leads to a }\textbf{diffusive regime}\textit{\ for all values of }$\gamma >0$%
\textit{\ and of }$K$. This point will be further discussed below.

\begin{figure}[tbph]
\centerline{\includegraphics[height=2.1759in]{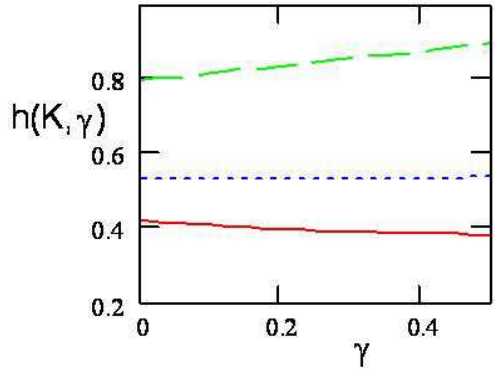}}
\caption{The coefficient $h(K,\protect%
\gamma )$ \textit{vs}. the Eulerian exponent $\protect\gamma $, for three
values of $K$: $K=0.1$ (solid red), $K=0.3$ (dotted blue), $K=0.5$ (dashed
green).}
\label{LFDE 14}
\end{figure}

We now consider the numerator $h$ of Eq. (\ref{eeq.110}), determined from
the logarithmic representation of $\mathcal{L}(\theta )$ (Fig. \ref{LFDE 12}%
). We first note (Fig. \ref{LFDE 14}) that, for given $K$, it is a very
slowly varying function of $\gamma $: we therefore consider it approximately
independent of $\gamma $.

%%%%%%%%%%%%%%%%%%%%%%%%%%%%%%%%%%%%%%%%%%%%%%%%%%%%%%%%%%%%%%%%%%%%% 
%%%%%%%%%%%%%%%%%%%%%%%%%%%%%%%%%%%%%%%%%%%%%%%%%%%%%%%%%%%%%%%%%%%%%

\begin{figure}[tbph]
\centerline{\includegraphics[height=2.4379in]{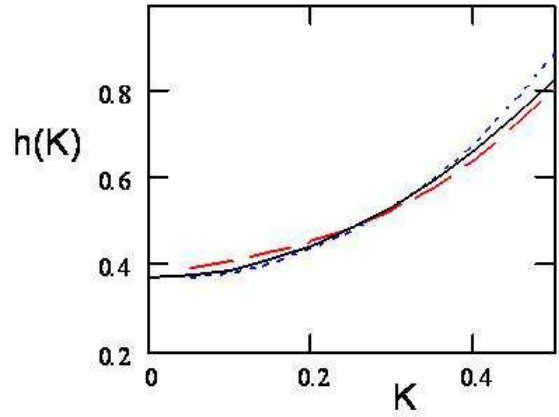}}
\caption{The coefficient $h$ as a function
of the Kubo number $K$. Red dashed: $\protect\gamma =0.1$, blue dotted: $%
\protect\gamma =0.5$.\ The solid black curve is the empirical fit (  \protect
\ref{eeq.146}).}
\label{LFDE 15}
\end{figure}

The dependence on $K$ appears to be quadratic, for relatively small values
of $K$. It is quite well fitted by the following empiric formula, valid for $%
\gamma \lesssim 0.5$ (Fig. \ref{LFDE 15}):

\begin{equation}
h(K)\simeq 0.37+1.85~K^{2}.  \label{eeq.144}
\end{equation}

%%%%%%%%%%%%%%%%%%%%%%%%%%%%%%%%%%%%%%%%%%%%%%%%%%%%%%%%%%%%%%%%%%%%% 
%%%%%%%%%%%%%%%%%%%%%%%%%%%%%%%%%%%%%%%%%%%%%%%%%%%%%%%%%%%%%%%%%%%%%

Finally, the value of the cut-off time is obtained by setting $\mathcal{L}%
(\theta _{L})=1$.

\begin{equation}
\theta _{L}(\gamma ,K)=h(K)^{1/(2+\gamma )}  \label{eeq.145}
\end{equation}

\begin{figure}[tbph]
\centerline{\includegraphics[height=2.9421in]{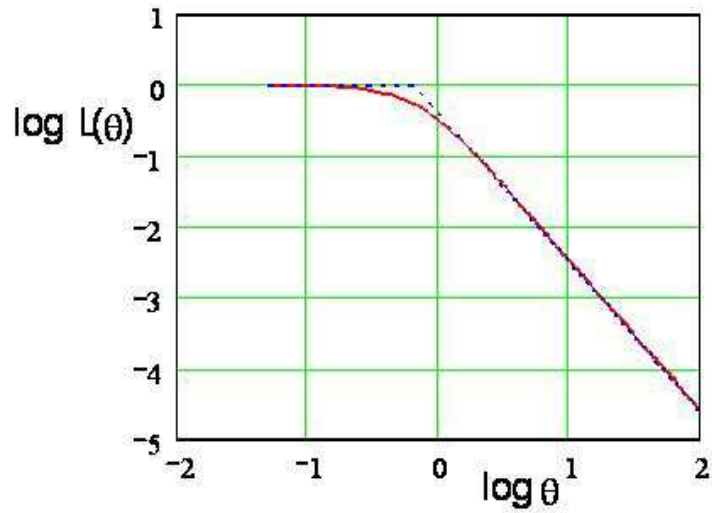}}
\caption{"Exact" numerical Lagrangian
velocity correlation (solid red) compared to the analytical approximation (%
\protect\ref{eeq.146}) (dashed black).}
\label{LFDE 16}
\end{figure}

To sum up, we found a reasonable analytical representation for the
Lagrangian velocity correlation in the Corrsin approximation. A simple
feature of this representation is the fact that (see Fig. 13 and Fig.\ 14)
each of the two output parameters, $\Lambda ,~h,$ is a function of a single
input parameter $\gamma $ or $K$:

\begin{equation}
\mathcal{L(}\theta )=\left\{ 
\begin{array}{c}
1,\ \ \ \ \ \ \theta <\theta _{L}(\gamma ,K), \\ 
\dfrac{h(K)}{\theta ^{2+\gamma }},\ \ \ \ \ \ \theta >\theta _{L}(\gamma ,K)%
\end{array}%
\right. ~\ ,  \label{eeq.146}
\end{equation}

combined with Eqs. (\ref{eeq.144}) and (\ref{eeq.145}). This analytic
approximation is compared to the "exact" (numerical) Lagrangian velocity
correlation in Fig. \ref{LFDE 16}.\smallskip

%%%%%%%%%%%%%%%%%%%%%%%%%%%%%%%%%%%%%%%%%%%%%%%%%%%%%%%%%%%%%%%%%%%%% 
%%%%%%%%%%%%%%%%%%%%%%%%%%%%%%%%%%%%%%%%%%%%%%%%%%%%%%%%%%%%%%%%%%%%%

\section{Running Diffusion Coefficient\label{Dif Coef}}

\begin{figure}[tbph]
\centerline{\includegraphics[height=2.6377in]{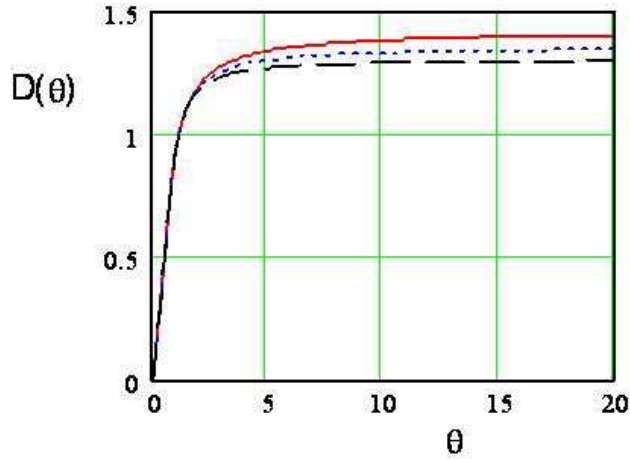}}
\caption{Running diffusion coefficient $%
\mathcal{D}(\protect\theta )$ for three values of $\protect\gamma =0.1$
(solid red), $\protect\gamma =0.3$ (dotted blue), $\protect\gamma =0.5$
(dashed black). $K=0.3.$}
\label{LFDE 17}
\end{figure}

The running diffusion coefficient for an input algebraic Eulerian
correlation is obtained from Eq.\ (\ref{eeq.108}). Using the algebraic form (%
\ref{eeq.111}) this integral was calculated analytically in Eqs. (\ref%
{eeq.110}) and (\ref{eeq.114}). Using also the forms (\ref{eeq.143}) - (\ref%
{eeq.145}) we obtain the result plotted in Fig. \ref{LFDE 17} for fixed $%
K=0.3$ and three values of $\gamma =0.1,0.3,0.5$. As we know that the regime
is always diffusive, the running diffusion coefficient tends toward a finite
positive diffusion constant for sufficiently large times (in practice, $%
\theta \gtrsim 10$). One sees that the asymptotic diffusion coefficient is a
weakly decreasing function of the Eulerian exponent.

%%%%%%%%%%%%%%%%%%%%%%%%%%%%%%%%%%%%%%%%%%%%%%%%%%%%%%%%%%%%%%%%%%%%% 
%%%%%%%%%%%%%%%%%%%%%%%%%%%%%%%%%%%%%%%%%%%%%%%%%%%%%%%%%%%%%%%%%%%%%

\begin{figure}[tbph]
\centerline{\includegraphics[height=2.4924in]{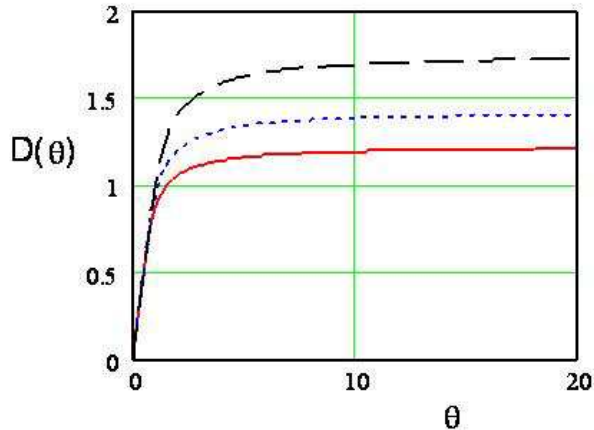}}
\caption{Running diffusion coefficient $%
\mathcal{D}(\protect\theta )$ for three values of $K=0.1$ (solid red), $%
K=0.3 $ (dotted blue), $K=0.5$ (dashed black). $\protect\gamma =0.1.$}
\label{LFDE 18}
\end{figure}

Fig. \ref{LFDE 18} shows the dependence of the diffusion coefficient on the
Kubo number for fixed $\gamma =0.1$. The asymptotic diffusion coefficient is
a rather strongly increasing function of $K$.

%%%%%%%%%%%%%%%%%%%%%%%%%%%%%%%%%%%%%%%%%%%%%%%%%%%%%%%%%%%%%%%%%%%%% 
%%%%%%%%%%%%%%%%%%%%%%%%%%%%%%%%%%%%%%%%%%%%%%%%%%%%%%%%%%%%%%%%%%%%%
\smallskip \smallskip \smallskip

In most existing works on the Langevin equation the Eulerian velocity
time-correlation is assumed to decay exponentially ($\beta =1$).\ This
produces a Lagrangian correlation that also has an exponential asymptotic
behavior, and thus automatically ensures the convergence of the integral (%
\ref{eeq.108}) defining the diffusion constant. It is interesting to compare
the diffusion coefficients obtained in the algebraic case and in the
exponential case. For the latter we use a Eulerian correlation that is
tailored in such a way as to resemble the algebraic one: we thus start the
exponential decay at the cut-off time of the algebraic one (Fig. \ref{LFDE
19}):

\begin{equation}
\mathcal{T}_{\exp }(\theta /K)=\left\{ 
\begin{array}{c}
1,\ \ \ \ \theta \leq K \\ 
\exp \left[ \dfrac{\theta -K}{K}\right] ,\ \ \ \ \ \theta >K%
\end{array}%
\right. .  \label{eeq.147}
\end{equation}

%%%%%%%%%%%%%%%%%%%%%%%%%%%%%%%%%%%%%%%%%%%%%%%%%%%%%%%%%%%%%%%%%%%%% 
\begin{figure}[tbph]
\centerline{\includegraphics[height=2.5728in]{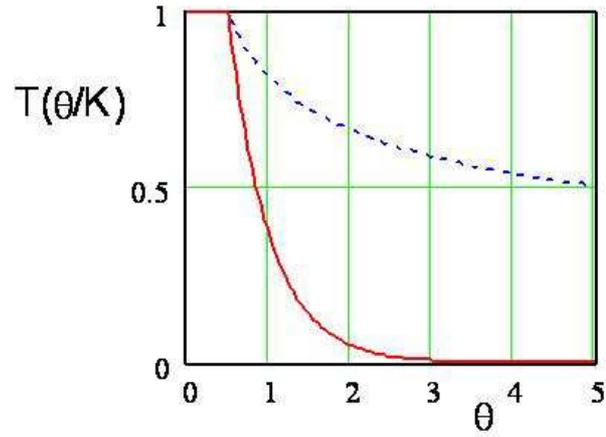}}
\caption{Temporal part of the Eulerian
velocity correlation. Solid red: exponential, Eq. (\protect\ref{eeq.147});
dotted blue: algebraic, Eq. (\protect\ref{eeq.109}). $K=0.5,\ \protect\gamma %
=0.3.$}
\label{LFDE 19}
\end{figure}
%%%%%%%%%%%%%%%%%%%%%%%%%%%%%%%%%%%%%%%%%%%%%%%%%%%%%%%%%%%%%%%%%%%%%

\begin{figure}[tbph]
\centerline{\includegraphics[height=2.514in]{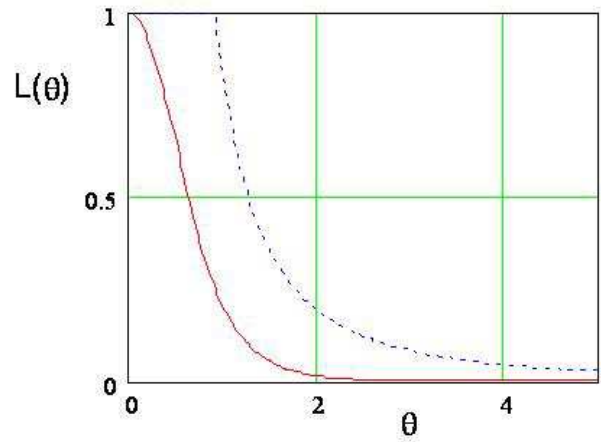}}
\caption{Lagrangian velocity correlation
for the case of an algebraic (solid red) [$K=0.5,\ \protect\gamma =0.1$] and
of an exponential (dotted blue) Eulerian correlation with the same $K$.}
\label{LFDE 20}
\end{figure}

The Lagrangian velocity correlation is calculated in the Corrsin
approximation up to the second iteration, by following exactly the same
steps as in Secs. \ref{Determ Lagr 1} and \ref{Determ Lagran 2}. The result
is shown in Fig. \ref{LFDE 20}. (Note that $\mathcal{L}_{\exp }(0)=1$, hence
it is not necessary to introduce a cut-off time $\theta _{L}$ as for the
algebraic case). As expected, $\mathcal{L}_{\exp }(\theta )$ decays much
more quickly: it has no long-time tail as the algebraic one.

%%%%%%%%%%%%%%%%%%%%%%%%%%%%%%%%%%%%%%%%%%%%%%%%%%%%%%%%%%%%%%%%%%%%% 
%%%%%%%%%%%%%%%%%%%%%%%%%%%%%%%%%%%%%%%%%%%%%%%%%%%%%%%%%%%%%%%%%%%%%

\begin{figure}[tbph]
\centerline{\includegraphics[height=2.5149in]{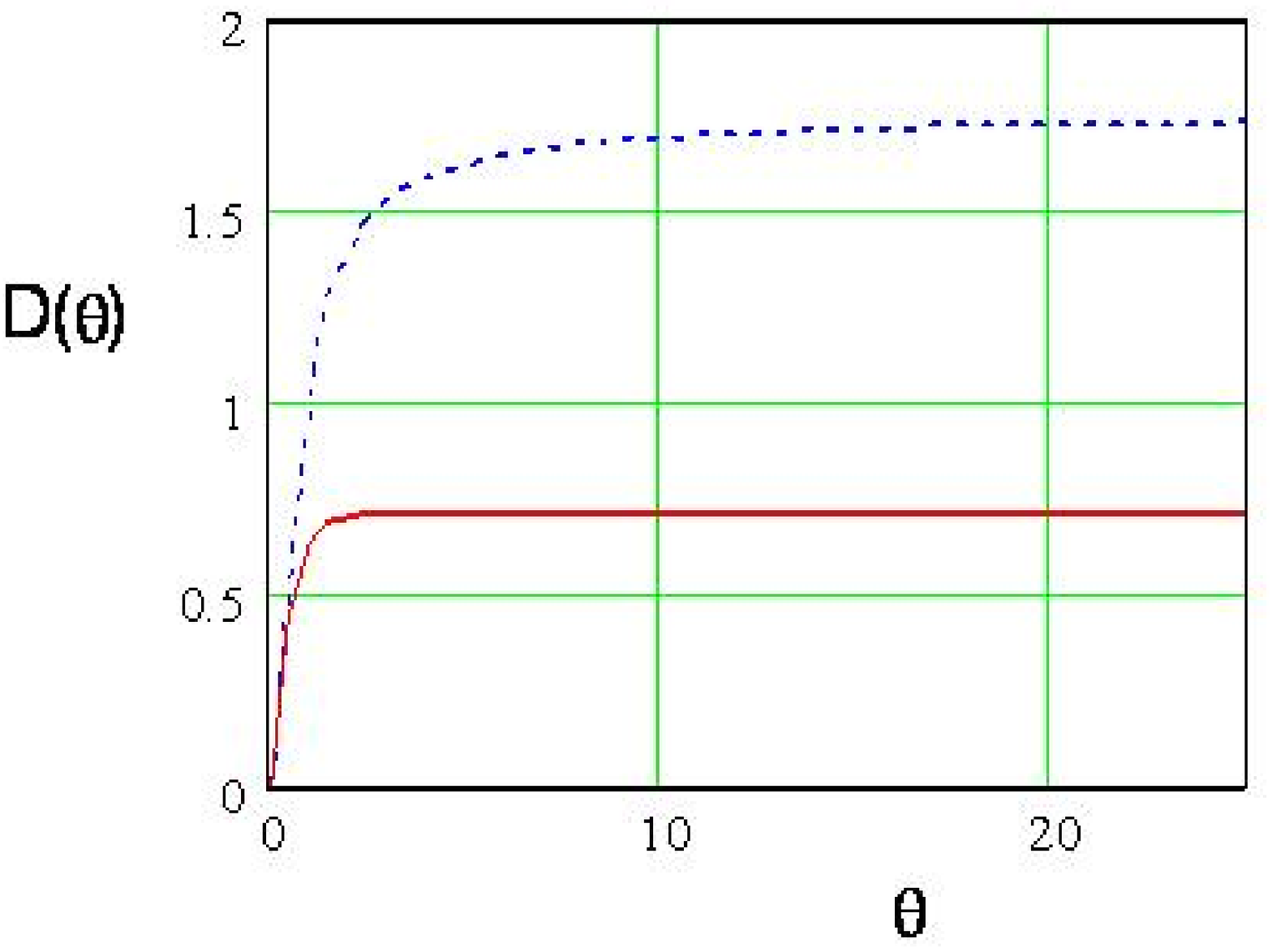}}
\caption{Running diffusion coefficient for
the case of an algebraic (solid red) [$K=0.5,\ \protect\gamma =0.1$] and of
an exponential (dotted blue) Eulerian correlation with the same $K$.}
\label{LFDE 21}
\end{figure}

The running diffusion coefficients corresponding to the same parameters are
compared in Fig. \ref{LFDE 21}. The algebraic correlation produces a running
diffusion coefficient that saturates at a longer time.\ As a result, the
asymptotic diffusion constant $\mathcal{D>D}_{\exp }.$ This results from the
fact that the area under the Lagrangian correlation is larger in the
algebraic case (Fig. \ref{LFDE 20}), due to the slow decay of the
correlation.

%%%%%%%%%%%%%%%%%%%%%%%%%%%%%%%%%%%%%%%%%%%%%%%%%%%%%%%%%%%%%%%%%%%%% 
%%%%%%%%%%%%%%%%%%%%%%%%%%%%%%%%%%%%%%%%%%%%%%%%%%%%%%%%%%%%%%%%%%%%%

\section{Conclusions\label{Conclus}}

Our main purpose in the present work was to study the relation between a
"semi-dynamical" description of a system of particles, based on the
V-Langevin equation and its associated hybrid kinetic equation, and on the
other hand on a purely stochastic description in terms of a continuous time
random walk and its limiting form as a fractional diffusion equation. We
wanted, in particular, to find out under what conditions on the velocity
correlation function a V-Langevin equation describes a "strange transport
regime", which could be equivalently represented by a CTRW or a FDE.

It was known from previous work \cite{RB} that a \textit{subdiffusive regime}
can appear in some limiting cases.\ This occurs, \textit{e.g.}, for a
collisional plasma in a strong magnetic field, in the limit of infinite
correlation length perpendicular to $\mathbf{B}$ \cite{Rech-R}, \cite{Wang}
or for electrostatic drift wave turbulence in the limit of infinite Kubo
number \cite{RB}. In these cases, subdiffusion is produced by sticking of
the particles to the magnetic field lines, or by their trapping in the
turbulent field. It might be stressed that in the latter case, a diffusive
or a subdiffusive ($K\rightarrow \infty )$ regime was found even when the
spatial correlation is long-range (Lorentzian), but the time correlation is
exponential \cite{Vlad}. A \textit{superdiffusive regime} is, however, much
more elusive. To the best of our knowledge, a V-Langevin equation producing L%
\'{e}vy-type superdiffusion was only found for a one-dimensional two-state
model (velocity = $\pm V$) \cite{West-Grigo}.\ The restriction to a velocity
constant in absolute value was necessary in order to replace the L\'{e}vy
flight CTRW by a L\'{e}vy walk.

In the present work we showed that, for one-dimensional systems, the results
of \cite{West-Grigo} concerning the \textit{time-fractional superdiffusion}
can be extended to the more physical case of a general random time-dependent
velocity function equipped with an algebraic time-correlation function (but
with a Gaussian space-correlation). We confirm that in this case we find a
process that is an "intermediate" between diffusion and wave propagation.\
This type of superdiffusive processes, distinct from the L\'{e}vy-type, has
been studied in detail in \cite{Main 2001} and \cite{Main 2003}.

We then generalized our investigation to the study of a random velocity 
\textit{field} in two dimensions. In this case one is faced with the
difficult problem of determining the \textit{Lagrangian velocity correlation}%
, a problem that was treated within the Corrsin approximation. It was first
shown that if the Eulerian correlation decays algebraically as $\theta
^{-\gamma }$ for long times (and its spatial part is Gaussian), the
Lagrangian correlation has the same property, with a different exponent: $%
\theta ^{-\Lambda }$.

We then established a general criterion for the existence of a
superdiffusive regime, based on the value of the Lagrangian exponent $%
\Lambda $. It was shown in Sec. \ref{Algeb Gen} that whenever $0<\Lambda <1$%
, the regime is superdiffusive and can be equivalently described by a
time-fractional FDE. When $\Lambda >1$, however, the semi-dynamical
description predicts a diffusive regime for all $\Lambda $, whereas the
formally associated FDE predicts subdiffusion, or even looses meaning, for $%
\Lambda >2$. Thus, the purely stochastic CTRW or FDE descriptions of the
turbulence are severely limited in this case by the existence of a
bifurcation at $\Lambda =1$.

In the final stage (Sec. \ref{Determ Lagran 2}) the Lagrangian exponent $%
\Lambda $ was determined numerically.\ The result shows that $\Lambda
=2+\gamma >2$, for all values of $\gamma $ and of $K$. Thus, for the present
model and with the approximations used in its treatment: [local
approximation of the non-Markovian diffusion equation (\ref{eeq.55}),
Corrsin approximation of the Lagrangian velocity correlation (\ref{eeq.101}%
)]:

\begin{itemize}
\item \textit{The regime is always }\textbf{normal diffusive,}\textit{\ for
all values of the input Eulerian parameters }$\gamma $ and $K$. There never
appears any superdiffusion, in spite of the long algebraic tail of the
temporal velocity correlation. This is to be contrasted with the
one-dimensional case of a purely time-dependent velocity, where the regime
is superdiffusive for all values $0<\gamma <1$.

\item \textit{The "formally associated" fractional differential equation,
has no meaning. }The evolution cannot be described by a purely stochastic
process.

\item \textit{Although the regime is diffusive,} \textit{the density profile
is not a Gaussian packet.} Indeed, although the Lagrangian correlation
decays sufficiently fast in order to ensure the existence of a finite
asymptotic diffusion constant, the non-Markovian character of the equation (%
\ref{eeq.107}) cannot be neglected for finite time.
\end{itemize}

The results obtained here may appear somewhat disappointing, in the sense
that no time-fractional superdiffusion was found for two-dimensional motion.
Nevertheless, we believe that the result of the exhaustive analysis
performed here is interesting in exhibiting the limits of a purely
stochastic description.\ It shows that a long-tailed algebraic temporal
correlation is not sufficient for producing superdiffusion (for
dimensionality $d>1$). In order to obtain a L\'{e}vy-type superdiffusive
regime, we must presumably act on the spatial part of the Eulerian velocity
correlation. This problem is, however, highly non-trivial. On one hand, in
the Corrsin approximation, Eq. (\ref{eeq.106}) is invalid, because it is
based on the Gaussian form of the spatial Eulerian correlation. But one
should go back even further in the chain of approximations. The local
approximation leading from Eq. (\ref{eeq.55}) to (\ref{eeq.56}) (in two
dimensions) is questionable in the presence of long-range spatial
correlations. We hope to come back to this problem in forthcoming work.

\newpage

\section*{Appendix: A Primer on Fractional Calculus\label{Fract Calc}}

Fractional Calculus has been known to mathematicians for a very long time.\
The problem appears to have been first raised by L'Hospital and by Leibniz
in 1695. The first systematic construction appears in the early 19-th
century and is due to Liouville (1832) (well known to every physicist from
the first lecture in statistical mechanics), who used this concept in
potential theory.\ It was further developed by B.\ Riemann in 1847. For a
long time the concept was studied only by mathematicians.\ It entered
physics only in the last half of the 20-th century, through the theory of
stochastic equations associated with the problems of random walks (Sec. \ref%
{CTRW FDE}).

The search for a natural generalization of the ordinary derivative $%
d^{n}/dx^{n}$, with $n$: an integer, is clearly attractive for a
mathematician. It is as natural as the passage from the integers to
rational, and later to real numbers. We must find a way to interpolate in a
consistent way the concept of differentiation between two successive
integers, $n$ and $n+1$. The way of doing this is however not obvious.\
Indeed, several different approaches have been devised by mathematicians.\
We shall only discuss here the one which is most widely used in the
literature (for a discussion of other approaches, see \cite{Podlubny}, \cite%
{Butzer} and \cite{West-B-Grigo}).\ One will find detailed expositions of
the subject in \cite{Gor-Main 1997}, \cite{Gor-Main 2005}, \cite{Podlubny}, 
\cite{Hilfer}, \cite{Metz-Klaf 2000}, \cite{Metz-Klaf 2004}, \cite{Zaslav}.\
Excellent tutorials are the papers by \cite{Main 1996} and especially \cite%
{Main 2001}, \cite{Main 2003}. A brief, but very clear summary is given in 
\cite{delCast 2004}.

The names of \textit{fractional} integral, \textit{fractional }derivative, 
\textit{fractional} calculus are actually rather improper. The object of
these concepts is to extend the notions of integration and of
differentiation from integer order $n$ to \textit{arbitrary real order }$%
\beta $, not necessarily a fraction. The name has, however, acquired general
practice.\bigskip\ 

\ \textbf{RIEMANN-LIOUVILLE FRACTIONAL INTEGRAL\bigskip }

Consider first the \textit{multiple integral of order }$n$\textit{\ (}$n$%
\textit{: integer)\footnote{%
In the forthcoming formula we will always use latin letters $m,n,r,...$ for
integer indices, and greek letters for real, non-integer indices: $\alpha
,\beta ,...$}} of a real-valued function $f(x)$ of the real variable $x$:

\begin{equation}
_{a}J_{x}^{n}~f(x)=\int_{a}^{x}dx_{1}\int_{a}^{x_{1}}dx_{2}...%
\int_{a}^{x_{n-1}}dx_{n}~f(x_{n}).  \label{eeq.148}
\end{equation}

The two lower subscripts relate to the limits of integration ($a$ is a fixed
real number); the superscript $n$ denotes the order (multiplicity). This
equation can be written in the following equivalent form, in terms of a 
\textit{single} integral:

\begin{equation}
_{a}J_{x}^{n}~f(x)=\frac{1}{(n-1)!}\int_{a}^{x}dy~\frac{f(y)}{(x-y)^{1-n}}.
\label{eeq.149}
\end{equation}

The equivalence between the two forms is easily demonstrated by partial
integration. We may put \ $_{a}J_{x}^{0}~f(x)=f(x)$ \cite{Podlubny}. For $%
n=1 $, the formula reduces to the single integration:

\begin{equation}
_{a}J_{x}^{1}~f(x)=\int_{a}^{x}dy~f(y)  \label{eeq.149a}
\end{equation}

\ Let us check $n=2$:

\begin{eqnarray*}
_{a}J_{x}^{2}~f(x) &=&\int_{a}^{x}dy~(x-y)~f(y) \\
&=&\left. \left\{ (x-y)\int_{a}^{y}dx_{1}f(x_{1})\right\} \right\vert
_{a}^{x}-\int_{a}^{x}(-1)\text{ }dy\int_{a}^{y}dx_{2}f(x_{2}) \\
&=&\int_{a}^{x}dx_{1}\int_{a}^{x_{1}}dx_{2}~f(x_{2}).
\end{eqnarray*}

In the form (\ref{eeq.149}) the way toward a generalization of the operation
of integration (\ref{eeq.149}) is obvious:

\begin{equation}
\begin{array}{l}
_{a}J_{x}^{\alpha }f(x)~{\large =~}\dfrac{1}{\Gamma (\alpha )}{\large ~}%
\int_{a}^{x}{\normalsize dy~}\dfrac{f(y)}{(x-y)^{1-\alpha }}{\normalsize ,\
\ \ \ \alpha >0}:\text{any positive real number.}%
\end{array}
\label{eeq.150}
\end{equation}

This is the definition of the \textbf{Riemann - Liouville (RL) fractional
integral of }$f(x)$ \textbf{of arbitrary real order }$\alpha $.

Some of the more important properties are given here:\bigskip

A) \textit{RL\ fractional integral of a monomial}

\begin{equation}
\begin{array}{l}
_{0}J_{x}^{\alpha }~x^{\mu }{\normalsize =}\dfrac{\Gamma (\mu +1)}{\Gamma
(\mu +\alpha +1)}{\normalsize ~x}^{\mu +\alpha }{\normalsize .}%
\end{array}
\label{eeq.151}
\end{equation}

B) \textit{Commutative composition rule:}

\begin{equation}
\begin{array}{l}
_{0}J_{x}^{\mu }\cdot ~_{0}J_{x}^{\nu }\cdot f(x)=~_{0}J_{x}^{\nu }\cdot
~_{0}J_{x}^{\mu }\cdot f(x)=~_{0}J_{x}^{\mu +\nu }\cdot f(x)%
\end{array}
\label{eeq.152}
\end{equation}

\qquad Proof:\textit{\ }%
\begin{eqnarray*}
_{0}J_{x}^{\mu }\cdot ~_{0}J_{x}^{\nu }\cdot f(x) &=&\frac{1}{\Gamma (\mu )}%
\int_{0}^{x}d\xi ~(x-\xi )^{-1+\mu }~\frac{1}{\Gamma (\nu )}\int_{0}^{\xi
}d\tau ~(\xi -\tau )^{-1+\nu }f(\tau ) \\
&=&\frac{1}{\Gamma (\mu )\Gamma (\nu )}\int_{0}^{x}d\tau ~f(\tau )\int_{\tau
}^{x}d\xi ~(x-\xi )^{-1+\mu }(\xi -\tau )^{-1+\nu }
\end{eqnarray*}

\qquad \qquad Substitute: \ $\xi =\tau +(x-\tau )\zeta $

\begin{eqnarray*}
&=&\frac{1}{\Gamma (\mu )\Gamma (\nu )}\int_{0}^{x}d\tau ~f(\tau
)\int_{0}^{1}d\zeta ~(x-\tau )~[\zeta (x-\tau )]^{-1+\nu }~[(1+\zeta
)(x-\tau )]^{-1+\mu } \\
&=&\frac{1}{\Gamma (\mu )\Gamma (\nu )}\int_{0}^{x}d\tau ~f(\tau )~(x-\tau
)^{-1+\mu +\nu }~\int_{0}^{1}d\zeta ~\zeta ^{-1+\nu }~(1+\zeta )^{-1+\mu }
\end{eqnarray*}

\qquad \qquad The $\zeta $-integral is well-known is expressed in terms of
the well-known $B$-function:

\begin{equation}
\int_{0}^{1}d\tau ~\tau ^{z-1}~(1-\tau )^{w-1}=B(z,w)=\frac{\Gamma (z)\Gamma
(w)}{\Gamma (z+w)}.  \label{eeq.152a}
\end{equation}

\qquad \qquad hence:

\begin{gather*}
_{0}J_{x}^{\mu }\cdot ~_{0}J_{x}^{\nu }\cdot f(x)=\frac{1}{\Gamma (\mu +\nu )%
}\int_{0}^{x}d\tau ~(x-\tau )^{-1+\mu +\nu }~f(\tau ) \\
=~_{0}J_{x}^{\mu +\nu }f(x)=~_{0}J_{x}^{\nu }\cdot ~_{0}J_{x}^{\mu }\cdot
f(x),\ \ \ \ \ QED
\end{gather*}%
\bigskip

\bigskip

\textbf{RIEMANN-LIOUVILLE FRACTIONAL DERIVATIVE\bigskip }

The fractional derivative is constructed by combining the operations of the
usual (integer-order) derivative with the fractional integral.

We define the \textit{fractional derivative} as follows:

\begin{equation}
_{a}D_{x}^{\alpha }~f(x)=\frac{d^{m}}{dx^{m}}\left[ _{a}J_{x}^{(m-\alpha
)}f(x)\right] ,\ \ \ \alpha >0,\ \ \ \ m-1<\alpha \leq m.  \label{eeq.154}
\end{equation}

where $\alpha $ is any positive real number, not an integer, and $m$ is the
smallest integer larger than $\alpha $.

In order to "understand intuitively" this rather unusual definition, we may
think, very casually, and without any rigour, as follows. We have a solid
reference point, \textit{viz}. a rigorously defined fractional object, the
RL\ integral. We then start with a RL integral of order $m-\alpha $, where $%
m $ is the smallest integer which makes $m-\alpha >0$, thus ensuring that $%
_{a}J_{x}^{(m-\alpha )}f(x)$ is a true fractional integral (\ref{eeq.150}).
Next, we "differentiate away" the integer $m$, by taking the $m$-th ordinary
derivative of this object. This looks like "lowering the order" of $%
_{a}J_{x}^{(m-\alpha )},$ thus "transforming" it into $_{a}J_{x}^{-\alpha }$%
, which could be viewed as a fractional integral of negative order, $-\alpha 
$, to be interpreted as a fractional derivative of order $+\alpha $. But,
again, this paragraph is just cavalier talk; in particular, viewing $%
J^{-\alpha }$ as an "inverse" to $D^{\alpha }$ is an ambiguous statement, as
will be shown below. We now go back to business, and consider Eq. (\ref%
{eeq.154}) as a given definition, and study some of its consequences.

A basic requirement is that the definition (\ref{eeq.154}) be compatible
with the ordinary (integer-order) derivative. When $\alpha $ is an integer, $%
\alpha =N$, we have $m-1=\alpha =N,$ thus $m=N+1$, and recalling Eq.\ (\ref%
{eeq.149a}), we have:

\begin{equation*}
_{a}D_{x}^{N}f(x)=\frac{d^{N+1}}{dx^{N+1}}\left[ _{a}J_{x}^{1}f(x)\right] =%
\frac{d^{N}}{dx^{N}}~\frac{d}{dx}~\int_{a}^{x}dy~f(y)=\frac{d^{N}}{dx^{N}}%
~f(x);
\end{equation*}

the operation thus reduces to the ordinary differentiation, as it should.

Combining Eqs. (\ref{eeq.150}) and (\ref{eeq.154}) we find:

\begin{equation}
\begin{array}{l}
_{a}D_{x}^{\alpha }f(x)=\dfrac{d^{m}}{dx^{m}}~\dfrac{1}{\Gamma (m-\alpha )}%
~\int_{a}^{x}dy~\dfrac{f(y)}{(x-y)^{\alpha +1-m}},\ \ \ \alpha >0,\ \ \ \ \
m-1\leq \alpha <m.%
\end{array}
\label{eeq.155}
\end{equation}

This is the definition of the \textbf{Riemann - Liouville (RL)
left-fractional derivative of order }$\alpha $.\bigskip

We may also define a \textbf{Riemann-Liouville (RL) right-fractional
derivative} in which it is the upper integration limit $b$ that is fixed:

\begin{equation}
_{x}D_{b}^{\alpha }\phi =\frac{d^{m}}{dx^{m}}~\frac{(-1)^{m}}{\Gamma
(m-\alpha )}~\int_{x}^{b}dy~\frac{\phi (y)}{(y-x)^{\alpha +1-m}},\ \ \ \ \
\alpha >0,\ \ \ \ \ \ m-1\leq \alpha <m.  \label{eeq.156}
\end{equation}%
\bigskip

We now explore a few properties of the fractional derivatives: we shall meet
some surprises.\ These are all due to a fundamental fact.\ \textit{The
fractional derivative (for any of the forms defined above) of a function }$%
f(x)$ \textit{is expressed as an }\textbf{integral }\textit{of }$f(x)$%
\textit{\ over a given range of }$x$ \footnote{%
For this reason, the fractional derivative is sometimes called "\textit{%
differintegral}".}\ It is therefore an intrinsically \textbf{non-local
operator}.\ In other words, \textit{the value of the fractional derivative
(for non-integer }$\alpha $) $_{a}D_{x}^{\alpha }f(x)$ \textit{at the point }%
$x$ \textit{depends on the values of }$f$\textit{\ taken in the whole range
of integration defining this quantity.} It is quite surprising that, upon
varying continuously the order $\alpha $, this non-local character
disappears \textit{suddenly} whenever $\alpha $ reaches any integer value!
What is not surprising, therefore, is that fractional calculus entered
physics when researchers began to be interested in evolutionary processes
endowed with memory ("non-Markovian processes") or in the influence of the
spatial neighborhood in the phenomena at a given point $(x,t)$. \bigskip

a) \textit{Linearity:}

\begin{equation}
_{0}D_{x}^{\alpha }~[\lambda ~f(x)+\mu ~g(x)]=\lambda ~_{0}D_{x}^{\alpha
}~f(x)+\mu ~_{0}D_{x}^{\alpha }g(x)  \label{eeq.156a}
\end{equation}

\qquad The proof follows trivially from the definition of the RL\
derivative.\bigskip

b) \textit{RL Fractional derivative of a monomial:}

\begin{equation}
\begin{array}{l}
_{0}D_{x}^{\alpha }~x^{\mu }=\dfrac{\Gamma (\mu +1)}{\Gamma (\mu -\alpha +1)}%
~x^{\mu -\alpha },\ \ \ \ \alpha >0,\ \ \ \ \ \mu >-1.%
\end{array}
\label{eeq.157}
\end{equation}

\qquad \textit{Proof. \ }Using the definition (\ref{eeq.155}):

\begin{equation}
_{0}D_{x}^{\alpha }~x^{\mu }=\frac{d^{m}}{dx^{m}}~\frac{1}{\Gamma (m-\alpha )%
}~\int_{0}^{x}dy~y^{\mu }~(x-y)^{-1+m-\alpha }.  \label{eeq.157a}
\end{equation}

\qquad Let us do this calculation in some detail.\ Substituting $%
y\rightarrow xt$ we find

\begin{equation*}
_{0}D_{x}^{\alpha }~x^{\mu }=\left( \frac{d^{m}}{dx^{m}}~x^{m+\mu -\alpha
}\right) ~\frac{1}{\Gamma (m-\alpha )}\int_{0}^{1}dt~t^{\mu
}~(1-t)^{-1+m-\alpha }.
\end{equation*}

\qquad The integral appearing here is expressed in terms of the $B$%
-function, (\ref{eeq.152a}), hence:

\begin{equation*}
_{0}D_{x}^{\alpha }~x^{\mu }=\left( \frac{\partial ^{m}}{\partial x^{m}}%
~x^{m+\mu -\alpha }\right) ~\frac{1}{\Gamma (m-\alpha )}~\frac{\Gamma
(m-\alpha )\Gamma (\mu +1)}{\Gamma (\mu +1-\alpha +1)}.
\end{equation*}

\qquad Next, we calculate (setting $\gamma =\mu -\alpha )$

\begin{equation*}
\frac{d^{m}}{dx^{m}}~x^{m+\gamma }=(m+\gamma )(m-1+\gamma )...(1+\gamma
)~x^{\gamma }.
\end{equation*}

\qquad Noting the identity:

\begin{equation*}
\Gamma (m+\gamma +1)=(m+\gamma )(m-1+\gamma )...(1+\gamma )~\Gamma
((1+\gamma ),
\end{equation*}

\qquad we combine all these partial results to obtain the final formula (\ref%
{eeq.157}).\bigskip

The limitation to $\mu >-1$ in Eq. (\ref{eeq.157}) is due to the factor $%
\Gamma (1+\mu )$ which has a pole in $\mu =-1$. Note that for any
non-integer value of $\alpha $ the RL derivative $_{0}D_{x}^{\alpha }~x^{\mu
}$ has a single zero (for any $x\neq 0)$ in the allowed region, namely for $%
\mu =\alpha -1$ (see below). Indeed, any zero of the right hand side of Eq. (%
\ref{eeq.157}) originates from the factor $1/\Gamma (\mu +1-\alpha )$ which
vanishes whenever its argument is zero or a negative integer, see Fig. \ref%
{LFDE 22a}; a single one of them is larger than $-1$.

\begin{equation}
_{0}D_{x}^{\alpha }~x^{\alpha -1}=0,\ \ \ \ \ \forall \alpha >0.
\label{eeq.157b}
\end{equation}

%%%%%%%%%%%%%%%%%%%%%%%%%%%%%%%%%%%%%%%%%%%%%%%%%%%%%%%%%%%%%%%%%%%%% 
\begin{figure}[tbph]
\centerline{\includegraphics[height=3.0485in]{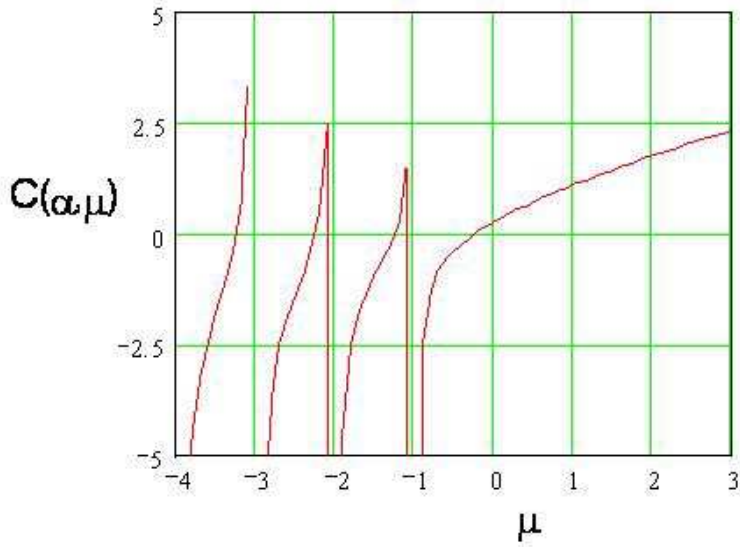}}
\caption{Coefficient $C(\protect\alpha ,%
\protect\mu )=\Gamma (\protect\mu +1)/\Gamma (-\protect\alpha +\protect\mu %
+1)$ of the RL derivative of $x^{\protect\mu }$ as a function of $\protect%
\mu .\ \protect\alpha =0.75.$}
\label{LFDE 22a}
\end{figure}
%%%%%%%%%%%%%%%%%%%%%%%%%%%%%%%%%%%%%%%%%%%%%%%%%%%%%%%%%%%%%%%%%%%%%

The formula (\ref{eeq.157}) is deceptively simple, and would correspond with
intuition.\ Just as the ordinary integral-order derivative ($%
d^{n}/dx^{n})x^{m}$ lowers the power of $x$ by $n$, the fractional
derivative transforms $x^{\mu }$ into $x^{\mu -\alpha }$, with a prefactor.
But a further reflection shows us that this result is not as innocent as it
appears.\bigskip

b1) \textit{RL Fractional derivative of a constant: }Consider, indeed, the
case $\mu =0$, i.e., the

\begin{equation*}
_{0}D_{x}^{\alpha }~cx^{0}=c~\frac{\Gamma (1)}{\Gamma (1-\alpha )}%
~x^{-\alpha },
\end{equation*}

i.e.,

\begin{equation}
_{0}D_{x}^{\alpha }~c=\frac{c}{\Gamma (1-\alpha )}~x^{-\alpha }\neq 0\ 
\label{eeq.158}
\end{equation}

Thus, the \textit{RL fractional derivative of non-integer order} $\alpha >0$%
\textit{\ of a constant is non-zero!} Is this result consistent with the
ordinary derivative? The answer comes from the properties of the gamma
function.\ Indeed, whenever $\alpha =n$, where $n$ is a non-zero integer, $%
n=1,2,3,...$ the argument of the $\Gamma $ function is a \textit{negative
integer, or zero}.\ But we know that

\begin{equation*}
\Gamma (0)=\Gamma (-1)=\Gamma (-2)=...=\infty
\end{equation*}

Hence, for any integer value of $\alpha $, Eq.\ (\ref{eeq.158}) reduces, as
it should, to:

\begin{equation}
_{0}D_{x}^{n}~c=0,\ \ \ \ \ n=1,2,...  \label{eeq.159}
\end{equation}

Another feature of Eq. (\ref{eeq.157}) seems to contradict our previous
statement: $_{0}D_{x}^{\alpha }~x^{\mu }$ is a perfectly \textit{local}
expression, depending only on the value of $x^{\mu }$ at $x$.\footnote{%
This feature is never mentioned in the literature, as far as we know.} In
order to understand what happens here, we consider the RL derivative of $%
x^{\mu }$ for a lower integration limit $0<a<x$ instead of zero. Starting
the calculation as above, we find after the first step:

\begin{equation}
_{a}D_{x}^{\alpha }~x^{\mu }=\frac{d^{m}}{dx^{m}}\left( ~x^{m+\mu -\alpha }~%
\frac{1}{\Gamma (m-\alpha )}\int_{a/x}^{1}d\xi ~\xi ^{\mu }~(1-\xi
)^{-1+m-\alpha }\right) ,\ \ \ x>a.  \label{eeq.160}
\end{equation}

There are two important differences with Eq.\ (\ref{eeq.157a}).\ First, the
integral depends on $x$, through its lower limit; hence, the $m$-th
derivative acts on the whole product, thus yielding a much more complicated $%
x$-dependence. Next, the integral with a non-zero lower limit is no longer a 
$B$-function, but rather a very complicated hypergeometric function. In
order to illustrate our case, we consider the simplest non-trivial case that
leads to simple calculations: $\mu =2$, $\alpha =\frac{1}{2}$, hence $m=1$.
Eq. (\ref{eeq.160}) then reduces to:

\begin{equation}
_{a}D_{x}^{1/2}~x^{2}=\frac{d}{dx}\left( ~x^{5/2}~\frac{1}{\Gamma (1/2)}%
\int_{a/x}^{1}d\xi ~\xi ^{2}~(1-\xi )^{-1/2}\right) .  \label{eeq.161}
\end{equation}

The integral is tabulated; a simple calculation then leads to the result:

\begin{equation}
_{a}D_{x}^{1/2}~x^{2}=\frac{1}{\sqrt{\pi }}\left[ \frac{x^{2}}{\sqrt{x-a}}+2x%
\sqrt{x-a}-\frac{(x-a)^{3/2}}{3}\right] ,\ \ \ x>a.  \label{eeq.162}
\end{equation}

The result reduces to Eq. (\ref{eeq.157}) when $a\rightarrow 0$:

\begin{equation}
_{0}D_{x}^{1/2}~x^{2}=\frac{8}{3\sqrt{\pi }}~x^{3/2}=\frac{\Gamma (3)}{%
\Gamma (5/2)}~x^{3/2}.  \label{eeq.163}
\end{equation}

\begin{figure}[tbph]
\centerline{\includegraphics[height=3.3806in]{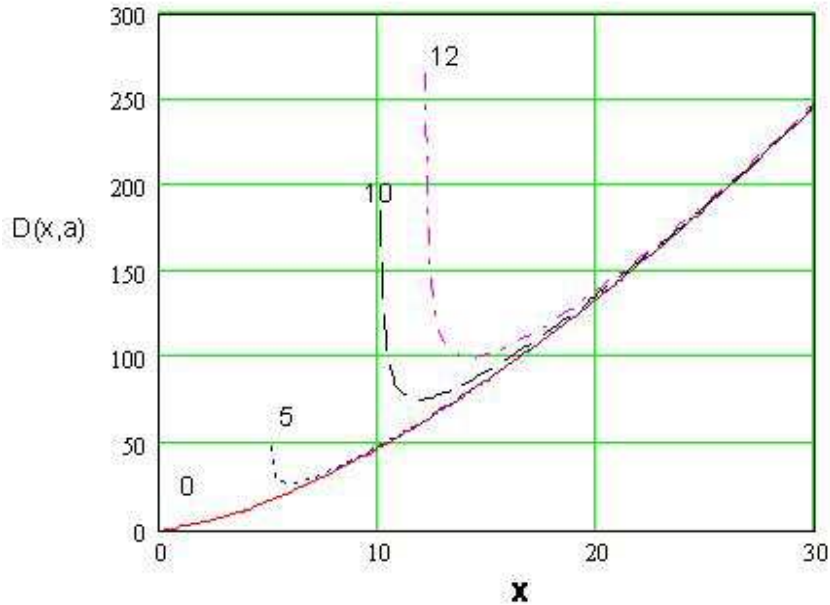}}
\caption{Fractional derivative of $x^{2}$:
\ $_{a}D_{x}^{1/2}x^{2}\equiv D(x,a)$, for $a=0,5,10,12$.}
\label{LFDE 22}
\end{figure}

The non-local character of the fractional derivative is now luminously
illustrated. The complicated function appearing in the right hand side of
Eq. (\ref{eeq.162}) is determined by the choice of the function $\xi ^{\mu }$
in Eq. (\ref{eeq.160}), hence by its values taken over the whole integration
range. In particular, the fractional derivative depends explicitly on the
lower integration limit $a$. A striking fact is the \textit{singularity of
the RL fractional derivative }when $x\rightarrow a$. This singularity is
present for any $a$, arbitrarily close to zero, but disappears when $a=0$.
When $x\gg a$, $_{a}D_{x}^{1/2}~x^{2}\rightarrow ~_{0}D_{x}^{1/2}~x^{2}$.
All this clearly shows that the value $a=0$ is not a generic one. The
situation is illustrated in Fig. \ref{LFDE 22}.

%%%%%%%%%%%%%%%%%%%%%%%%%%%%%%%%%%%%%%%%%%%%%%%%%%%%%%%%%%%%%%%%%%%%% 
%%%%%%%%%%%%%%%%%%%%%%%%%%%%%%%%%%%%%%%%%%%%%%%%%%%%%%%%%%%%%%%%%%%%%

c)\textit{Inversion of the RL fractional derivative.\bigskip }

Many authors state rather loosely that "the RL fractional derivative is the
inverse of the RL\ fractional integral". They even adopt the notation: \ $%
_{a}J_{x}^{\alpha }\equiv ~_{a}D_{x}^{-\alpha }$ \cite{Podlubny}. It will be
shown now that this statement is, for the least, ambiguous.\ Hence
Podlubny's notation should preferably be avoided in order to avoid
misunderstandings (Mainardi).\bigskip

c1) \textit{RL Fractional derivative of a fractional integral of same order.}

\begin{equation}
\begin{array}{l}
_{0}D_{x}^{\alpha }~[_{0}J_{x}^{\alpha }~f(x)]=f(x).%
\end{array}
\label{eeq.164}
\end{equation}

\qquad \textit{Proof}:

\qquad \qquad For integer $\alpha =n\geq 1$:

\begin{eqnarray*}
_{0}D_{x}^{n}~[_{0}J_{x}^{n}~f(x)] &=&\frac{d^{n}}{dx^{n}}~\frac{1}{(n-1)!}%
~\int_{0}^{x}dy~(x-y)^{n-1}f(y) \\
&=&\frac{d}{dx}~\int_{0}^{x}dy~f(y)=f(x)
\end{eqnarray*}

\qquad \qquad For non-integer $\alpha $, we enclose this number again
between its nearest neighbor integers, defining the integer $k$ as: $k-1\leq
\alpha <k.$

\qquad \qquad Using now the composition rule (\ref{eeq.152}) for the
fractional integrals, we note the identity:

\begin{equation*}
_{0}J_{x}^{k-\alpha }\cdot \lbrack ~_{0}J_{x}^{\alpha
}~f(x)]=~_{0}J_{x}^{k}~f(x)
\end{equation*}

\qquad \qquad Using now the definition (\ref{eeq.154}) we have:

\begin{equation*}
_{0}D_{x}^{\alpha }~[_{0}J_{x}^{\alpha }~f(x)]=\frac{d^{k}}{dx^{k}}~\left\{
_{0}J_{x}^{k-\alpha }~[_{0}J_{x}^{\alpha }~f(x)]\right\} =\frac{d^{k}}{dx^{k}%
}~_{0}J_{x}^{k}~f(x)=f(x).
\end{equation*}

\qquad \qquad (The result follows from the fact that the last step only
involves operations of integer order $k$, i.e., ordinary derivative and
integral.\bigskip )

From the result (\ref{eeq.164}) one is tempted to conclude that, indeed,
"the RL fractional derivative is the inverse of the RL\ fractional
integral": the fractional RL\ derivative annuls the action of the fractional
integral. The situation is, however, not so clear-cut. It appears from the
following property that the double operation \ $D~J$ is \textit{not
commutative}! Indeed:\bigskip

c2) \textit{Fractional integral of a RL fractional derivative of the same
order.}

\begin{equation}
\begin{array}{l}
_{0}J_{x}^{\alpha }~[_{0}D_{x}^{\alpha }~f(x)]=f(x)-\sum\limits_{j=1}^{k}~ 
\left[ _{0}D_{x}^{\alpha -j}~f(x)\right] _{x=0}~\dfrac{x^{\alpha -j}}{\Gamma
(\alpha -j+1)},\ \ \ \ k-1\leq \alpha <k.%
\end{array}
\label{eeq.165}
\end{equation}

\qquad \textit{Proof.}

\qquad \qquad 
\begin{gather*}
_{0}J_{x}^{\alpha }~\left[ _{0}D_{x}^{\alpha }~f(x)\right] =~\frac{1}{\Gamma
(\alpha )}~\int_{0}^{x}dy~(x-y)^{\alpha -1}~_{0}D_{y}^{\alpha }~f(y) \\
=\frac{1}{\Gamma (\alpha )}~\int_{0}^{x}dy~\frac{1}{\alpha }~\left[ \frac{d}{%
dx}(x-y)^{\alpha }\right] ~_{0}D_{y}^{\alpha }~f(y) \\
=\frac{d}{dx}~\left\{ \frac{1}{\Gamma (\alpha +1)}~\int_{0}^{x}dy~(x-y)^{%
\alpha }~_{0}D_{y}^{\alpha }~f(y)\right\} \\
=\frac{d}{dx}~\left\{ \frac{1}{\Gamma (\alpha +1)}~\int_{0}^{x}dy~(x-y)^{%
\alpha }~\frac{d^{k}}{dy^{k}}~_{0}J_{y}^{k-\alpha }~f(y)\right\} .~
\end{gather*}%
\qquad

\qquad \qquad A partial integration yields:

\begin{eqnarray*}
&=&\frac{d}{dx}\left\{ \frac{1}{\Gamma (\alpha )}\int_{0}^{x}dy~(x-y)^{%
\alpha -1}~\frac{d^{k-1}}{dy^{k-1}}~_{0}J_{y}^{k-\alpha }~f(y)\right. \\
&&\left. -\frac{1}{\Gamma (\alpha +1)}~x^{\alpha }~\left[ \frac{d^{k-1}}{%
dx^{k-1}}~_{0}J_{x}^{k-\alpha }f(x)\right] _{x=0}\right\} .~
\end{eqnarray*}

\qquad \qquad Continuing the partial integrations, and using Eqs. (\ref%
{eeq.150}) and (\ref{eeq.152}), we end up with:

\begin{gather*}
=\frac{d}{dx}\left\{ \frac{1}{\Gamma (\alpha -k+1)}~\int_{0}^{x}dy~(x-y)^{%
\alpha -k}~_{0}J_{y}^{k-\alpha }~f(y)\right. \\
\left. -\sum\limits_{j=1}^{k}\left[ \frac{d^{k-j}}{dx^{k-j}}%
~_{0}J_{x}^{k-\alpha }f(x)\right] _{x=0}~\frac{x^{\alpha -j+1}}{\Gamma
(\alpha -j+2)}\right\} \\
=\frac{d}{dx}\left\{ ~_{0}J_{x}^{\alpha -k+1}\cdot ~_{0}J_{x}^{k-\alpha
}f(x)\right. \\
\left. -\sum\limits_{j=1}^{k}\left[ \frac{d^{k-j}}{dx^{k-j}}%
~_{0}J_{x}^{k-\alpha }f(x)\right] _{x=0}~\frac{x^{\alpha -j+1}}{\Gamma
(\alpha -j+2)}\right\} \\
=~\frac{d}{dx}\left\{ _{0}J_{x}^{1}f(x)-\sum\limits_{j=1}^{k}\left[ \frac{%
d^{k-j}}{dx^{k-j}}~_{0}J_{x}^{k-\alpha }f(x)\right] _{x=0}~\frac{x^{\alpha
-j+1}}{\Gamma (\alpha -j+2)}\right\} \\
=f(x)-\sum\limits_{j=1}^{k}\left[ \frac{d^{k-j}}{dx^{k-j}}%
~_{0}J_{x}^{k-\alpha }f(x)\right] _{x=0}~\frac{x^{\alpha -j}}{\Gamma (\alpha
-j+1)},\ \ \ \ QED
\end{gather*}

It is precisely because of this non-commutativity of integration and
differentiation that one cannot state unambiguously that one operation is
the inverse of the other.\bigskip

d) \textit{General composition rules.}

\qquad In all subsequent formulae we take the lower limit \ $a=0$.\ We
simplify the notations as follows: \ $_{0}J_{x}^{\alpha }=J^{\alpha },$ \ \ $%
_{0}D_{x}^{\alpha }=D^{\alpha }.$ The proofs of the following results are
very similar to the previous ones \cite{Podlubny}.\bigskip

d1) \textit{Composition of two fractional integrals. \ }[see (\ref{eeq.152})]

\begin{equation}
\begin{array}{c}
J^{\alpha }~J^{\beta }~f(x)=J^{\beta }~J^{\alpha }~f(x)=J^{\alpha +\beta
}~f(x)%
\end{array}
\label{eeq.166}
\end{equation}

d2) \textit{RL derivative of fractional integral.}

\begin{equation}
\begin{array}{c}
D^{\alpha }~J^{\beta }~f(x)=J^{\beta -\alpha }~f(x),\ \ \ \ \ \ 0<\alpha
<\beta \\ 
D^{\alpha }~J^{\beta }~f(x)=D^{\alpha -\beta }~f(x),\ \ \ \ \ 0<\beta <\alpha%
\end{array}
\label{eeq.167}
\end{equation}%
\bigskip

d3) \textit{Fractional integral of RL derivative.}

\begin{equation}
\begin{array}{c}
J^{\alpha }~D^{\beta }~f(x)=D^{\beta -\alpha }~f(x)-\sum\limits_{j=1}^{n}~ 
\left[ D^{\beta -j}~f(x)\right] _{x=0}~\dfrac{x^{\alpha -j}}{\Gamma (\alpha
-j+1)},\  \\ 
0<\alpha <\beta ,\ \ \ \ n-1\leq \beta <n%
\end{array}
\label{eeq.168}
\end{equation}

\begin{equation}
\begin{array}{c}
J^{\alpha }~D^{\beta }~f(x)=J^{\alpha -\beta }~f(x)-\sum\limits_{j=1}^{n}~ 
\left[ D^{\beta -j}~f(x)\right] _{x=0}~\dfrac{x^{\alpha -j}}{\Gamma (\alpha
-j+1)},\  \\ 
0<\beta <\alpha ,\ \ \ n-1\leq \beta <n%
\end{array}
\label{eeq.168a}
\end{equation}%
\bigskip

d4) \textit{Composition of RL derivatives.}

\begin{equation}
\begin{array}{l}
\begin{array}{c}
_{0}D_{x}^{\alpha }\cdot ~_{0}D_{x}^{\beta }~f(x)=~_{0}D_{x}^{\alpha +\beta
}~f(x)-\sum\limits_{j=1}^{n}\left[ ~_{0}D_{x}^{\beta -j}f(x)\right] _{x=0}~%
\dfrac{x^{-\alpha -j}}{\Gamma (-\alpha -j+1)}, \\ 
_{0}D_{x}^{\beta }\cdot ~_{0}D_{x}^{\alpha }~f(x)=~_{0}D_{x}^{\alpha +\beta
}~f(x)-\sum\limits_{j=1}^{m}\left[ ~_{0}D_{x}^{\alpha -j}f(x)\right] _{x=0}~%
\dfrac{x^{-\beta -j}}{\Gamma (-\beta -j+1)} \\ 
\ m-1\leq \alpha <m,\ \ \ \ n-1\leq \beta <n%
\end{array}%
\ \ \ 
\end{array}
\label{eeq.169}
\end{equation}%
\bigskip

\newpage

\qquad \textbf{Integral transforms of fractional operators.\bigskip }

\qquad \textbf{A) \ Fourier transform\bigskip\ }

\qquad The Fourier transform involves integrals ranging from $-\infty $ to $%
+\infty $. This implies a number of technical assumptions about the
regularity of the functions appearing in the theory: we do not discuss them
here, assuming that all conditions are satisfied for the existence of the
objects of the forthcoming equations.\ We begin with the following result:

\begin{equation}
\begin{array}{c}
_{-\infty }D_{x}^{\alpha }~e^{ikx}=(ik)^{\alpha }~e^{ikx}.%
\end{array}
\label{eeq.170}
\end{equation}

\qquad \textit{Proof:}

\qquad \qquad 
\begin{gather*}
F_{+}(x)\equiv ~_{-\infty }D_{x}^{\alpha }~e^{ikx}=\frac{1}{\Gamma (m-\alpha
)}~\frac{d^{m}}{dx^{m}}~\int_{-\infty }^{x}dy~(x-y)^{-1+m-\alpha }~e^{iky} \\
=\frac{1}{\Gamma (m-\alpha )}~\frac{d^{m}}{dx^{m}}~e^{ikx}\int_{0}^{\infty
}dt~t^{-1+m-\alpha }~e^{-ikt} \\
,\ \ \ \ m-1\leq \alpha <m
\end{gather*}

\qquad \qquad where we set $x-y=t$. We now note that, from the definition of 
$m$, we have: \ $0<m-\alpha \leq 1$. We may thus use the tabulated integral:

\begin{equation*}
\int_{0}^{\infty }d\xi ~\xi ^{\gamma -1}~e^{-ik\xi }=(ik)^{-\gamma }~\Gamma
(\gamma ),\ \ \ \ \ 0<\gamma <1
\end{equation*}

\qquad \qquad Hence:

\begin{eqnarray*}
_{-\infty }D_{x}^{\alpha }~e^{ikx} &=&\frac{1}{\Gamma (m-\alpha )}~\frac{%
d^{m}}{dx^{m}}~e^{ikx}~(ik)^{-m+\alpha }~\Gamma (m-\alpha ) \\
&=&(ik)^{\alpha }~e^{ikx},\ \ \ \ \ \ \ QED
\end{eqnarray*}

This expression is not very convenient in applications.\ On one hand, the
appearance of $i^{\alpha }$ is somewhat awkward.\ But more important, if we
wish to use this result in relation with the Fourier transform, we need to
cover the whole range ($-\infty ,\ \infty )$, and not only the left range ($%
-\infty ,\ x$). We therefore consider also the right RL\ fractional
derivative (\ref{eeq.156}):

\begin{equation}
\begin{array}{c}
_{x}D_{\infty }^{\alpha }~e^{ikx}=(-ik)^{\alpha }~e^{ikx}.%
\end{array}
\label{eeq.171}
\end{equation}

\qquad \textit{Proof:}

\qquad \qquad 
\begin{eqnarray*}
F_{-}(x) &\equiv &~_{x}D_{\infty }^{\alpha }~e^{ikx}=\frac{(-1)^{m}}{\Gamma
(m-\alpha )}~\frac{d^{m}}{dx^{m}}~\int_{x}^{\infty }dy~(y-x)^{-1+m-\alpha
}~e^{iky} \\
&=&\frac{(-1)^{m}}{\Gamma (m-\alpha )}~\frac{d^{m}}{dx^{m}}~\int_{-\infty
}^{-x}d\eta ~(-\eta -x)^{-1+m-\alpha }~e^{-ik\eta }
\end{eqnarray*}

\qquad \qquad Setting $\xi =-x$, we have:

\begin{equation*}
F_{-}(x)=\frac{1}{\Gamma (m-\alpha )}~\frac{d^{m}}{d\xi ^{m}}~\int_{-\infty
}^{\xi }d\eta ~(-\eta +\xi )^{-1+m-\alpha }~\left( e^{ik\eta }\right) ^{\ast
}=F_{+}^{\ast }(\xi )
\end{equation*}

\qquad \qquad Thus:

\begin{equation*}
F_{-}(x)=F_{+}^{\ast }(\xi )=(-ik)^{\alpha }~e^{-ik\xi }=(-ik)^{\alpha
}~e^{ikx}\ \ \ \ \ \ \ \ \ QED
\end{equation*}

In order to adapt the previous results to the Fourier transform, in
particular to cover the whole range $(-\infty ,~+\infty )$, we define a new
type of fractional derivative, which is totally symmeric.\ It is justified
by the linearity property of the RL\ derivative (\ref{eeq.156a}):

\begin{equation}
\begin{array}{c}
D_{|x|}^{\alpha }~f(x)=-\dfrac{1}{2~\cos \dfrac{\pi \alpha }{2}}~\left[
~_{-\infty }D_{x}^{\alpha }+~_{x}D_{\infty }^{\alpha }\right] ~f(x)%
\end{array}
\label{eeq.172}
\end{equation}

This is called the \textbf{Riesz fractional derivative.}

The usefulness of this concept appears obviously in the following result,
expressing the \textit{Fourier transform of the Riesz derivative}:

\begin{equation}
\begin{array}{c}
\mathcal{F}~\left\{ D_{|x|}^{\alpha }~f(x)\right\} =-|k|^{\alpha }~\widehat{f%
}(k)%
\end{array}
\label{eeq.173}
\end{equation}

where \ $\widehat{f}(k)$ denotes the Fourier transform of \ $f(x)$.

\qquad \textit{Proof:}

\qquad \qquad 
\begin{eqnarray*}
\mathcal{F}~\left\{ D_{|x|}^{\alpha }~f(x)\right\} &=&\mathcal{F}~\left\{
D_{|x|}^{\alpha }~\int dk^{\prime }~e^{-ik^{\prime }x}~\widehat{f}(k^{\prime
})\right\} \\
&=&-\mathcal{F~}\left\{ \int dk^{\prime }~|k^{\prime }|^{\alpha
}~e^{-ik^{\prime }x}~\widehat{f}(k^{\prime })\right\} \\
&=&-\frac{1}{2\pi }~\int dx~e^{ikx}~\int dk^{\prime }~|k^{\prime }|^{\alpha
}~e^{-ik^{\prime }x}~\widehat{f}(k^{\prime }) \\
&=&-~|k|^{\alpha }~\widehat{f}(k),\ \ \ \ \ QED
\end{eqnarray*}

The result (\ref{eeq.173}) ends our quest of the Fourier transform of the
(Riesz) fractional derivative of a function, expressed in terms of the
Fourier transform of that function.\bigskip

\textbf{B) Laplace transform\bigskip }

The (direct) Laplace transform of a function involves an integral over the
variable from $0$ to $\infty .$ It is therefore particularly well adapted to
the study of functions of \textit{time}, when the latter is restricted to
positive values. The two properties of the Laplace transform that are most
useful in applications (especially for the solution of differential
equations are:

\textit{The convolution theorem: }for any two functions\textit{\ }$f(t)$, $%
g(t)$ of time, defined in the range ($0,\ \infty $), their convolution is
defined as:

\begin{equation}
(f\ast g)=\int_{0}^{t}d\tau ~f(t-\tau )~g(\tau )  \label{eeq.174}
\end{equation}

\textit{Laplace transform of the (ordinary) derivative:}

\begin{equation}
\mathcal{L~}\left\{ f^{(n)}(t)\right\} =s^{n}~\widetilde{f}%
(s)-\sum\limits_{j=0}^{n-1}~s^{n-j-1}~f^{(j)}(0)  \label{eeq.175}
\end{equation}

where $f^{(n)}(t)$ is the $n$-th time derivative of $f(t)$, and $\widetilde{f%
}(s)$ is the Laplace transform of $f(t)$. The explicit appearance of the
functions $f^{(j)}(0)$ in this formula makes the Laplace transform a
particularly useful tool for the solution of initial value problems. We may
now note that the value $j=0$ in the sum of the right hand side is
"special": it corresponds to $f^{(0)}(t)=f(t)$, whereas all other terms in
the sum contain true derivatives:

\begin{equation}
\mathcal{L~}\left\{ f^{(n)}(t)\right\} =s^{n}~\widetilde{f}%
(s)-s^{n-1}~f(0)-\sum\limits_{j=1}^{n-1}~s^{n-j-1}~f^{(j)}(0)
\label{eeq.175a}
\end{equation}

We now consider the Laplace transforms of the various objects of fractional
calculus.\bigskip

\textit{a) Laplace transform of the fractional integral.\bigskip }

This is easily obtained by noting that the RL integral can be viewed as a
convolution:

\begin{equation}
_{0}J_{t}^{\beta }~f(t)=\frac{1}{\Gamma (\beta )}~\int_{0}^{t}d\tau ~(t-\tau
)^{\beta -1}~f(\tau )=t^{\beta -1}\ast \frac{1}{\Gamma (\beta )}~f(t)~
\label{eeq.176}
\end{equation}

Hence, noting the well-known result:

\begin{equation}
\mathcal{L~}(t^{\beta -1})=\Gamma (\beta )~s^{-\beta },\ \ \ \ \beta >0,
\label{eeq.177a}
\end{equation}

we find immediately, using the convolution theorem (\ref{eeq.174}):

\begin{equation}
\begin{array}{c}
\mathcal{L~}\left\{ ~_{0}J_{t}^{\beta }~f(t)\right\} =s^{-\beta }~\widetilde{%
f}(s).%
\end{array}
\label{eeq.178}
\end{equation}%
\bigskip

\textit{b) Laplace transform of the RL derivative.\bigskip }

We write the RL derivative (\ref{eeq.154}) in the form:

\begin{equation}
_{0}D_{t}^{\beta }~f(t)\equiv \frac{d^{m}}{dt^{m}}~g(t),\ \ \ \ m-1\leq
\beta <m  \label{eeq.179}
\end{equation}

with:

\begin{equation}
g(t)=~_{0}J_{t}^{m-\beta }~f(t)  \label{eeq.180}
\end{equation}

Using Eq. (\ref{eeq.175}) for the $m$-th (ordinary) derivative, we have: \ \ 

\begin{equation}
\mathcal{L}~\left\{ ~_{0}D_{t}^{\beta }~f(t)\right\} =s^{m}~\widetilde{g}%
(s)-\sum\limits_{j=0}^{m-1}~s^{j}~g^{(m-j-1)}(0)  \label{eeq.181}
\end{equation}

Next, we calculate:

\begin{equation}
g^{(m-j-1)}=\frac{d^{m-j-1}}{dt^{m-j-1}}~\frac{1}{\Gamma (m-\beta )}%
~\int_{0}^{t}d\tau ~(t-\tau )^{m-\beta -1}~f(\tau )~  \label{eeq.182}
\end{equation}

Using Eq. (\ref{eeq.167}), we find:

\begin{equation}
g^{(m-j-1)}=~_{0}D_{t}^{m-j-1}~_{0}J_{t}^{m-\beta }~f(t)=\mathcal{O}^{-\beta
+j+1}~f(t),  \label{eeq.183}
\end{equation}

where:

\begin{equation}
\mathcal{O}^{\mu }=\left\{ 
\begin{array}{c}
_{0}J_{t}^{\mu },\ \ \ \ \mu >0 \\ 
_{0}D_{t}^{\mu },\ \ \ \ \mu <0%
\end{array}%
\right. .  \label{eeq.184}
\end{equation}

From the two constraints: $m-1<\beta <m$ and \ $0\leq j\leq m-1$, we deduce:

\begin{itemize}
\item when $j=m-1$, we have $-\beta +j+1=m-\beta >0$, hence: \ $\mathcal{O}%
^{\mu }=~_{0}J_{t}^{\mu }$ ,

\item when $j<m-1$, we have $-\beta +j+1<0$, hence: \ $\mathcal{O}^{\mu
}=~_{0}D_{t}^{\mu }.$
\end{itemize}

Note also, from Eq. (\ref{eeq.178}):

\begin{equation*}
\widetilde{g}(s)=\mathcal{L}~\left\{ ~_{0}J_{t}^{-\beta +m}f(t)\right\}
=s^{\beta -m}~\widetilde{f}(s)
\end{equation*}

We thus obtain the final form of the Laplace transform of the RL fractional
derivative:

\begin{equation}
\begin{array}{c}
\mathcal{L}~\left\{ ~_{0}D_{t}^{\beta }~f(t)\right\} =s^{\beta }~\widetilde{f%
}(s)-s^{m-1}~\left[ ~_{0}J_{t}^{-\beta +m}~f(t)\right] _{t=0}-\sum%
\limits_{j=0}^{m-2}s^{j}~\left[ ~_{0}D_{t}^{\beta -j-1}~f(t)\right] _{t=0}.%
\end{array}
\label{eeq.185}
\end{equation}

The analogy with Eq. (\ref{eeq.175a}) is striking. The first term has the
same form. In the second term the initial value of the function $f(0)$ is
replaced, for non-integer $\beta ,$ by the initial value of the fractional 
\textit{integral }$_{0}J_{t}^{-\beta +m}~f(t)$, \ whereas in the terms under
the sum, the initial values of the ordinary derivatives are replaced by the
initial values of the corresponding RL fractional derivatives. In spite of
this analogy, the RL time-derivative is not easily adapted to the physical
applications. The Laplace transform is well known to be a valuable tool for
the solution of \textit{initial value problems of differential equations},
because it incorporates these initial values in Eq. (\ref{eeq.175a}).\ If we
now consider \textit{initial value problems of fractional differential
equations involving \textbf{RL} fractional derivatives with respect to time}%
, we are faced with a serious difficulty if we try to use a Laplace
transformation. The specification of (physical) values of the unknown
function and of a sufficient number of its derivatives at $t=0$, this does
not allow us to determine the correponding initial valued of the fractional
integrals and derivatives appearing in Eq. (\ref{eeq.185}\textit{)}. This
difficulty has been overcome by M.\ Caputo in \cite{Caputo 1}, \cite{Caputo
2} by introducing an alternative definition of the fractional derivative,
that is particularly useful for functions of a variable restricted to a
finite range, such as time $\in (0,\ t)$.\bigskip

\textbf{The Caputo fractional derivative.\bigskip }

The Caputo fractional derivative is defined as follows:

\begin{equation}
\begin{array}{c}
_{0}^{C}D_{t}^{\beta }~f(t)=\dfrac{1}{\Gamma (n-\beta )}~\int%
\nolimits_{0}^{t}d\tau ~\dfrac{1}{(t-\tau )^{\beta +1-n}}~\dfrac{d^{n}}{%
d\tau ^{n}}~f(\tau ),\ \ \ \ n-1\leq \beta <n.%
\end{array}
\label{eeq.186}
\end{equation}

Thus, the only difference with the RL derivative Eq.(156) is the commutation
of the integration and the differentiation. The Caputo derivative results
from a composition of a fractional integral with an ordinary derivative:

\begin{equation}
_{0}^{C}D_{t}^{\beta }~f(t)=~_{0}J_{t}^{n-\beta }~\left[ \frac{d^{n}}{dt^{n}}%
~f(t)\right]  \label{eeq.186a}
\end{equation}

Let us introduce a simplified notation:

\begin{equation}
_{0}J_{t}^{\beta }\rightarrow J^{\beta },\ \ \ \ _{0}D_{t}^{\beta
}\rightarrow D^{\beta },\ \ \ \ _{0}^{C}D_{t}^{\beta }\rightarrow D_{\ast
}^{\beta }  \label{eeq.187}
\end{equation}

The Caputo derivative is related to the Riemann-Liouville one as follows 
\cite{Gor-Main 1997}:

\begin{equation}
\begin{array}{c}
D_{\ast }^{\beta }~f(t)=D^{\beta }~f(t)-\sum\limits_{j=0}^{n-1}~f^{(j)}(0)~%
\dfrac{t^{j-\beta }}{\Gamma (j-\beta +1)}.%
\end{array}
\label{eeq.188}
\end{equation}

We give here the proof for the case $0<\beta <1$, which implies $n=1$. The
proof of the general case is similar, but longer.\bigskip

\qquad 
\begin{eqnarray*}
\Gamma (1-\beta )~D^{\beta }f(t) &=&\frac{d}{dt}~\int_{0}^{t}d\tau ~\frac{%
f(\tau )}{(t-\tau )^{\beta }}=\left. \left[ \frac{f(\tau )}{(t-\tau )}\right]
\right\vert _{\tau =t}+\int_{0}^{t}d\tau ~\left[ \frac{d}{dt}~(t-\tau
)^{-\beta }\right] ~f(\tau ) \\
&=&\left. \left[ \frac{f(\tau )}{(t-\tau )}\right] \right\vert _{\tau
=t}-\left. \left[ \frac{f(\tau )}{(t-\tau )}\right] \right\vert _{\tau
=0}^{\tau =t}+\int_{0}^{t}d\tau ~\frac{1}{(t-\tau )^{\beta }}~\frac{d}{d\tau 
}f(\tau ) \\
&=&t^{-\beta }~f(0)+\int_{0}^{t}d\tau ~\frac{1}{(t-\tau )^{\beta }}~\frac{d}{%
d\tau }f(\tau );
\end{eqnarray*}

\qquad Hence, finally:

\begin{equation}
D_{\ast }^{\beta }~f(t)=D^{\beta }~f(t)-\frac{1}{\Gamma (1-\beta )}%
~t^{-\beta },\ \ \ \ \ 0<\beta <1  \label{eeq.189}
\end{equation}

\qquad which agrees with Eq. (\ref{eeq.188}).\bigskip

An important property, which is obvious from Eq. (\ref{eeq.186}), and also
from (\ref{eeq.189}), combined with (\ref{eeq.158}), is:

\begin{equation}
D_{\ast }^{\beta }~1=0,  \label{eeq.190}
\end{equation}

which makes the Caputo derivative closer to the ordinary derivative.\bigskip

We now consider the composition of the Caputo derivative $D_{\ast }^{\beta }$
with the fractional integral of the same order $J^{\beta }$. A first
relation is immediately obtained from the combination of Eqs. (\ref{eeq.164}%
) and (\ref{eeq.188}):

\begin{equation}
\begin{array}{c}
D_{\ast }^{\beta }~J^{\beta }~f(t)=f(t)-\sum\limits_{j=1}^{n-1}~\left[
J^{\beta -j}~f(t)\right] _{t=0}~\dfrac{t^{j-\beta }}{\Gamma (j-\beta +1)}.%
\end{array}
\label{eeq.191}
\end{equation}

Commuting the integral and the derivative and using successively Eqs. (\ref%
{eeq.188}), (\ref{eeq.165}) and (\ref{eeq.151}) we obtain:

\begin{gather*}
J^{\beta }~D_{\ast }^{\beta }~f(t)=J^{\beta }~D^{\beta
}~f(t)-\sum\limits_{j=0}^{n-1}~\frac{1}{\Gamma (j-\beta +1)}~\left[ \frac{%
d^{j}}{dt^{j}}~f(t)\right] _{t=0}~J^{\beta }~t^{j-\beta } \\
=f(t)-\sum\limits_{k=1}^{n}~\left[ D^{\beta -k}~f(t)\right] _{t=0}~\frac{%
t^{\beta -k}}{\Gamma (\beta -k+1)}-\sum\limits_{j=0}^{n-1}~\left[ \frac{%
d^{j}}{dt^{j}}~f(t)\right] _{t=0}~\frac{t^{j}}{\Gamma (j+1)}
\end{gather*}

and finally (with $k\rightarrow j+1$):

\begin{equation}
\begin{array}{c}
J^{\beta }~D_{\ast }^{\beta }~f(t)=f(t)-\sum\limits_{j=0}^{n-1}~\left\{ 
\left[ D^{\beta -j-1}f(t)\right] _{t=0}~\dfrac{t^{\beta -j-1}}{\Gamma (\beta
-j)}+f^{(j)}(0)~\dfrac{t^{j}}{\Gamma (j+1)}\right\}%
\end{array}
\label{eeq.192}
\end{equation}

Eqs. (\ref{eeq.191}) and (\ref{eeq.192}) show that the Caputo derivative is 
\textbf{not} an inverse of the fractional integral, either to the left or to
the right. General composition rules of the type of Eqs. (\ref{eeq.166}) - (%
\ref{eeq.169}) can be derived for the Caputo derivative, using the previous
results; they are rather complicated, and will not be written down
explicitly. We now consider the most important property of the Caputo
derivative.\bigskip

\textit{Laplace transform of the Caputo derivative.\bigskip }

The result is easily obtained by combining the definition (\ref{eeq.186a})
with Eqs. (\ref{eeq.178}) and (\ref{eeq.175a}):

\begin{eqnarray*}
\mathcal{L}~\left[ ~_{0}^{C}D_{t}^{\beta }~f(t)\right] &=&\mathcal{L}~\left[
~_{0}J_{t}^{n-\beta }\left( \frac{d^{n}f(t)}{dt^{n}}\right) \right]
=s^{\beta -n}~\mathcal{L}~\left( \frac{d^{n}f(t)}{dt^{n}}\right) \\
&=&s^{\beta -n}~\left[ s^{n}~\widetilde{f}(s)-s^{n-1}~f(0)-\sum%
\limits_{k=0}^{n-2}~s^{j}~f^{(n-k-1)}(0)\right] ,
\end{eqnarray*}

and finally, setting $n-k-1=j$:

\begin{equation}
\begin{array}{c}
\mathcal{L}~\left[ ~_{0}^{C}D_{t}^{\beta }~f(t)\right] =s^{\beta }~%
\widetilde{f}(s)-s^{\beta -1}~f(0)-\sum\limits_{j=1}^{n-1}~s^{\beta
-j-1}~f^{(j)}(0),\ \ \ \ \ n-1<\beta <n.%
\end{array}
\label{eeq.193}
\end{equation}

This result has exactly the same form as the Laplace transform of an
ordinary derivative, Eq. (\ref{eeq.175a}), with $n\rightarrow \beta $.
Contrary to the Laplace transform of the RL derivative (\ref{eeq.185}), the
Laplace transform of the Caputo derivative only involves the initial values
of the function $f(t)$ and of its ordinary derivatives, instead of the
initial values of the fractional integral and of the RL fractional
derivatives. In any physical differential equation, the former are the given
input. This property makes the Caputo derivative an invaluable tool for
solving initial value problems of fractional differential equations.

\section*{Acknowledgments}

We gratefully acknowledge extremely fruitful discussions with Profs. J.\
Klafter, F.\ Mainardi, R.\ Gorenflo, P.\ Grigolini and R.\ Sanchez.

\end{document}